\documentclass[10pt]{article}
\usepackage{epstopdf,amsfonts,amsmath,stmaryrd,mathabx,amssymb,mathdots}

\newcommand\encadremath[1]{\vbox{\hrule\hbox{\vrule\kern8pt
\vbox{\kern8pt \hbox{$\displaystyle #1$}\kern8pt}
\kern8pt\vrule}\hrule}}
\def\enca#1{\vbox{\hrule\hbox{
\vrule\kern8pt\vbox{\kern8pt \hbox{$\displaystyle #1$}
\kern8pt} \kern8pt\vrule}\hrule}}

\numberwithin{equation}{section}

\newcommand\figureframex[3]{
\begin{figure}[bth]
\hrule\hbox{\vrule\kern8pt
\vbox{\kern8pt \vbox{
\begin{center}
{\mbox{\epsfxsize=#1.truecm\epsfbox{#2}}}
\end{center}
\caption{#3}
}\kern8pt}
\kern8pt\vrule}\hrule
\end{figure}
}
\newcommand\figureframey[3]{
\begin{figure}[bth]
\hrule\hbox{\vrule\kern8pt
\vbox{\kern8pt \vbox{
\begin{center}
{\mbox{\epsfysize=#1.truecm\epsfbox{#2}}}
\end{center}
\caption{#3}
}\kern8pt}
\kern8pt\vrule}\hrule
\end{figure}
}

\makeatletter
\@addtoreset{equation}{section}
\makeatother
\newtheorem{theorem}{Theorem}[section]

\newtheorem{remark}{Remark}[section]
\newtheorem{proposition}{Proposition}[section]
\newtheorem{lemma}{Lemma}[section]
\newtheorem{corollary}{Corollary}[section]
\newtheorem{definition}{Definition}[section]
\newtheorem{assumption}{Assumption}[section]
\def\br{\begin{remark}\rm\small}
\def\er{\end{remark}}
\def\bt{\begin{theorem}}
\def\et{\end{theorem}}
\def\bd{\begin{definition}}
\def\ed{\end{definition}}
\def\bp{\begin{proposition}}
\def\ep{\end{proposition}}
\def\bl{\begin{lemma}}
\def\el{\end{lemma}}
\def\bc{\begin{corollary}}
\def\ec{\end{corollary}}
\def\beaq{\begin{eqnarray}}
\def\eeaq{\end{eqnarray}}

\newcommand{\dd}{\mathrm{d}}

\newcommand{\beq}{\begin{equation}}
\newcommand{\eeq}{\end{equation}}
\newcommand{\bea}{\begin{eqnarray}}
\newcommand{\eea}{\end{eqnarray}}

\newcommand{\Tr}{{\rm Tr}\,}
\newcommand{\e}{{\rm e}}

\newcommand{\Tau}{{\cal T}}

\newcommand{\sheet}[2]{\stackrel{{#2}}{#1}}

\newcommand{\Res}{\mathop{\,\rm Res\,}}
\usepackage[pdftex]{hyperref}
\hypersetup{colorlinks,urlcolor=magenta,citecolor=red,linkcolor=blue,filecolor=black}

\begin{document}

\sloppy

\pagestyle{empty}
\addtolength{\baselineskip}{0.20\baselineskip}

\hfill IPHT t13/281, CRM 3332

\vspace{0.5cm}

\begin{center}
{\Large \textbf{Rational differential systems, loop equations, and application to the $q$-th reductions of KP}}
\end{center}

\vspace{15pt}

\begin{center}
{\sl Michel Berg\`ere}\footnote{Institut de Physique Th\'eorique, CEA Saclay, \href{mailto:michel.bergere@cea.fr}{\textsf{michel.bergere@cea.fr}}.},\hspace*{0.05cm}{\sl Ga\"etan Borot}\footnote{Section de Math\'ematiques, Universit\'e de Gen\`eve ; MPI f\"ur Mathematik, Bonn ; MIT Maths Department, \href{mailto:gbort@mpim-bonn.mpg.de}{\textsf{gborot@mpim-bonn.mpg.de}}. This work benefited from the support of the Fonds Europ\'een S16905 (UE7--CONFRA), the Swiss NSF, a Forschungsstipendium of the Max-Planck-Gesellschaft, and the Simons foundation.},\hspace*{0.05cm}{\sl Bertrand Eynard}\footnote{Institut de Physique Th\'eorique, CEA Saclay ; Centre de Re\-cher\-ches Ma\-th\'e\-ma\-ti\-ques, Montr\'eal, QC, Canada, \href{mailto:bertrand.eynard@cea.fr}{\textsf{bertrand.eynard@cea.fr}}. B.E. thanks the CRM, the FQRNT grant from the Qu\'ebec government, Piotr Su\l kowski and the ERC starting grant Fields-Knots.}

\vspace{10pt}

\end{center}

\vspace{20pt}
\begin{center}
{\bf Abstract}
\end{center}

%

\vspace{0.5cm}
To any solution of a linear system of differential equations, we associate a kernel, correlators satisfying a set of loop equations, and in presence of isomonodromic parameters, a Tau function. We then study their semiclassical expansion (WKB type expansion in powers of the weight $\hbar$ per derivative) of these quantities. When this expansion is of topological type (TT), the coefficients of expansions are computed by the topological recursion with initial data given by the semiclassical spectral curve of the linear system. This provides an efficient algorithm to compute them at least when the semiclassical spectral curve is of genus $0$. TT is a non trivial property, and it is an open problem to find a criterion which guarantees it is satisfied. We prove TT and illustrate our construction for the linear systems associated to the $q$-th reductions of KP -- which contain the $(p,q)$ models as a specialization. 

\vspace{26pt}
\pagestyle{plain}
\setcounter{page}{1}


\section{Introduction}

Let $\mathbf{L}(x)$ be a $d \times d$ matrix with entries being rational functions of $x$, and $\mathcal{P}$ the set of poles of $\mathbf{L}$. We consider matrix $\mathbf{\Psi}(x)$ (whose columns form a basis of solutions) of the differential system:
\beq
\label{1} \hbar\,\partial_x \mathbf{\Psi}(x) = \mathbf{L}(x)\mathbf{\Psi}(x),
\eeq
i.e. $\mathbf{\Psi}(x)$ is a $d \times d$ invertible matrix solving $\eqref{1}$. It is well-known that $\mathbf{\Psi}(x)$ is locally holomorphic in $\widehat{\mathbb{C}}\setminus\mathcal{P}$. The matrix $\mathbf{L}$ (and thus $\mathbf{\Psi}$) may depend on $\hbar$, and on extra parameters $t_{\alpha}$. The goal of this article is to establish a set of loop equations satisfied by some quantities built out of $\mathbf{\Psi}$, and analyze their consequences, especially for small $\hbar$ expansions -- whether at the formal level, or at the level of asymptotics. Very often, if one wishes to study the asymptotic behavior in some parameter $x$ or $t_{\alpha}$ of a differential system, one can introduce by hand a parameter $\hbar$ to put the system in the form \eqref{1}, so that the asymptotic regime of interest correspond to $\hbar \rightarrow 0$.

\subsection{Outline}

The paper is organized in three parts.

\vspace{0.2cm}

Firstly, in Section~\ref{S1}, we associate to any $d \times d$ invertible matrix $\mathbf\Psi(x)$ solution of a linear differential system:
\begin{itemize}
\item[$\bullet$] a $d \times d$ matrix $\mathbf{K}(x,y)$, called \emph{kernel}.
\item[$\bullet$] an infinite family of functions $\mathcal{W}_k(\sheet{x_1}{a_1},\ldots,\sheet{x_k}{a_k})$, indexed by integer $a_1,\ldots,a_k \in \ldbrack 1,d\rdbrack$, called \emph{k-points correlators}, or shortly, correlators.
\end{itemize}
and we show that the $k$-point correlators satisfy a set of linear equations (Theorem~\ref{linearL}) and a set of quadratic equations (Theorem~\ref{quda}).  We use the name \emph{loop equations} to refer collectively to those set of equations. We also introduce a notion of "insertion operator" (Definition~\ref{defdeltay}) allowing the derivation of $k$-linear loop equations for $k \leq d$ (the size of the differential system) from the master ones. These results are of purely algebraic nature and hold for any system \eqref{1}. When $\mathbf{L}$ depends on a set of parameters $\vec{t}$ preserving the monodromy of the solutions, we can also associate to $\mathbf{\Psi}(x,\vec{t})$ a Tau function $\mathcal{T}(\vec{t})$, defined up to a constant prefactor.

Secondly, in Section~\ref{S3}, we study the semiclassical expansion in powers of $\hbar$ and describe in detail the monodromy of its coefficients (Section~\ref{S32}-\ref{S34}). We introduce in Definition~\ref{detype} the notion of "expansion of topological type" -- also referred to as the TT property -- and show that the expansion can be computed by the topological recursion of \cite{EOFg} when the TT property holds. In practice, the main consequence of our theory is Theorem~\ref{topro}, and  in presence of isomonodromic times, this also allows the computation of the expansion of $\ln \mathcal{T}(\vec{t})$ (Corollary \ref{co42}).

Finally, in Section~\ref{S4}, we apply our theory to the linear system associated to the $q$-th reduction of KP, and illustrate it more specifically with examples of the $(p,q)$ models (Section~\ref{S6}). As a motivation, those hierarchies are believed to describe the algebraic critical edge behavior that can be reached in the two hermitian matrix model, and universality classes of $2d$ quantum gravity coupled to conformal field theories \cite{GM90,doug,GrosMig,dFZJ}. In any $q$-th reduction of KP, we show (\S~\ref{sysm}-\ref{Spropole}) that the TT property holds, and that our Theorem~\ref{topro} can be applied.

\subsubsection{Comments}

The earlier work \cite{BEdet} described the construction of Section~\ref{S1} for general $2 \times 2$ rational systems, but implicitly assumed the TT property. It was illustrated for $(2m + 1,2)$ systems in \cite{BEpq}, and entails a rigorous proof -- modulo checking the TT property, which had not been performed so far -- of an equivalence between the three usual approaches of quantum gravity, namely topological gravity (in relation with intersection theory on the moduli space of curves), random maps, and $(2m + 1,2)$ models (see \cite{dFZJ} for a review on those equivalences in physics). Again taking the TT property as an assumption, \cite{CafassoMarchal} treated the models $(2m,1)$, in relation with the merging of two cuts in random matrix theory. The TT property was made explicit and checked by integrability arguments in \cite{BETW} for a $2 \times 2$ linear system associated to the Painlev\'e II equation \cite{FN80}, justifying the computation of asymptotics of the GUE Tracy-Widom law by the topological recursion. The same approach -- with a justification of the TT property -- was applied more recently \cite{BEMarchal} to the $2 \times 2$ linear system of associated to Painlev\'e V \cite{Jimbo}, relevant to get the GUE sine kernel law. So far, this concerned only $2 \times 2$ systems.

The present work aims at presenting a complete theory for general $d \times d$ rational systems, and developing tools to study the TT property. Its application to the $(p,q)$ models can then be used to establish rigorously the equivalence between the three quantum gravities for all $(p,q)$ models. For clarity, this will appear in a separate work \cite{BEcartespq}.

In \cite{BEInt}, the two last authors have made a conjecture to construct an integrable system out of the topological recursion of a given spectral curve. The present work aims at the converse: showing that the semiclassical expansion of linear differential systems satisfying the TT property are computed by the topological recursion of their semiclassical spectral curve.

The TT property is neither expected to hold in general -- even among integrable systems -- nor obvious to establish for a given system. Our proof that it holds for the $q$-th reduction of the KP hierarchy depends in an essential way on the integrability of the latter, i.e. on the existence of another system $\hbar\,\partial_{t} \mathbf{\Psi}(x,t) = \mathbf{M}(x,t)\mathbf{\Psi}(x,t)$ with rational coefficients in $x$, which is compatible with \eqref{1}, but also on the specific form of $\mathbf{M}(x,t)$. This is clear from the technical results of Section~\ref{sysm2} and \ref{Spropole}. 

Within the TT property, the structure of the asymptotic expansion of correlators is identified in Theorem~\ref{topro}, but when the semiclassical spectral curve has genus $\mathfrak{g} > 0$, it features an unknown "holomorphic part" $H_n^{(g)}(z_1,\ldots,z_n)$, which are basically the moduli of the space of solutions of loop equations. A given solution $\mathbf{\Psi}(x)$ knows which $H_n^{(g)}(z_1,\ldots,z_n)$ is chosen. It thus remains to investigate which conditions have to be added to the loop equations to determine completely the holomorphic part. They probably should take the form of period conditions. Actually, for many interesting solutions $\mathbf{\Psi}(x)$, we expect the TT property to breakdown if the semiclassical spectral curve has genus $\mathfrak{g} > 0$.

We stress that, even when the TT property does not hold, the loop equations of Theorem~\ref{linearL} and \ref{quda} are still satisfied and provide some constraints on the asymptotic expansion of $\mathbf{\Psi}(x)$. In particular, the existence of a non-trivial moduli space of solutions of loop equations -- which, in the context of expansion in powers of $\hbar$, can be parametrized by a "holomorphic part" at each order in $\hbar$ -- can be responsible for the breakdown of expansion in powers of $\hbar$, since the moduli may depend on $\hbar$ in a complicated way. This mechanism explains for instance the oscillatory asymptotics in matrix models \cite{Ecv,BGmulti}. It is also implicit in \cite{BEInt}, where the candidate Tau function is constructed as a sum over a lattice of step $\hbar$ in the moduli space of solutions of the loop equations: the interferences between the terms of the sum create in general an oscillatory $\hbar \rightarrow 0$, described by Theta functions evaluated at an argument proportional to $1/\hbar$. This suggest that in general when $\hbar \rightarrow 0$, the "fast variables" live in the moduli space of solutions of loop equations, whereas the dependence in the "adiabatic variables" is captured by the loop equations themselves.

An important, open problem, would be to show that the asymptotics of (bi)orthogonal polynomials are given by certain Baker-Akhiezer functions of an integrable system, which depend on the universality class. Around a point where the density of zeroes vanishes like a power $p/q$, the integrable system should be related to the $(p,q)$ models. This remains beyond the scope of the present investigation.

\section{Linear differential systems and loop equations}
\label{S1}
\subsection{Kernel, determinantal formulae and correlators}
\label{S2}
\bd
\label{D1}The \emph{kernel} is a $d \times d$ matrix depending on two variables $x_1,x_2 \in \widehat{\mathbb{C}}\setminus\mathcal{P}$, defined by:
\beq
\mathbf{K}(x_1,x_2) = \frac{\mathbf{\Psi}^{-1}(x_1)\mathbf{\Psi}(x_2)}{x_1 - x_2},
\eeq
It is often referred to as the "parallel transport" operator, because it satisfies
$\mathbf \Psi(x_2) = (x_1-x_2)\,\mathbf\Psi(x_1)\,\mathbf K(x_1,x_2)$, i.e. it transports $\mathbf\Psi(x_1)$ to $\mathbf\Psi(x_2)$.
\ed
It obviously satisfies a replication formula:
\beq
\label{repK}\mathbf{K}(x_1,x_2)\mathbf{K}(x_2,x_3) = \frac{x_1 - x_3}{(x_1 - x_2)(x_2 - x_3)}\,\mathbf{K}(x_1,x_3),
\eeq
and it has a simple pole at coinciding points:
\beq
\mathbf{K}(x_1,x_2) \mathop{\sim}_{x_1 \rightarrow x_2} \frac{\mathbf{1}_{d}}{x_1 - x_2}.
\eeq
 
\bd
\label{D2} The \emph{$n$-point correlators} are a family of symmetric functions in $n$ variables, indexed by $a_1,\ldots,a_n \in \ldbrack 1,d \rdbrack$, defined as follows:
\bea
\mathcal{W}_1(\sheet{x}{a}) & = & \lim_{x' \rightarrow x}\left(\mathbf{K}_{a,a}(x,x') - \frac{1}{x - x'} \right), \\
\label{eq:dee}\forall n \geq 2,\qquad \mathcal{W}_n(\sheet{x_1}{a_1},\ldots,\sheet{x_n}{a_n}) & = & (-1)^{n + 1} \sum_{\sigma = n\textrm{-cycles}} \prod_{i = 1}^n \mathbf{K}_{a_i,a_{\sigma(i)}}(x_i,x_{\sigma(i)}),
\eea
and the \emph{"non--connected" $n$-point correlators} by:
\beq
\label{noncW}\overline{\mathcal{W}}_n(\sheet{x_1}{a_1},\ldots,\sheet{x_n}{a_n}) = "\mathrm{det}"\,\mathbf{K}_{a_i,a_j}(x_i,x_j),
\eeq
where "det" means that each occurrence of $\mathbf{K}_{a_i,a_i}(x_i,x_i)$ in the determinant should be replaced by $\mathcal{W}_1(\sheet{x_i}{a_i})$.
\ed
For instance, we have for any $a,b \in \ldbrack 1,d \rdbrack$, with $a \neq b$:
\bea
\mathcal{W}_1(\sheet{x}{a})  & = &  - \hbar^{-1}\big(\mathbf{\Psi}^{-1}(x)\mathbf{L}(x)\mathbf{\Psi}(x)\big)_{a,a},\\
\mathcal{W}_2(\sheet{x_1}{a_1},\sheet{x_2}{a_2}) & = & - \mathbf{K}_{a_1,a_2}(x_1,x_2)\mathbf{K}_{a_2,a_1}(x_2,x_1), \\
\label{Wdouble0} \lim_{x_1 \rightarrow x} \mathcal{W}_2(\sheet{x_1}{a},\sheet{x}{b}) & = & -\hbar^{-2} \big(\mathbf{\Psi}^{-1}(x)\mathbf{L}(x)\mathbf{\Psi}(x)\big)_{a,b}\,\big(\mathbf{\Psi}^{-1}(x)\mathbf{L}(x)\mathbf{\Psi}(x)\big)_{b,a}.
\eea
We may give another representation for the correlators, by:
\bd We define the projectors on state $a$:
\beq
\mathbf{P}(\sheet{x}{a}) = \mathbf{\Psi}(x)\,\mathbf{E}_a\,\mathbf{\Psi}^{-1}(x),
\eeq
where $\mathbf{E}_a={\rm diag}(0,\dots,0,\stackrel{a}{1},0,\dots,0)$ denotes the diagonal matrix with $a^{\mathrm{th}}$-entry $1$, and zero entries elsewhere. 
\ed
We observe that $\mathbf{P}(\sheet{x}{a})$ form a basis of rank one projectors:
\beq
\label{Projprop}\mathbf{P}(\sheet{x}{a})\mathbf{P}(\sheet{x}{b}) = \delta_{a,b}\mathbf{P}(\sheet{x}{a}), \qquad \Tr\,\mathbf{P}(\sheet{x}{a}) = 1, \qquad \sum_{a = 1}^{d} \mathbf{P}(\sheet{x}{a}) = \mathbf{1}_{d},
\eeq
which satisfies a Lax equation
\beq
\label{projLax}\hbar\,\partial_x\mathbf{P}(\sheet{x}{a}) = [\mathbf{L}(x),\mathbf{P}(\sheet{x}{a})].
\eeq
\bp
\label{propW}The correlators can be written:
\bea
\label{W1r}\mathcal{W}_1(\sheet{x}{a}) & = & - \hbar^{-1}\Tr \mathbf{P}(\sheet{x}{a})\mathbf{L}(x), \\
\label{W2r}\mathcal{W}_2(\sheet{x_1}{a_1},\sheet{x_2}{a_2}) & = & \frac{\Tr \mathbf{P}(\sheet{x_1}{a_1})\mathbf{P}(\sheet{x_2}{a_2})}{(x_1-x_2)^2} 
= -\, \frac{\Tr (\mathbf{P}(\sheet{x_1}{a_1})-\mathbf{P}(\sheet{x_2}{a_2}))^2}{2\,(x_1-x_2)^2} + \frac{1}{(x_1-x_2)^2},
\eea
and for $n\geq 3$
\beq
\label{Wkr}\mathcal{W}_n(\sheet{x_1}{a_1},\dots,\sheet{x_n}{a_n}) = (-1)^{n + 1}\sum_{\sigma = n\textrm{-cycles}} \frac{\Tr \mathbf{P}(\sheet{x_1}{a_1})\mathbf{P}(\sheet{x_{\sigma(1)}}{a_{\sigma(1)}})\mathbf{P}(\sheet{x_{\sigma^2(1)}}{a_{\sigma^2(1)}})\cdots \mathbf{P}(\sheet{x_{\sigma^{n-1}(1)}}{a_{\sigma^{n - 1}(1)}})}{\prod_{i = 1}^n (x_i - x_{\sigma(i)})}.
\eeq
\hfill $\Box$
\ep
For instance, we can deduce if $a_1 \neq a_2$:
\bea
\label{Wdouble}\lim_{x_1 \to x_2}\mathcal{W}_2(\sheet{x_1}{a_1},\sheet{x_2}{a_2}) & = &   - \hbar^{-2}\,\Tr \mathbf{P}(\sheet{x_2}{a_1})\mathbf{L}(x_2)\mathbf{P}(\sheet{x_2}{a_2})\mathbf{L}(x_2), \\
\mathcal{W}_3(\sheet{x_1}{a_1},\sheet{x_2}{a_2},\sheet{x_3}{a_3}) & = & \frac{\Tr\mathbf{P}(\sheet{x_1}{a_1})[\mathbf{P}(\sheet{x_2}{a_2}),\mathbf{P}(\sheet{x_3}{a_3})]}{(x_1 - x_2)(x_2 - x_3)(x_3 - x_1)}.
\eea
Although it is not clear from the definition, the $n$-point correlators do not have poles at coinciding points when $n \geq 3$. If $I = \ldbrack 1,n \rdbrack$, $(x_i)_{i \in I}$ and $(a_i)_i \in \ldbrack 1,d \rdbrack^{I}$, we denote $\sheet{x_I}{a_I}$ the family $(\sheet{x_i}{a_i})_{i \in I}$.

\begin{proposition}
\label{nopole} For any $n \geq 3$, any $a_1,\ldots,a_n \in \ldbrack 1,d \rdbrack$, and $1 \leq i \neq j \leq n$, $\mathcal{W}_n(\sheet{x_I}{a_I})$ is regular when $x_i \rightarrow x_j$.
\end{proposition}
\textbf{Proof.} By symmetry, it is enough to consider $i = 1$ and $j = 2$. The definition of $\mathcal{W}_k(x_I^{a_I})$ implies that it can have at most simple poles when $x_1 \rightarrow x_2$. Let us compute its residue from \eqref{Wkr}:
\bea
\Res_{x_1 \rightarrow x_2} \mathcal{W}_n(\sheet{x_I}{a_I}) & = & (-1)^{n + 1}\Big\{\sum_{\substack{\sigma = n\mathrm{-cycle} \\ \sigma(1) = 2}} \frac{\Tr \mathbf{P}(\sheet{x_2}{a_1})\mathbf{P}(\sheet{x_2}{a_2})\mathbf{P}(\sheet{x_{\sigma(2)}}{a_{\sigma(2)}})\cdots \mathbf{P}(\sheet{x_{\sigma^{n - 3}(2)}}{a_{\sigma^{n - 3}(2)}})\mathbf{P}(\sheet{x_{\sigma^{n - 2}(2)}}{a_{\sigma^{n - 2}(2)}})}{(x_{\sigma^{n - 2}(2)} - x_2)(x_2 - x_{\sigma(2)})\cdots (x_{\sigma^{n - 3}(2)} - x_{\sigma^{n - 2}(2)})} \nonumber \\
& & - \sum_{\substack{\sigma = n\mathrm{-cycle} \\ \sigma^{-1}(1) = 2}} \frac{\Tr \mathbf{P}(\sheet{x_2}{a_2})\mathbf{P}(\sheet{x_2}{a_1})\mathbf{P}(\sheet{x_{\sigma^2(2)}}{a_{\sigma^2(2)}})\cdots\mathbf{P}(\sheet{x_{\sigma^{n - 2}(2)}}{a_{\sigma^{n - 2}(2)}})\mathbf{P}(\sheet{x_{\sigma^{n - 1}(2)}}{a_{\sigma^{n - 1}(2)}})}{(x_{\sigma^{n - 1}(2)} - x_2)(x_2 - x_{\sigma^2(2)})\cdots (x_{\sigma^{n - 2}(2)} - x_{\sigma^{n - 1}(2)})}\Big\}.
\eea
Using the relation $\mathbf{P}(\sheet{x_2}{a_1})\mathbf{P}(\sheet{x_2}{a_2}) = \delta_{a_1,a_2}\mathbf{P}(\sheet{x_2}{a_2})$, we can rewrite:
\bea
\Res_{x_1 \rightarrow x_2} \mathcal{W}_n(\sheet{x_I}{a_I}) & = & (-1)^{n + 1}\delta_{a_1,a_2}\Big\{\sum_{\substack{\sigma = n\textrm{-cycle} \\ \sigma(1) = 2}} \frac{\Tr \mathbf{P}(\sheet{x_2}{a_2})\mathbf{P}(\sheet{x_{\sigma(2)}}{a_{\sigma(2)}})\cdots \mathbf{P}(\sheet{x_{\sigma^{n - 3}(2)}}{a_{\sigma^{n - 3}(2)}})\mathbf{P}(\sheet{x_{\sigma^{n - 2}(2)}}{a_{\sigma^{n - 2}(2)}})}{(x_{\sigma^{n - 2}(2)} - x_2)(x_2 - x_{\sigma(2)})\cdots (x_{\sigma^{n - 3}(2)} - x_{\sigma^{n - 2}(2)})} \nonumber \\
& & - \sum_{\substack{\sigma = n\textrm{-cycles} \\ \sigma^{-1}(1) = 2}} \frac{\Tr \mathbf{P}(\sheet{x_2}{a_2})\mathbf{P}(\sheet{x_{\sigma^2(2)}}{a_{\sigma^2(2)}})\cdots\mathbf{P}(\sheet{x_{\sigma^{n - 2}(2)}}{a_{\sigma^{n - 2}(2)}})\mathbf{P}(\sheet{x_{\sigma^{n - 1}(2)}}{a_{\sigma^{n - 1}(2)}})}{(x_{\sigma^{n - 1}(2)} - x_2)(x_2 - x_{\sigma^2(2)})\cdots (x_{\sigma^{n - 2}(2)} - x_{\sigma^{n - 1}(2)})}\Big\}.
\eea
The two sums range over the set of $(n - 1)$-cycles, and are actually equal. We conclude that $\mathcal{W}_k(\sheet{x_I}{a_I})$ is regular when $x_1 \rightarrow x_2$. \hfill $\Box$

\subsection{Loop equations}

We first show that the correlators satisfy a set of linear equations. 
\bt \label{linearL}(\textbf{Linear loop equation}) For any $n \geq 1$, any $c_2,\ldots,c_n \in \ldbrack 1,d \rdbrack$, we have:
\beq
\sum_{a = 1}^d \mathcal{W}_n(\sheet{x}{a},\sheet{y_2}{c_2},\ldots,\sheet{y_n}{c_n}) = -\delta_{n,1} \hbar^{-1} \,\Tr \mathbf{L}(x) + \frac{\delta_{n,2}}{(x - y_2)^2}.
\eeq
\et
\textbf{Proof.} We first address the cases $n = 1,2$ by direct computation starting from \eqref{W1r}-\eqref{W2r}, and use the properties \eqref{Projprop} of the projectors:
\bea
\label{220}\sum_{a = 1}^d \mathcal{W}_1(\sheet{x}{a}) & = & - \hbar^{-1}\,\Tr\Big(\sum_{a = 1}^d \mathbf{P}(\sheet{x}{a})\Big)\mathbf{L}(x) = -\hbar^{-1}\,\Tr \mathbf{L}(x), \\
\sum_{a = 1}^d \mathcal{W}_2(\sheet{x}{a},\sheet{y}{c}) & = & \frac{\Tr \big(\sum_{a = 1}^d \mathbf{P}(\sheet{x}{a})\big)\mathbf{P}(\sheet{y}{c})}{(x - y)^2}  = \frac{\Tr \mathbf{P}(\sheet{y}{c})}{(x - y)^2} = \frac{1}{(x - y)^2}.
\eea
For $n \geq 3$, combining the representation \eqref{Wkr} and the fact that $\sum_{a = 1}^d \mathbf{P}(\sheet{x}{a}) = \mathbf{1}_{d}$, we find that:
\beq
\sum_{a = 1}^d \mathcal{W}_n(\sheet{x}{a},\sheet{y_I}{c_I}) = (-1)^{n + 1}\sum_{\sigma = n\textrm{-cycle}} \frac{1}{(x - y_{\sigma(1)})(y_{\sigma^{-1}(1)} - x)}\,\frac{\Tr \mathbf{P}(\sheet{y_{\sigma(1)}}{c_{\sigma(1)}})\cdots\mathbf{P}(\sheet{y_{\sigma^{n - 1}(1)}}{c_{\sigma^{n - 1}(1)}})}{\prod_{i = 1}^{n - 2} (y_{\sigma^i(1)} - y_{\sigma^{i + 1}(1)})}
\eeq
is a rational function of $x$, which vanishes in the limit $x \rightarrow \infty$. Singularities can only arise as simple poles at $x = y_i$ for $i \in I$, but their residue is $0$ according to Proposition~\ref{nopole}. Hence, the left hand side vanishes identically. \hfill $\Box$

\bt\label{quda} (\textbf{Quadratic loop equations}) For any $n \geq 1$, any $c_2,\ldots,c_n \in \ldbrack 1,d \rdbrack$, 
\beq
\sum_{1 \leq a < b \leq d}\Big(\mathcal{W}_{n + 1}(\sheet{x}{a},\sheet{x}{b},\sheet{y_I}{c_I}) + \sum_{J \subseteq I} \mathcal{W}_{|J| + 1}(\sheet{x}{a},\sheet{y_J}{c_J})\mathcal{W}_{n - |J|}(\sheet{x}{b},\sheet{y_{I\setminus J}}{c_{I\setminus J}})\Big) = P_n(x;\sheet{y_I}{c_I})
\eeq
is a rational function of $x$, with possible poles at $x = x_i$ for $i \in I$ and poles of $\mathbf{L}$.
\et
As illustration, we give the formulas for $P_n$ up to $n = 3$:
\bea
\label{224}P_1(x) & = & \frac{1}{2\hbar^2}\big(-\Tr \mathbf{L}^2(x) + [\Tr \mathbf{L}(x)]^2\big),  \\
P_2(x;\sheet{y}{c}) & = & \frac{1}{\hbar}\,\frac{\Tr \mathbf{L}(x)[\mathbf{P}(\sheet{y}{c}) - \mathbf{1}_{d}]}{(x - y)^2}, \\
P_3(x;\sheet{y_1}{c_1},\sheet{y_2}{c_2}) & = & -\frac{1}{\hbar} \frac{\Tr[\mathbf{P}(\sheet{y_1}{c_1})\mathbf{P}(\sheet{y_2}{c_2}) + \mathbf{P}(\sheet{y_2}{c_2})\mathbf{P}(\sheet{y_1}{c_1})]\mathbf{L}(x)}{(x - y_1)(x - y_2)} + \frac{(y_1 - y_2)^2\mathcal{W}_2(\sheet{y_1}{c_1},\sheet{y_2}{c_2}) + 1}{(x - y_1)^2(x - y_2)^2}.
\eea

\noindent \textbf{Proof.} Notice that the left hand side makes sense even if $n = 1$, because the function $\mathcal{W}_2(\sheet{x}{a},\sheet{x}{b}) = \lim_{y \rightarrow x} \mathcal{W}_2(\sheet{y}{a},\sheet{x}{b})$ is well-defined when $a\neq b$, and given by \eqref{Wdouble0}. When $a \neq b$, $\mathcal{W}_n(\sheet{x}{a},\sheet{x}{b},\sheet{y_I}{c_I})$ can be computed from Definition~\ref{D2}, using $\mathbf{K}_{a,b}(x,x) = -\hbar^{-1}(\mathbf{\Psi}^{-1}\mathbf{L}\mathbf{\Psi})_{a,b}(x)$. We introduce a new quantity $\widetilde{\mathcal{W}}_n(\sheet{x}{a},\sheet{x}{b},\sheet{y_I}{c_I})$, as follows:
\begin{itemize}
\item[$\bullet$] when $a = b$, it is computed from Definition~\ref{D2} where each occurrence of $\mathbf{K}_{a,a}(x,x)$ is replaced by $-\hbar^{-1}(\mathbf{\Psi}^{-1}\mathbf{L}\mathbf{\Psi})_{a,b}(x)$ (which is also equal to $\mathcal{W}_1(\sheet{x}{a})$),
\item[$\bullet$] when $a \neq b$, it is equal to $\mathcal{W}_n(\sheet{x}{a},\sheet{x}{b},\sheet{y_I}{c_I})$.
\end{itemize}
We claim:
\begin{lemma}
\label{L1}\beq
\forall n \geq 1,\quad \forall a \in \ldbrack 1,d \rdbrack,\qquad \widetilde{\mathcal{W}}_{n + 1}(\sheet{x}{a},\sheet{x}{a},\sheet{y_I}{c_I}) + \sum_{J \subseteq I} \mathcal{W}_{|J| + 1}(\sheet{x}{a},\sheet{y_J}{c_J})\mathcal{W}_{n - |J|}(\sheet{x}{a},\sheet{y_{I\setminus J}}{c_{I\setminus J}}) = 0.
\eeq
\end{lemma}
The proof of the lemma will be given below. We deduce that:
\beq
P_k(x;\sheet{y_I}{c_I}) = \frac{1}{2}\sum_{a,b = 1}^d \widetilde{\mathcal{W}}_{n + 1}(\sheet{x}{a},\sheet{x}{b},\sheet{y_I}{c_I}) + \sum_{J \subseteq I} \frac{1}{2}\Big(\sum_{a = 1}^d \mathcal{W}_{|J| + 1}(\sheet{x}{a},\sheet{y_J}{c_J})\Big)\Big(\sum_{b = 1}^d \mathcal{W}_{n - |J|}(\sheet{x}{b},\sheet{y_{I\setminus J}}{c_{I\setminus J}})\Big).
\eeq
The last term is given by the linear loop equations (Theorem~\ref{linearL}): it vanishes when $n \geq 5$, and is a rational function of $x$ with poles at $x = x_i$ for some $i \in I$, or at poles of $\mathbf{L}$. We now focus on the first term, which is by definition:
\bea
\label{328}& & Q_k(x;\sheet{y_I}{c_I}) \equiv \frac{1}{2}\sum_{a,b = 1}^d \widetilde{\mathcal{W}}_{n + 1}(\sheet{x}{a},\sheet{x}{b},\sheet{y_I}{c_I}) = \frac{(-1)^{n}}{2}\sum_{a,b = 1}^d \Big\{ \\
& & \!\!\!\!\!-\frac{(\mathbf{\Psi}^{-1}\mathbf{L}\mathbf{\Psi})_{a,b}(x)}{\hbar}\!\!\!\!\!\!\!\sum_{\substack{\sigma = (n + 1)\textrm{-cycle} \\ \sigma(1) = 2}}\!\!\!\!\!\!\!\mathbf{K}_{c_{\sigma^{-1}(1)},a}(y_{\sigma^{-1}(1)},x)\mathbf{K}_{b,c_{\sigma(2)}}(x,y_{\sigma(2)}) \prod_{i = 1}^{n - 2} \mathbf{K}_{c_{\sigma^i(2)},c_{\sigma^{i + 1}(2)}}(y_{\sigma^i(2)},y_{\sigma^{i + 1}(2)}) \nonumber \\
& & \!\!\!\!\! -\frac{(\mathbf{\Psi}^{-1}\mathbf{L}\mathbf{\Psi})_{b,a}(x)}{\hbar} \!\!\!\!\!\!\!\!\sum_{\substack{\sigma = (n + 1)\textrm{-cycle} \\ \sigma(2) = 1}} \!\!\!\!\!\!\!\mathbf{K}_{a,c_{\sigma(1)}}(x,y_{\sigma(1)})\mathbf{K}_{c_{\sigma^{-1}(2)},b}(y_{\sigma^{-1}(2)},x)\prod_{i = 1}^{n - 2} \mathbf{K}_{c_{\sigma^i(1)},c_{\sigma^{i + 1}(1)}}(y_{\sigma^i(1)},y_{\sigma^{i + 1}(1)}) \nonumber \\
& & \!\!\!\!\! + \sum_{\substack{\sigma = (n + 1)\textrm{-cycle} \\ \sigma(1) \neq 2,\,\,\sigma(2) \neq 1}} \mathbf{K}_{a,c_{\sigma(1)}}(x,y_{\sigma(1)})\cdots\mathbf{K}_{c_{\sigma^{-1}(2)},b}(y_{\sigma^{-1}(2)},x)\mathbf{K}_{b,c_{\sigma(2)}}(x,y_{\sigma(2)})\cdots\mathbf{K}_{c_{\sigma^{-1}(1)},a}(y_{\sigma^{-1}(1)},x)\Big\}. \nonumber
\eea
The two first lines are equal by symmetry. Performing the sum over $a$ and $b$, and replacing the kernels involving the variable $x$ by their definition, we find:
\bea
& & Q_{n}(x;\sheet{y_I}{c_I}) \\
& = & \!\!\!\! (-1)^{n + 1}\Big\{\sum_{\substack{\sigma = (n + 1)\textrm{-cycle} \\ \sigma(1) = 2}} \!\!\!\!\!\!\! - \frac{[\mathbf{\Psi}^{-1}(x_{\sigma^{-1}(1)})\mathbf{L}(x)\mathbf{\Psi}(x_{\sigma(2)})]_{c_{\sigma^{-1}(1)},c_{\sigma(2)}}}{\hbar} \prod_{i = 1}^{n - 3} \mathbf{K}_{c_{\sigma^{i}(2)},c_{\sigma^{i + 1}(2)}}(y_{\sigma^{i}(2)},y_{\sigma^{i + 1}(2)}) \nonumber \\
& & + \sum_{\substack{\sigma = (n + 1)\textrm{-cycle} \\ \sigma(1) \neq 2,\,\,\sigma(2) \neq 1}} \prod_{j = 1,2} \frac{(y_{\sigma^{-1}(j)} - y_{\sigma(j)}) \mathbf{K}_{c_{\sigma^{-1}(j)},c_{\sigma(j)}}(y_{\sigma^{-1}(j)},y_{\sigma(j)})}{2(x - y_{\sigma^{-1}(j)})(x - y_{\sigma(j)})} \nonumber \\
& & \phantom{\sum_{\sigma^{i + 1}(1) \neq 1,2}} \times \prod_{\substack{i = 1 \\ \sigma^{i + 1}(1) \neq 1,2}}^{n - 2} \mathbf{K}_{c_{\sigma^{i}(1)},c_{\sigma^{i + 1}(1)}}(x_{\sigma^i(1)},x_{\sigma^{i + 1}(1)})\Big\}. \nonumber 
\eea
This expression is a rational function of $x$ which can have poles only at $x_i$ for $i \in I$, and at poles of $\mathbf{L}$. Therefore, we proved that $P_n(x;\sheet{y_I}{c_I})$ is a rational function of $x$ which can have poles only at those very points.

\vspace{0.2cm}

\noindent \textbf{Proof of Lemma~\ref{L1}.}
We have the analog of \eqref{328} for $a = b$:
\bea
& & \widetilde{\mathcal{W}}_{n + 1}(\sheet{x}{a},\sheet{x}{a},\sheet{y_I}{c_I}) = (-1)^n\Big\{-\frac{2(\mathbf{\Psi}^{-1}\mathbf{L}\mathbf{\Psi})_{a,a}(x)}{\hbar} \\
& & \times \sum_{\substack{\sigma = (n + 1)\textrm{-cycles} \\ \sigma(1) = 2}} \mathbf{K}_{a,c_{\sigma(2)}}(x,y_{\sigma(2)})\Big[\prod_{i = 1}^{n - 1} \mathbf{K}_{c_{\sigma^i(2)},c_{\sigma^{i + 1}(2)}}(y_{\sigma^i(2)},y_{\sigma^{i + 1}(2)})\Big]\mathbf{K}_{c_{\sigma^{n - 1}(2)},a}(y_{\sigma^{n - 1}(2)},x) \nonumber \\
& & + \sum_{\substack{1 \leq j,k\leq n \\ j + k = n}} \sum_{\substack{\sigma = (n + 1)\textrm{-cycles} \\ \sigma^{j + 1}(1) = 2}} \mathbf{K}_{a,c_{\sigma(1)}}(x,y_{\sigma(1)})\Big[\prod_{i = 1}^{j - 1} \mathbf{K}_{c_{\sigma^j(1)},c_{\sigma^{i + 1}(1)}}(y_{\sigma^{i}(1)},y_{\sigma^{i + 1}(1)})\Big]\mathbf{K}_{c_{\sigma^j(1)},a}(y_{\sigma^j(1)},x) \nonumber \\
& & \phantom{\sum_{j = 1}^{n - 1} \sum_{\sigma = (n + 1)\textrm{-cycles}}} \times \mathbf{K}_{a,c_{\sigma(2)}}(x,y_{\sigma(2)})\Big[\prod_{i = 1}^{k - 1} \mathbf{K}_{c_{\sigma^i(2)},c_{\sigma^{i + 1}(2)}}(y_{\sigma^{i}(2)},y_{\sigma^{i + 1}(2)})\Big]\mathbf{K}_{c_{\sigma^{k}(2)},a}(y_{\sigma^{k}(2)},x)\Big\}. \nonumber
\eea
We recognize in the first line $-2\mathcal{W}_1(\sheet{x}{a})\mathcal{W}_{n}(\sheet{x}{a},\sheet{y_I}{c_I})$. Besides, the two last lines amounts to a sum over two disjoint cycles of length $(j + 1)$ and $(k + 1)$, and we recognize each term correlators up to a sign factor. Namely:
\beq
\widetilde{\mathcal{W}}_{n + 1}(\sheet{x}{a},\sheet{x}{a},\sheet{x_I}{a_I}) = -2\mathcal{W}_1(\sheet{x}{a})\mathcal{W}_{n}(\sheet{x}{a},\sheet{y_I}{c_I}) - \sum_{\varnothing \subset J \subset I} \mathcal{W}_{|J| + 1}(\sheet{x}{a},\sheet{y_J}{c_J})\mathcal{W}_{n - |J|}(\sheet{x}{a},\sheet{y_{I\setminus J}}{c_{I\setminus J}}).
\eeq
The first term completes the sum with the terms $J = \varnothing$ and $J = I$, hence the result. \hfill $\Box$

\vspace{0.2cm}

\noindent \textbf{Detailed example.} Let us redo the computation in the case $n = 1$ to illustrate the method of the proof. We have:
\bea
P_1(x) & = & \frac{1}{\hbar^2}\sum_{1 \leq a < b \leq d} -\big(\mathbf{\Psi}^{-1}\mathbf{L}\mathbf{\Psi}\big)_{a,b}(x)\big(\mathbf{\Psi}^{-1}\mathbf{L}\mathbf{\Psi}\big)_{b,a}(x) + \big(\mathbf{\Psi}^{-1}\mathbf{L}\mathbf{\Psi}\big)_{a,a}\big(\mathbf{\Psi}^{-1}\mathbf{L}\mathbf{\Psi}\big)_{b,b}(x).
\eea
Notice that the summand vanish if $a = b$. We can thus write:
\bea
P_1(x) & = & \frac{1}{2\hbar^2}\sum_{a,b = 1}^d -\big(\mathbf{\Psi}^{-1}\mathbf{L}\mathbf{\Psi}\big)_{a,b}(x)\big(\mathbf{\Psi}^{-1}\mathbf{L}\mathbf{\Psi}\big)_{b,a}(x) + \big(\mathbf{\Psi}^{-1}\mathbf{L}\mathbf{\Psi}\big)_{a,a}(x)\big(\mathbf{\Psi}^{-1}\mathbf{L}\mathbf{\Psi}\big)_{b,b}(x) \nonumber \\
\label{234} & = & \frac{1}{2\hbar^2}\big(- \Tr \mathbf{L}^2(x) + [\Tr \mathbf{L}(x)]^2\big).
\eea
\hfill $\Box$ 

\subsection{Spectral curve}

\begin{definition}
The \emph{spectral curve} is the plane curve $\mathcal{S}$ of equation $\mathrm{det}(y - \mathbf{L}(x)) = 0$.
\end{definition}
The eigenvalues of $\mathbf{L}(x)$ are algebraic functions.

\begin{proposition}\label{propspcurve1}
The spectral curve can be expressed in terms of correlators:
\beq
\mathrm{det}(y - \mathbf{L}(x)) = \sum_{k = 0}^{d} y^{d - k} \sum_{1 \leq a_1 < \ldots < a_k \leq d} \overline{\mathcal{W}}_k(\sheet{x}{a_1},\ldots,\sheet{x}{a_k}).
\eeq
\end{proposition}

\noindent \textbf{Proof.} We first write the coefficients of a characteristic polynomial as a sum over minors:
\bea
\mathrm{det}(y - \mathbf{L}(x)) & = & \mathrm{det}(y - \mathbf{\Psi}^{-1}(x)\mathbf{L}(x)\mathbf{\Psi}(x)) \nonumber \\
& = & \sum_{k = 0}^d y^{d - k} \sum_{1 \leq a_1 < \ldots < a_k \leq d} \mathop{\det}_{1 \leq i,j \leq k}[-\mathbf{\Psi}^{-1}\mathbf{L}\mathbf{\Psi}]_{a_i,a_j}(x) \nonumber \\
& = & \sum_{k = 0}^d y^{d - k} \hbar^k \sum_{1 \leq a_1 < \ldots < a_k \leq d} \mathop{\det}_{1 \leq i,j \leq k} \widetilde{\mathbf{K}}_{a_i,a_j}(x,x),
\eea
where we have defined $\widetilde{\mathbf{K}}_{a,b}(x,x) = -\hbar^{-1}(\mathbf{\Psi}^{-1}\mathbf{L}\mathbf{\Psi})_{a,b}(x)$. Notice that $\widetilde{\mathbf{K}}_{a,b}(x,x) = \mathbf{K}_{a,b}(x,x)$ when $a \neq b$, whereas $\widetilde{\mathbf{K}}_{a,a}(x,x) = \mathcal{W}_1(\sheet{x}{a})$. And, the specialization of the definition of non-connected correlators \eqref{noncW} to $x_i \equiv x$ for $i \in \ldbrack 1,d \rdbrack$ and $a_1 < \ldots < a_k$ yields:
\beq
\overline{\mathcal{W}}_k(\sheet{x}{a_1},\ldots,\sheet{x}{a_k}) = \mathop{\det}_{1 \leq i,j \leq k}\,\widetilde{\mathbf{K}}_{a_i,a_j}(x,x),
\eeq
whence the announced formula. \hfill $\Box$

\vspace{0.2cm}

\noindent We remark that the coefficients of $y^{d-2}$ was already identified in Eqn.~\ref{224}.

\subsection{Gauge transformations}

If $\mathbf{\Psi}$ is a solution of \eqref{1}, and $\mathbf{G}$ is a matrix depending on $x$, $\widetilde{\mathbf{\Psi}} = \mathbf{G}\mathbf{\Psi}$ will also be solution of similar equation, with:
\bea
\widetilde{\mathbf{L}} & = & (\hbar\,\partial_x \mathbf{G})\mathbf{G}^{-1} + \mathbf{G}\mathbf{L}\mathbf{G}^{-1}. \nonumber
\eea
Any two arbitrary $d \times d$ matrices $\mathbf{\Psi}(x)$ and $\widetilde{\mathbf{\Psi}}(x)$ can be related by a gauge transformation $\mathbf{G}(x) = \widetilde{\mathbf{\Psi}}(x)\mathbf{\Psi}(x)^{-1}$. Therefore, the concept of gauge transformations is only meaningful if we impose some restriction on the form of $\mathbf{G}(x)$. Here, the natural restriction to impose is that $(\hbar\,\partial_x \mathbf{G})\mathbf{G}^{-1}$ is rational, and its poles should occur at poles of $\mathbf{L}$ with a lower (or equal) degree than in $\mathbf{L}$.

Gauge transformations in general completely change the kernel and the correlators. However, there are two special gauge transformations under which the correlators do not change. If $\mathbf{G}$ is independent of $x$:
\beq
\widetilde{\mathbf{L}} = \mathbf{G}\mathbf{L}\mathbf{G}^{-1}, \qquad \widetilde{\mathbf{P}} = \mathbf{G}\mathbf{P}\mathbf{G}^{-1},\qquad \widetilde{\mathbf{K}} = \mathbf{K},\qquad \widetilde{\mathcal{W}}_n = \mathcal{W}_n.
\eeq
(where, for bookkeeping, we included the gauge transformation of matrix  $\Gamma$ defined in section \ref{SSym}).
If $\mathbf{G}$ depends on $x$ but is scalar $\mathbf{G} = G\mathbf{1}_{d}$:
\beq
\widetilde{\mathbf{L}} = \mathbf{L} + \hbar\,\partial_{x}\ln G, \qquad \widetilde{\mathbf{P}} = \mathbf{P},\qquad \widetilde{\mathbf{K}}(x,y) = \frac{G(y)}{G(x)}\,\mathbf{K}(x,y),\qquad
\widetilde{\mathcal{W}}_n  = \mathcal{W}_n.
\eeq

\subsection{Insertion operator}

Let $(\mathbb C(x),\partial_x)$ be the differential ring generated by rational functions.
We consider a Picard-Vessiot ring $\mathbb{B}$ of the differential system $\hbar\,\partial_x\mathbf{\Psi}(x) = \mathbf{L}(x)\mathbf{\Psi}(x)$ \cite{vdPS}. It is is a simple extension of $(\mathbb C(x),\partial_x)$ by the matrix elements of $\mathbf{\Psi}(x)$ and $\big(\mathrm{det}\,\mathbf{\Psi}(x)\big)^{-1}$. Let $\mathbb{B}_n$ the $n$-variable analog of $\mathbb{B}$, i.e. the differential ring generated by rational functions in $n$ variables $x_1,\ldots,x_n$ and by the matrix elements of $\mathbf{\Psi}(x_i)$ and $\big(\mathrm{det}\,\mathbf{\Psi}(x_i)\big)^{-1}$. We denote the projective limit $\mathbb{B}_{\infty} = \lim_{n\to\infty} \mathbb{B}_n$. By construction, the matrix elements of $\mathbf{P}(\sheet{x}{a})$ or of $\mathbf L(x)$ are in $\mathbb{B}$, those of $\mathbf{K}(x_1,x_2)$ are in $\mathbb{B}_2$, and the $n$-point correlators $\mathcal{W}_n(\sheet{x_1}{a_1},\ldots,\sheet{x_n}{a_n})$ are in $\mathbb{B}_n$.

\begin{definition}\label{defdeltay}
An \emph{insertion operator} is a collection of derivations $(\delta_{y}^{a})_{1 \leq a \leq d}$ over $\mathbb{B}_{\infty}$, commuting with $\partial_{x_i}$, with the following properties:
\begin{itemize}
\item[$\bullet$] $\delta_y^{a}(\mathbb{B}_n) \subseteq \mathbb{B}_{n + 1}$.
\item[$\bullet$] $\delta_{y}^a(\mathbb{C}(x_i)) = 0$.
\item[$\bullet$] there exists matrices $\mathbf{U}(\sheet{y}{a})$ with entries in $\mathbb B$, so that:
\beq
\delta_{y}^a\mathbf{\Psi}(x) = \Big(\frac{\mathbf{P}(\sheet{y}{a})}{x - y} + \mathbf{U}(\sheet{y}{a})\Big)\mathbf{\Psi}(x),
\eeq
and such that $\mathbf U$ satisfies
\beq
\delta_x^a \mathbf U(\sheet{y}{b})-\delta_y^b \mathbf U(\sheet{x}{a})=[\mathbf U(\sheet{x}{a}),\mathbf U(\sheet{y}{b})].
\eeq
\end{itemize}
\end{definition}

\begin{lemma}
\label{L0a} If $\delta_{y}^{a}$ is an insertion operator, for any $n \geq 1$, any $a,b,a_1,\ldots,a_n \in \ldbrack 1,d \rdbrack$,
\bea
\label{deltaKkk} \delta_{y}^{a}\mathbf{K}(x_1,x_2) & = & -\mathbf{K}(x_1,y)\mathbf{E}_a\mathbf{K}(y,x_2),  \\
\label{Pder}\delta_{y}^{a} \mathbf{P}(\sheet{x}{b}) & = & \Big[\frac{\mathbf{P}(\sheet{y}{a})}{x - y} + \mathbf{U}(\sheet{y}{a}),\mathbf{P}(\sheet{x}{b})\Big], \nonumber \\
\label{Lder}\delta_{y}^{a} \mathbf{L}(x) & = & \Big[\frac{\mathbf{P}(\sheet{y}{a})}{x - y} + \mathbf{U}(\sheet{y}{a}),\mathbf{L}(x)\Big] - \frac{\mathbf{P}(\sheet{y}{a})}{(x - y)^2}, \nonumber \\
\label{trLder}\delta_{y}^{a} \Tr \mathbf{L}(x) & = &  - \frac{1}{(x - y)^2}, \nonumber \\
\label{detPsider}\delta_{y}^{a} \ln\det \mathbf{\Psi}(x) & = &   \frac{1}{x - y} + \Tr U(\sheet{y}{a}), \qquad
\delta_{y}^{a} \ln\left(\frac{\det\mathbf{\Psi}(x)}{\det\mathbf{\Psi}(z)}\right) =   \frac{1}{x - y} -\frac{1}{z-y},  \\
\label{djq} \delta_{y}^{a} \mathcal{W}_{n}(\sheet{x_1}{a_1},\ldots,\sheet{x_n}{a_n}) & = & \mathcal{W}_{n + 1}(\sheet{y}{a},\sheet{x_1}{a_1},\ldots,\sheet{x_n}{a_n}).
\eea
\end{lemma}

\noindent {\bf Proof.} Easy computations, done in appendix \ref{appproofdeltas}. \hfill $\Box$

\vspace{0.2cm}

\noindent The fact that the insertion operator sends ${\cal W}_n$ to ${\cal W}_{n+1}$  justifies the name "insertion operator". We remark that equations \eqref{deltaKkk} and \eqref{djq} are independent of $\mathbf{U}$.

\vspace{0.2cm}

\noindent\textbf{Remark}. Because of relation \eqref{detPsider}, $\det\mathbf{\Psi}$ is not constant regarding the action of the  insertion operator. Notice that in general, up to a scalar gauge transformation, one can always chose $\det\mathbf{\Psi}(x)$ to be a constant. What this means here, is that the insertion operator $\delta_y^a$ doesn't commute with gauge transformations.

\vspace{0.2cm}

Let us define the semi-connected correlators:
\bea
\mathcal{W}_{k;n}(\sheet{x_1}{a_1},\sheet{x_2}{a_2},\dots,\sheet{x_k}{a_k}\,;\,\sheet{y_1}{b_1},\dots,\sheet{y_n}{b_n})
= \sum_{I \vdash \ldbrack 1,k\rdbrack} \sum_{J_1 \dot{\cup} \cdots \dot{\cup} J_{\ell(\mu)} = \ldbrack 1,n \rdbrack} \prod_{j=1}^{\ell(I)}
\mathcal{W}_{|I_j|+ |J_j|}(\sheet{x_{I_j}}{a_{I_j}},\sheet{y_{J_j}}{b_{J_j}}).
\eea
Here, $I$ is a partition of $\ldbrack 1,k \rdbrack$, i.e. a set of $\ell(I)$ non-empty, pairwise disjoint subsets $I_i \subseteq \ldbrack 1,k \rdbrack$ whose union is $\ldbrack 1,k \rdbrack$, whereas the subsets $J_i \subset \ldbrack 1,n \rdbrack$ could be empty.

\bp[Most general loop equations]

For every $k \leq d$ and every $\{\sheet{y_1}{b_1},\ldots,\sheet{y_n}{b_n}\}$,
\beq
P_{k,n}(x;\sheet{y_1}{b_1},\ldots,\sheet{y_n}{b_n})
=\sum_{1 \leq a_1<a_2<\cdots < a_k \leq d} \mathcal{W}_{k;n}(\sheet{x}{a_1},\ldots,\sheet{x}{a_k};\sheet{y_1}{b_1},\ldots,\sheet{y_n}{b_n})
\eeq
is a rational function of $x$, with poles  at $x = y_j$ for some $j$ and at poles of $\mathbf{L}$.
\ep
\textbf{Proof.} The case $n=0$ is Proposition~\ref{propspcurve1}. The cases $n \geq 1$ are obtained by recursively applying $\delta_{y_j}^{b_j}$, for any insertion operator $\delta$.\hfill $\Box$

\section{Asymptotics and topological expansion}
\label{S3}

Loop equations form an infinite system of equations, in general difficult to solve.  In many applications, correlators have an asymptotic expansion (or are formal series) in powers of $\hbar$, and if this expansion is of "topological type" (Definition~\ref{detype} below), loop equations can be solved recursively in powers of $\hbar$, by the topological recursion of \cite{EOFg}. This claim is justified in this section.

We assume that $\mathbf{L}(x)$ has an asymptotic expansion in powers of $\hbar$, of the form:
\beq
\mathbf{L}(x) = \sum_{k \geq 0} \hbar^{k}\,\mathbf{L}^{[k]}(x),
\eeq
which is uniform for $x$ in some domain of the complex plane, or alternatively, $\mathbf{L}(x)\in \mathbb C[[\hbar]]$ is defined as a formal power series in $\hbar$. Let us denote
\beq
\mathbf{\Lambda}(x) = \mathrm{diag}(\lambda_1(x),\ldots,\lambda_d(x))
\eeq
the diagonal matrix of eigenvalues of $\mathbf{L}(x)$ counted with multiplicities and ordered arbitrarily.  $\mathbf{\Lambda}(x)$ also has an expansion in powers of $\hbar$:
\beq
\mathbf{\Lambda}(x) = \sum_{k \geq 0} \hbar^{k}\,\mathbf{\Lambda}^{[k]}(x).
\eeq

\subsection{The semiclassical spectral curve}

The semiclassical spectral curve is the locus of leading order eigenvalues:
\bd
The semiclassical spectral curve is defined as:
\beq
\mathcal{S}^{[0]} =\overline{\big\{(x,y) \in \mathbb{C}^2\,|\,\,\det(y\,\mathbf{1}_{d} - \mathbf L^{[0]}(x))=0\big\}}.
\eeq
\ed
It can be seen as the immersion of a compact Riemann surface ${\cal S}^{[0]}$ into $\mathbb{C}\times\mathbb{C}$, through the maps  $x\,:\,\mathcal{S}^{[0]} \to \mathbb{C}$ and $y\,:\,\mathcal{S}^{[0]} \to \mathbb{C}$. If $x$ is of degree $d$ (the degree in $y$ of the algebraic equation defining $\mathcal{S}^{[0]}$, i.e. the size of the matrix $\mathbf L^{[0]}(x)$), then the preimage of $x_0 \in \mathbb{C}$ is denoted:
\beq
x^{-1}(\{x_0\}) = \{z^0(x_0),\dots,z^{d-1}(x_0)\} \subseteq {\cal S}^{[0]}.
\eeq
In other words, $\mathcal{S}^{[0]}$ is realized as a branch covering of $\mathbb{C}$ of degree $d$ by the projection $x\,:\,\mathcal{S}^{[0]} \to \mathbb{C}$. The zeroes of $\dd x$ in $\mathcal{S}^{[0]}$ are the \emph{ramification points}, and their $x$-coordinate are the \emph{branchpoints}. Branchpoints $\beta_i \in \mathbb{C}$ occur when $z^a(\beta_i) = z^b(\beta_i)$ for at least two distinct indices $a$ and $b$, and we then denote $r_i = z^a(\beta_i) = z^b(\beta_i)$. Let us call $\mathbf{r}$ the set of ramification points.

$\lambda_{a}^{[0]}(x)$ are the eigenvalues of $\mathbf{L}^{[0]}(x)$, i.e. by definition they are the $y$ coordinates of points of ${\cal S}^{[0]}$, i.e. they are the $y$ image of some $z^a(x)$:
\beq\label{yzala}
\big\{y(z^{a}(x)) \quad a \in \ldbrack 1,d \rdbrack \big\} = \big\{\lambda_{a}^{[0]}(x) \quad a \in \ldbrack 1,d \rdbrack\big\}.
\eeq
Double points $\alpha_i \in \mathbb{C}$ occur where two or more eigenvalues collide, i.e.
$$y(z^a(\alpha_i))=\lambda^{[0]}_{a}(\alpha_i) = \lambda^{[0]}_{b}(\alpha_i)=y(z^b(\alpha_i))$$
for at least two distinct indices $a \neq b$, but $\dd x(z^a(\alpha_i)) \neq 0$ and $\dd x(z^b(\alpha_i))\neq 0$ -- a fortiori,  $z^a(\alpha_i)$ and $z^b(\alpha_i)$ must be distinct points in $\mathcal{S}^{[0]}$.

The space $H^1(\mathcal{S}^{[0]})$ of holomorphic $1$-forms on $\mathcal{S}^{[0]}$ is a complex vector space of dimension $\mathfrak{g}$, where $\mathfrak{g}$ is the genus of $\mathcal{S}^{[0]}$. In particular, if $\mathfrak{g} = 0$, $H^1(\mathcal{S}^{[0]}) = \{0\}$ and a meromorphic form on $\mathcal{C}$ is completely determined by the singular behavior at its poles.

\begin{definition}
\label{defbid} Let $\mathcal{B}(\mathcal{S}^{[0]})$ the set of \emph{fundamental bidifferentials of the second kind}, i.e. $B(z_1,z_2)$ which are symmetric $2$-form in $(\mathcal{S}^{[0]})^2$, with no residues, and a double pole at $z_1 = z_2$ with behavior in any local coordinate $\xi$:
\beq
B(z_1,z_2) \mathop{=}_{z_1 \rightarrow z_2} \frac{\dd\xi(z_1)\,\dd\xi(z_2)}{\big(\xi(z_1) - \xi(z_2)\big)^2} + O(1).
\eeq
\end{definition}
Since one can add to $B$ any symmetric bilinear combination of holomorphic forms, $\mathcal{B}(\mathcal{S}^{[0]})$ is an affine space, whose underlying vector space is $\mathrm{Sym}^2[H^1(\mathcal{S}^{[0]})]$, so it has complex dimension $\mathfrak{g}(\mathfrak{g} + 1)/2$.

\subsection{Expansions in powers of $\hbar$}
\label{S32}

We now assume that $\mathcal{S}^{[0]}$ is a regular plane curve, i.e. $\dd x$ and $\dd y$ do not have common zeroes. Therefore, $\mathbf{L}^{[0]}(x)$ has simple eigenvalues for any $x$ which is not a branchpoint or double point, hence is diagonalizable. So must be $\mathbf{L}(x)$ at least when $\hbar$ is small and $x$ stays away from the branchpoints or double points. We can thus find a matrix of eigenvectors $\mathbf{V}(x)$:
\beq
\mathbf{L}(x) = \mathbf{V}(x)\mathbf{\Lambda}(x)\mathbf{V}^{-1}(x),
\eeq
which admits an expansion in powers of $\hbar$:
\beq
\label{Uexp}\mathbf{V}(x) = \sum_{k \geq 0} \hbar^k\,\mathbf{V}^{[k]}(x).
\eeq
Such a matrix is defined up to transformations $\mathbf{V}(x) \rightarrow \mathbf{V}(x)\mathbf{D}(x)\mathbf{\Sigma}$, where $\mathbf{D}(x)$ is a diagonal matrix and $\mathbf{\Sigma}$ a permutation matrix. We can use the first freedom to impose:
\beq
\label{norm}\forall a \in \ldbrack 1,d \rdbrack,\qquad \big(\mathbf{V}^{-1}(x)\,\partial_x\mathbf{V}(x)\big)_{a,a} = 0.
\eeq
and we then say that $\mathbf{V}(x)$ is a normalized matrix of eigenvectors. Any two such matrices are related by a transformation $\mathbf{V}(x) \rightarrow \mathbf{V}(x)\mathbf{D}\mathbf{\Sigma}$, where $\mathbf{D}$ is a constant diagonal matrix and $\mathbf{\Sigma}$ a permutation matrix.

We would like to study solutions of \eqref{1} which have an expansion in powers of $\hbar$. For this purpose, we fix a base point $o$, an invertible matrix of constants $\mathbf{C}$, and introduce a matrix $\widehat{\mathbf{\Psi}}(x)$ such that:
\beq
\label{dsi}\mathbf{\Psi}(x) = \mathbf{V}(x)\widehat{\mathbf{\Psi}}(x)\exp\Big(\frac{1}{\hbar}\int^{x}_{o} \mathbf{\Lambda}(x')\dd x'\Big)\mathbf{C}.
\eeq
$\mathbf{\Psi}(x)$ is a solution of \eqref{1} if and only if:
\beq
\label{1u}\hbar\,\partial_x \widehat{\mathbf{\Psi}}(x) = - \hbar\,\mathbf{T}(x)\widehat{\mathbf{\Psi}}(x) + [\mathbf{\Lambda}(x),\widehat{\mathbf{\Psi}}(x)],
\eeq
where $\mathbf{T}(x) = \mathbf{V}(x)^{-1}\partial_x\mathbf{V}(x)$ also has an expansion in powers of $\hbar$ derived from \eqref{Uexp}:
\beq
\mathbf{T}(x) = \sum_{k \geq 0} \hbar^k\,\mathbf{T}^{[k]}(x).
\eeq

\begin{proposition}
\label{22a}Eqn.~\ref{1u} has a unique solution which is a formal power series in $\hbar$ of the form:
\beq
\label{psija}\widehat{\mathbf{\Psi}}(x) = \mathbf{1}_{d} +\sum_{k \geq 1} \hbar^{k}\,\widehat{\mathbf{\Psi}}^{[k]}(x) 
\eeq
up to transformations $\widehat{\mathbf{\Psi}}^{[k]}(x) \rightarrow \widehat{\mathbf{\Psi}}^{[k]}(x) + \widehat{\mathbf{C}}^{[k]}$, where $\widehat{\mathbf{C}}^{[k]}$ is a diagonal matrix of constants. A priori, the entries of $\widehat{\mathbf{\Psi}}^{[k]}(x)$ are multivalued functions of $x$ with monodromies around branchpoints, double points, and poles at the poles of $(\mathbf{L}^{[j]}(x))_{j \geq 0}$.
\end{proposition}
\textbf{Proof.} Inserting the ansatz \eqref{psija} in \eqref{1u} and collecting the terms of order $\hbar^{k + 1}$ yields, for any $a,b \in \ldbrack 1,d \rdbrack$:
\beq
\label{la}\partial_x\widehat{\mathbf{\Psi}}^{[k]}_{a,b} = - \sum_{j = 0}^k (\mathbf{T}^{[k - j]}\widehat{\mathbf{\Psi}}^{[j]})_{a,b} + (\lambda_{a}^{[0]} - \lambda_{b}^{[0]})\widehat{\mathbf{\Psi}}^{[k + 1]}_{a,b} + \sum_{j = 0}^{k} (\lambda_a^{[k - j]} - \lambda_b^{[k - j]})\widehat{\mathbf{\Psi}}^{[j]}_{a,b}.
\eeq
Since we assume that $\mathcal{S}^{[0]}$ is regular and $x$ is away from a branchpoint or a double point, we have $\lambda^{[0]}_{a}(x) \neq \lambda^{[0]}_{b}(x)$ when $a \neq b$, which allows to write:
\beq
\label{squ1}\widehat{\mathbf{\Psi}}^{[k + 1]}_{a,b} = \frac{1}{\lambda^{[0]}_{a} - \lambda^{[0]}_{b}}\Big(\partial_{x}\widehat{\mathbf{\Psi}}^{[k]}_{a,b} + \sum_{j = 0}^{k} (\mathbf{T}^{[k - j]}\widehat{\mathbf{\Psi}}^{[j]})_{a,b} - (\lambda_a^{[k - j]} - \lambda_b^{[k - j]})\widehat{\mathbf{\Psi}}^{[j]}_{a,b}\Big).
\eeq
This equation determines the off-diagonal part of $\widehat{\mathbf{\Psi}}^{[k + 1]}$ in terms of $\widehat{\mathbf{\Psi}}^{[j]}$ for $j \in \ldbrack 0,k \rdbrack$. For $a = b$ in \eqref{la}, we rather find:
\beq
\label{squ2}\partial_x\widehat{\mathbf{\Psi}}^{[k+1]}_{a,a} = \sum_{\substack{c = 1 \\ c \neq a}} \mathbf{T}^{[0]}_{a,c}\widehat{\mathbf{\Psi}}_{c,a}^{[k+1]} + \sum_{j = 0}^{k} (\mathbf{T}^{[k+1 - j]}\widehat{\mathbf{\Psi}}^{[j]})_{a,a}.
\eeq
We took into account the normalization\footnote{Notice that we only need \eqref{norm} at leading order here.} \eqref{norm}, so that the right hand side involves only off-diagonal entries of $\widehat{\mathbf{\Psi}}^{[k+1]}$, or the entries of $\widehat{\mathbf{\Psi}}^{[j]}$ for $j \in \ldbrack 0,k\rdbrack$.

We proceed by recursion starting from the initial condition $\widehat{\mathbf{\Psi}}^{[0]} = \mathbf{1}_{d}$. Assuming that $\widehat{\mathbf{\Psi}}^{[j]}$ are completely known for $j \in \ldbrack 0,k \rdbrack$, we obtain the off-diagonal part of $\widehat{\mathbf{\Psi}}^{[k + 1]}$ from \eqref{squ1}, and solving the first order differential equation \eqref{squ2} we then obtain the diagonal part of $\widehat{\mathbf{\Psi}}^{[k + 1]}$ up to a diagonal matrix of integration constants $\widehat{\mathbf{C}}^{[k + 1]}$. It is clear that the singularities of $\widehat{\mathbf{\Psi}}^{[k]}$ can only occur at singularities of $\lambda_a^{[j]}(x)$ and $\mathbf{T}^{[j]}(x)$, i.e. either at semiclassical branchpoints or poles of $(\mathbf{L}^{[j]}(x))_{j \geq 0}$, or at double points where $\lambda_a^{[0]}=\lambda_b^{[0]}$.
\hfill $\Box$

\bp[Analytic continuation]\label{propanalyticPhi}
The matrices $\mathbf V(x)$, $\mathbf\Lambda(x)$ and $\widetilde{\mathbf{\Psi}}(x) = \mathbf V(x)\widehat{\mathbf\Psi}(x)$, all have a power series expansion in $\hbar$, whose coefficients are such that their $a^{\rm th}$-column vector is the evaluation of meromorphic function on $\mathcal{S}^{[0]}$ at $z^a(x)$.
In particular, there exists a vector $\tilde{\mathbf{\psi}}^{[k]}(z)$ such that:
\beq
\label{eq:316}\widetilde{\mathbf{\Psi}}_{i,a}(x)=\left(\mathbf V(x)\,\widehat{\mathbf\Psi}(x)\right)_{i,a} =\sum_{k \geq 0} \hbar^k\,\tilde{\mathbf\psi}^{[k]}_i(z^a(x)).
\eeq
\ep
\noindent \textbf{Proof.} For the diagonal matrix $\mathbf\Lambda$, we have already seen in \eqref{yzala} that $\lambda_a^{[0]}(x) = y(z^a(x))$.
Solving $\det(\lambda_a(x)\,\mathbf 1_{d}-\mathbf L(x))=0$ with $\mathbf L(x)=\sum_{k \geq 0} \hbar^{k}\,\mathbf{L}^{[k]}(x)$ and $\lambda_a(x)=\sum_{k \geq 0} \hbar^{k}\,\lambda_a^{[k]}(x)$, by recursion on $k$, shows easily that each $\lambda_a^{[k]}(x)$ is a meromorphic function $\lambda^{[k]}(z^a(x))$ for all $k$.
Similarly, Kramers formula for computing the eigenvectors of $\mathbf L(x)$, shows that up to a normalization factor, the eigenvector corresponding to the $a^{\rm th}$-eigenvalue $\lambda_a(x)$, has also a power series expansion in $\hbar$ whose coefficients are meromorphic functions of $z^a(x)$ at each order.
In other words, one can chose a matrix $\widehat{\mathbf V}(x)$ of eigenvectors of $\mathbf L(x)$ satisfying
\beq
\mathbf L(x) = \widehat{\mathbf V}(x) \mathbf \Lambda(x) \widehat{\mathbf V}^{-1}(x)
\eeq
of the form
\beq
\widehat{\mathbf V}(x) = \sum_{k \geq 0} \hbar^k\,\widehat{\mathbf V}^{[k]}(x)
\qquad , \quad
\widehat{\mathbf V}^{[k]}(x)_{i,a} = \hat v^{[k]}_i(z^a(x)).
\eeq
Then, notice that any symmetric meromorphic function of $(z^1(x),\ldots,z^d(x))$ is a meromorphic function of $x$, and thus a meromorphic function of any $z^a(x)$. And, any symmetric meromorphic function of $(z^1(x),\ldots,z^d(x))_{\hat{a}}$ (i.e. all $z^j(x)$'s except $z^a(x)$), is a meromorphic function of $x$ and of $z^a(x)$, and thus is a meromorphic function of $z^a(x)$.
In particular, this implies that the determinant of $\widehat{\mathbf V}(x)$ is a power series of $\hbar$ whose coefficients are meromorphic function of $z^a(x)$,
and the inverse matrix $\widehat{\mathbf V}^{-1}(x)$ takes the form:
\beq
\widehat{\mathbf V}^{-1}_{a,i}(x) = \sum_{k \geq 0} \hbar^k\,\hat{v}^{[k]}_i(z^a(x)).
\eeq
This implies that
\beq
\left( \widehat{\mathbf V}^{-1}(x) \,\partial_x \widehat{\mathbf V}(x)\right)_{a,a}  = \sum_{k \geq 0} \hbar^k\,\hat{t}^{[k]}(z^a(x)),
\eeq
where each $\hat{t}^{[k]}(z)$ is a meromorphic function on the semi-classical spectral curve ${\cal S}^{[0]}$.

\smallskip
We chose to normalize our basis of eigenvectors $\mathbf V(x) = \widehat{\mathbf V}(x)\,\mathbf{D}(x)$ where $\mathbf{D}(x)$ is some diagonal matrix, so that \eqref{norm} is satisfied, i.e. we have to choose $D(x)$ satisfying:
\beq
\mathbf{D}_{a,a}^{-1}(x)\,\partial_x \mathbf{D}_{a,a}(x) = -\,\left( \widehat{\mathbf V}^{-1}(x) \,\partial_x \widehat{\mathbf V}(x)\right)_{a,a} = - \sum_{k \geq 0} \hbar^k\,\hat{t}^{[k]}(z^a(x)).
\eeq
This shows that $D_{a,a}(x)$ also has a power series expansion in $\hbar$ whose coefficients are meromorphic functions of $z^a(x)$.
Finally, this shows that $\mathbf V(x)$ has the form:
\beq
\mathbf V_{i,a}(x) = \sum_{k \geq 0} \hbar^k\,v^{[k]}_i(z^a(x)),
\eeq
where each $v^{[k]}_i(z)$ is a meromorphic function on the semi-classical spectral curve.

\smallskip

If we choose $\mathbf{C}$ to be diagonal, we see that:
\beq
\widetilde{\mathbf{\Psi}}(x)=\mathbf V(x)\,\widehat{\mathbf{\Psi}}(x) = \mathbf \Psi(x)\,\mathbf{C}^{-1}\,\exp\Big(-\frac{1}{\hbar}\int_\alpha^x \mathbf \Lambda(x')\,\dd x'\Big)
\eeq
obeys:
\beq
\hbar\,\partial_x \widetilde{\mathbf{\Psi}}(x) = \mathbf L(x)\,\widetilde{\mathbf{\Psi}}(x) - \widetilde{\mathbf{\Psi}}(x)\,\mathbf\Lambda(x).
\eeq
The equation for the $a^{\rm th}$-column of $\widetilde{\mathbf{\Psi}}(x)$ involves only $\mathbf{\Lambda}_{a,a}(x)$, and thus is order by order in $\hbar$ analytical in $z^a(x)$, and since we know that $\widetilde{\mathbf\Psi}(x)$ has only meromorphic singularities, we see again that the column vectors of $\widetilde{\mathbf{\Psi}}(x)$ have an $\hbar$ expansion such that the coefficients are meromorphic functions of $z^a(x)$.
\hfill $\Box$

\bc\label{propanalyticalstruct}
The coefficients $\tilde{\psi}_{i}^{[k]}(z)$ appearing in the expansion of $\widetilde{\mathbf{\Psi}}_{i,a}(x)$, are meromorphic functions of $z\in{\cal S}^{[0]}$ whose poles occur only at values of $z$ such that $\exists a\neq b$ and $x\in\mathbb C$ with $z=z^a(x)=z^b(x)$, or at poles of $\mathbf{L}^{[l]}(x)$ for $l \leq k$. In other words, $\phi_{i}^{[k]}(z)$ can be singular only at ramification points, at preimages in $\mathcal{S}^{[0]}$ of double points, or at poles of $\mathbf L^{[l]}$ on the semi--classical spectral curve ${\cal S}^{[0]}$.
\ec
\noindent\textbf{Proof.} 
$\widetilde{\mathbf\Psi}(x)$ was constructed so that it has at most meromorphic singularities  at poles of $\mathbf L(x)$.
Then, one can see in \eqref{squ1} that singularities can occur only when $\lambda_a^{[0]}(x)=\lambda_b^{[0]}(x)$ for some $a\neq b$, i.e. at branchpoints or double points.
\hfill $\Box$

\subsection{Expansion of the correlators}

In this section, we consider the projectors, the correlators, etc. (see Section~\ref{S2}) associated to the solution $\mathbf{\Psi}(x)$ deduced from Proposition~\ref{22a} via \eqref{dsi}. 

\begin{lemma}
Assume that the constant matrix $\mathbf{C}$ in \eqref{dsi} is diagonal. Then, the projectors have an expansion in powers of $\hbar$, of the form:
\beq
\mathbf{P}(\sheet{x}{a}) = \sum_{k \geq 0} \hbar^k\,\mathbf{P}^{[k]}(\sheet{x}{a}),
\eeq
and there exists a sequence of  matrices $\mathbf{p}^{[k]}(z)$ of meromorphic functions in $z \in \mathcal{S}^{[0]}$, with poles at ramification points, at preimages in $\mathcal{S}^{[0]}$ of double points, and at poles of $(\mathbf{L}^{[j]}(x))_{j \geq 0}$, such that $\mathbf{p}^{[k]}(z^{a}(x)) = \mathbf{P}^{[k]}(\sheet{x}{a})$.
\end{lemma}
\textbf{Proof.} Since we assume $\mathbf{C}$ to be diagonal, the exponentials -- which might have prevented the existence of an expansion in powers of $\hbar$ -- disappear:
\bea
\mathbf{P}(\sheet{x}{a}) & = & \mathbf{V}(x)\widehat{\mathbf{\Psi}}(x)\exp\Big(\frac{1}{\hbar}\int_o^x \mathbf{\Lambda}(x')\dd x'\Big) \mathbf{C}\mathbf{E}_a\mathbf{C}^{-1}\exp\Big(-\frac{1}{\hbar}\int_0^x \mathbf{\Lambda}(x')\dd x'\Big)\widehat{\mathbf{\Psi}}^{-1}(x)\mathbf{V}^{-1}(x) \nonumber \\
& = & \mathbf{V}(x)\widehat{\mathbf{\Psi}}(x)\mathbf{E}_a\widehat{\mathbf{\Psi}}^{-1}(x)\mathbf{V}^{-1}(x) \cr
&=& \widetilde{\mathbf\Psi}(x)\,\mathbf E_a \widetilde{\mathbf\Psi}^{-1}(x).
\eea
From Proposition~\ref{propanalyticPhi}, $\widetilde{\mathbf{\Psi}}(x)$  has an expansion in $\hbar$, so $\mathbf{P}(\sheet{x}{a})$ has an expansion in $\hbar$. Moreover, $\widetilde{\mathbf{\Psi}}(x)\,\mathbf E_a \widetilde{\mathbf{\Psi}}^{-1}(x)$ involves only the $a^{\rm th}$ column of $\widetilde{\mathbf{\Psi}}(x)$ and the $a^{\rm th}$ line of $\widetilde{\mathbf{\Psi}}^{-1}(x)$, i.e. the coefficients of the expansion are meromorphic functions of $z^a(x)$. From Corollary~\ref{propanalyticalstruct}, those meromorphic functions can be singular only at ramification points, at preimages in $\mathcal{S}^{[0]}$ of double points, or at poles of $\mathbf L(x)$ in $\mathcal{S}^{[0]}$. \hfill $\Box$

\vspace{0.2cm}

\noindent Notice that to leading order, $\widetilde{\mathbf{\Psi}}(x) =\mathbf 1_{d} + O(\hbar)$ and:
\beq
\label{Pleading}\mathbf{P}^{[0]}(\sheet{x}{a}) = (\mathbf{V}^{[0]}(x))^{-1}\mathbf{E}_a\mathbf{V}^{[0]}(x)
\eeq
is the projection on the $a$-th eigenspace of $\mathbf{L}^{[0]}(x)$.  From the expression of the correlators in terms of the projectors, we deduce:

\begin{corollary}
\label{W1cora}For any $a \in \ldbrack 1,d \rdbrack$, $\mathcal{W}_1(\sheet{x}{a})$ has an expansion in powers of $\hbar$, of the form:
\beq
\mathcal{W}_1(\sheet{x}{a}) = \sum_{k \geq -1} \hbar^{k}\,\mathcal{W}_1^{[k]}(\sheet{x}{a}),
\eeq
and there exist meromorphic functions $w_{1}^{[k]}(z)$ in $z \in \mathcal{S}^{[0]}$, with poles at the ramification points, or at preimages in $\mathcal{S}^{[0]}$ of double points, or at poles of $(\mathbf{L}^{[j]}(x))_{j \geq 0}$, so that $w_{1}^{[k]}(z^{a}(x)) = \mathcal{W}_1^{[k]}(\sheet{x}{a})$.
\end{corollary}

\noindent For example we have:
\beq
\mathcal{W}_1^{[0]}(\sheet{x}{a})  =  -\lambda_{a}^{[0]}(x), 
\eeq

\begin{corollary}
For any $n \geq 2$, any $a_1,\ldots,a_n \in \ldbrack 1,d \rdbrack$, the correlators have an expansion in powers of $\hbar$:
\beq
\mathcal{W}_n(\sheet{x_1}{a_1},\ldots,\sheet{x_n}{a_n}) = \sum_{k \geq 0} \hbar^{k}\,\mathcal{W}_n^{[k]}(\sheet{x_1}{a_1},\ldots,\sheet{x_n}{a_n})
\eeq
and there exist symmetric meromorphic functions $w_{n}^{[k]}(z_1,\ldots,z_n)$ in $(z_1,\ldots,z_n) \in (\mathcal{S}^{[0]})^n$, with poles when $z_i$ is at a ramification point or at a double pole or at a pole of $(\mathbf{L}^{[j]}(x))_{j \geq 0}$, and so that $$w_{n}^{[k]}(z^{a_1}(x_1),\ldots,z^{a_n}(x_n)) = \mathcal{W}_{n}^{[k]}(\sheet{x_1}{a_1},\ldots,\sheet{x_n}{a_n}).$$ On top of that, $w_{2}^{[0]}(z_1,z_2)$ has a double pole at $z_1 = z_2$, and behaves as:
\beq
w_{2}^{[0]}(z_1,z_2) \mathop{=}_{z_1 \rightarrow z_2} \frac{x'(z_1)x'(z_2)}{\big(x(z_1) - x(z_2)\big)^2} + O(1).
\eeq
\end{corollary}

\subsection{Expansion in $\hbar$ with poles  assumptions}
\label{S34} 
Many interesting systems have the property that their leading asymptotic behavior at the poles of $\mathbf{L}(x)$ is governed by the $\hbar\to 0$ limit, i.e. in some sense that $\mathbf{L}^{[j]}(x)$ for $j>0$ is somewhat "smaller" than $\mathbf{L}^{[0]}(x)$.
When this holds, only $\mathcal{W}_1^{[0]}(\sheet{x}{a})$ can have poles at the poles of  $\mathbf{L}(x)$, all other $\mathcal{W}_n^{[g]}$ have no poles at the poles of $\mathbf{L}(x)$. Let us make it precise.

\begin{assumption}\label{asumptionpolesL}
Let us assume that $\mathbf{L}(x)=\sum_{j \geq 0} \hbar^j\,\mathbf{L}^{[j]}(x)$ has the property that for any $j>0$ the poles of $\mathbf{L}^{[j]}(x)$ are a subset of the poles of $\mathbf{L}^{[0]}(x)$, and the expansion of its eigenvalues
\beq
\lambda_a(x) = \sum_{j \geq 0} \hbar^j\,\lambda_a^{[j]}(x)
\eeq
 is such that, for any $j > 0$, $\lambda_a^{[j]}(x) \rightarrow 0$ when $x$ approaches a pole of $\mathbf{L}(x)$. Equivalently, this means that the characteristic polynomial of $\mathbf{L}(x)$ satisfies
\beq
Q(x,y) = \det\big(y\,\mathbf{1}_{d} -\mathbf{L}(x)\big) = \sum_{j \geq 0} \hbar^j\,Q^{[j]}(x,y),
\eeq
where the coefficients, for $j > 0$, are such that:
\beq
D^{[0]}(x)\,Q^{[j]}(x,y)=\sum_{(m,n) \in \mathrm{interior}(\mathcal{N})} \hat{Q}^{[j]}_{m,n-1}\,x^{m}\,y^{n-1},
\eeq
where $D^{[0]}(x)$ is the common denominator of all coefficients of $Q^{[0]}(x,y)$, $\mathcal{N}$ is the envelope of the Newton's polytope of $D^{[0]}(x)\,Q^{[0]}(x,y)$.
\end{assumption}
\begin{corollary}
When assumption \ref{asumptionpolesL} is satisfied,
only $\mathcal{W}_1^{[0]}(\sheet{x}{a})$ can have poles at the poles of 
$\mathbf{L}(x)$, all other $\mathcal{W}_n^{[k]}$ are regular at the poles of $\mathbf{L}(x)$.

\end{corollary}

\begin{corollary}
\label{cor43}$\omega_2^{(0)}(z_1,z_2) = w_2^{[0]}(z_1,z_2)\dd x(z_1)\dd x(z_2)$ defines an element of $\mathcal{B}(\mathcal{S}^{[0]})$ (see Definition~\ref{defbid}).
\end{corollary}

\vspace{0.2cm}

\noindent For instance, we have from Proposition~\ref{propW} and \eqref{Pleading}:
\bea
\mathcal{W}_2^{[0]}(\sheet{x_1}{a_1},\sheet{x_2}{a_2}) & = & \frac{[(\mathbf{V}^{[0]})^{-1}(x_1)\mathbf{V}^{[0]}(x_2)]_{a_1,a_2}[(\mathbf{V}^{[0]})^{-1}(x_2)\mathbf{V}^{[0]}(x_1)]_{a_2,a_1}}{(x_1 - x_2)^2}.
\eea

\subsection{Expansion of topological type and topological recursion}

\begin{definition}[TT property]
\label{detype}We say that the correlators have an expansion of topological type (or have the TT property) when they have:
\begin{itemize}
\item[$\bullet$] the $\hbar \leftrightarrow -\hbar$ symmetry: $(\mathcal{W}_n)_{-\hbar} = (-1)^n(\mathcal{W}_n)_{\hbar}$.
\item[$\bullet$] the $\hbar^{n - 2}$ property: for any $n \geq 2$, $\mathcal{W}_n \in O(\hbar^{n - 2})$.
When these two properties are satisfied, the $\hbar$ expansion of the correlators looks like:
\beq
\label{wnde}\forall n \geq 1,\qquad \mathcal{W}_n = \sum_{g \geq 0} \hbar^{2g - 2 + n}\,\mathcal{W}_n^{(g)}.
\eeq
\item the pole property: when $(g,n)\neq (0,1),(0,2)$, the $\omega_n^{(g)}$ have poles only at the ramification points.
In particular they must have no pole at the preimages in ${\cal S}^{[0]}$ of double points, or at the poles of $\mathbf{L}^{[k]}(x)$. And $\omega_2^{(0)}(z_1,z_2)$ has a double pole at $z_1=z_2$, and no other pole.
\end{itemize}
\end{definition}

In the Section~\ref{sectionintegrable}, we shall study some sufficient conditions (related to integrable systems) to have the TT property, and in Section~\ref{S4}, we shall show that $q$-th reductions of the KP hierarchy, have the TT property.
We believe that the TT property is closely related to integrability, but we do not have a proof of such a statement. Let us just mention that the $\hbar^{n-2}$ property is a highly non--trivial one. For example large random matrices, it is related to the "concentration" property \cite{BG11}.
\medskip

When the TT property is satisfied, one can plug the $\hbar$ expansion \eqref{wnde} into the loop equations to obtain a set of equations satisfied by $\mathcal{W}_n^{(g)}$. The key point is that those equations can be solved recursively on $2g - 2 + n$. The prototype of such a result is known since \cite{ACM,ACKM,ACKMe}. The solution is given by the topological recursion developed in \cite{EORev}. The topological recursion associates to a plane curve $(\mathcal{S}^{[0]},x,y)$ (algebraic in our case) and $\omega_2^{(0)} \in \mathcal{B}(\mathcal{S}^{[0]})$, a sequence of symmetric meromorphic $n$-forms  $\omega^{(g)}_n$ on $({\mathcal{S}^{[0]}})^n$, defined by a recursion on $2g - 2 + n$ in terms of the geometry of the curve $\mathcal{S}^{[0]}$. It was first presented under the assumption that ramification points are simple \cite{EOFg}, and extended to arbitrary ramification points in \cite{BHLMR}. Then, it was shown \cite{BEthink} that the general formula of \cite{BHLMR} is a limiting case of the formula of \cite{EOFg} for simple ramification points. For instance, the semiclassical spectral curve of $r$-KdV has one ramification point of order $r$. For readability, we present now the case of simple ramification points, and refer to \cite{BEthink} for the case of arbitrary ramifications.

\begin{theorem}
\label{topro} If the correlators have an expansion of topological type, and $\dd x$ has only simple zeroes on the semiclassical spectral curve $\mathcal{S}^{[0]}\,:\,\mathrm{det}(y\,\mathbf{1}_{d} - \mathbf{L}^{[0]}(x)) = 0$, then the coefficients of \eqref{wnde} are given by:
\beq
\mathcal{W}_n^{(g)}(\sheet{x_1}{a_1},\ldots,\sheet{x_n}{a_n})\dd x_1\cdots\dd x_n = \omega_n^{(g)}(z^{a_1}(x_1),\ldots,z^{a_n}(x_n))
\eeq
and $\omega^{(g)}_n$ satisfy: 
\bea
& & \omega_n^{(g)}(z_1,z_2,\dots,z_n)   \\
&=& \sum_{r \in \mathbf{r}} \Res_{z\to r} K_r(z_1,z)\,\Big[ \omega_{n+1}^{(g-1)}(z,\sigma_r(z),z_2,\dots,z_n)  + \sum'_{\substack{h+h'=g \\ I \dot{\cup} I'= \ldbrack 2,n \rdbrack}} \omega_{1+|I|}^{(h)}(z,z_I)\,\omega_{1+|I'|}^{(h')}(\sigma_r(z),z_{I'})\,\Big] \nonumber \\
\label{eq340}&& + H_{n}^{(g)}(z_1,\ldots,z_n),
\eea
where $H_{n}^{g}(z_1,\ldots,z_n)$ is some symmetric holomorphic $n$-form on $({\mathcal{S}^{[0]}})^n$, $\sum'$ means that we exclude $(h,I)=(0,\emptyset)$ and $(h',I')=(0,\emptyset)$, $r$ are the ramification points (i.e. the zeroes of $\dd x$), $\sigma_r$ is the local Galois involution near the ramification point $r$, i.e. the holomorphic map defined in the vicinity of $r$, such that $x\circ \sigma_r=x$ and $\sigma_r \neq \mathrm{id}$. And, the recursion kernel is:
\beq
K_r(z_1,z)  = \frac{\frac{1}{2}\int_{\sigma_r(z)}^z \omega_{2}^{(0)}(z_1,\cdot)}{\omega_1^{(0)}(z)-\omega_1^{(0)}(\sigma_r(z))}
\eeq
where $\omega_1^{(0)} = -y\dd x$ on $\mathcal{S}^{[0]}$.
\end{theorem}

\begin{corollary}
\label{toproc} If furthermore ${\mathcal{S}^{[0]}}$ has genus $0$, $H_{n}^{(g)} \equiv 0$ (since there are no holomorphic $1$-forms on $\mathcal{S}^{[0]}$) and $\omega_n^{(g)}$ are exactly given by the topological recursion of \cite{EOFg} applied to the initial data $\omega_1^{(0)} = -y\dd x$ and $\omega_2^{(0)}$ (see Corollary~\ref{cor43}).
\end{corollary}

\noindent \textbf{Proof.} The proof is essentially done in \cite{EOFg,BEO}. To be self-contained, we redo it in Appendix~\ref{apptoprec}. \hfill $\Box$

\subsection{Symmetry $\hbar \leftrightarrow -\hbar$}
\label{SSym}

Here we give a sufficient condition for the existence of an $\hbar \leftrightarrow -\hbar$ symmetry. We do not know whether this criterion is also a necessary condition.

\begin{proposition}
\label{pSym} Assume there exists an invertible matrix $\mathbf{\Gamma}$, independent of $x$, such that:
\beq
\mathbf{\Gamma}\mathbf{L}_{\hbar}^{T}(x)\mathbf{\Gamma}^{-1} = \mathbf{L}_{-\hbar}(x).
\eeq
Then, if $\mathbf{\Psi}_{+}$ is a solution of \eqref{1}, $\mathbf{\Psi}_{-} = \mathbf{\Gamma}(\mathbf{\Psi}_{+}^{-1})^{T}$ is a solution of \eqref{1} with $\hbar \rightarrow -\hbar$. The projector associated to the two solutions are related by $\mathbf{P}_+ = \mathbf{\Gamma}\mathbf{P}_-^{T}\mathbf{\Gamma}^{-1}$, and the correlators by $(\mathcal{W}_n)_+ = (-1)^{n}\,(\mathcal{W}_n)_-$ for any $n \geq 1$.
\end{proposition}
\textbf{Proof.} The relation between the projectors is an easy computation, and given Proposition~\ref{propW} for the $n$-point correlators, we deduce $(\mathcal{W}_n)_+ = (-1)^n(\mathcal{W}_n)_-$ for any $n \geq 2$. For $n = 1$, we check it directly:
\bea
(\mathcal{W}_1)_-(\sheet{x}{a}) & = & \hbar\,[\mathbf{\Psi}_{-}^{-1}(x)\mathbf{L}_{-\hbar}(x)\mathbf{\Psi}_{-}(x)]_{a,a} \nonumber \\
& = & \hbar\,\Tr \mathbf{\Psi}_{-}^{-1}(x)\mathbf{L}_{-\hbar}(x)\mathbf{\Psi}_{-}(x)\mathbf{E}_a = \Tr \mathbf{P}_-(\sheet{x}{a})\mathbf{L}_{-\hbar}(x) \nonumber \\
& = & \hbar\,\Tr \mathbf{\Gamma}\mathbf{P}_+^{T}(\sheet{x}{a})\mathbf{\Gamma}^{-1}\mathbf{L}_{\hbar}(x) = \Tr \mathbf{P}_+^T(\sheet{x}{a})\mathbf{L}_{\hbar}^{T}(x) \nonumber \\
& = & \hbar\,\Tr \mathbf{P}(\sheet{x}{a})\mathbf{L}_{\hbar}(x) = -(\mathcal{W}_1)_+(\sheet{x}{a}).
\eea
\hfill $\Box$

\section{Case of isomonodromic integrable systems}\label{sectionintegrable}

We believe that integrable systems is the good setting to have the TT property satisfied. We give some arguments here, and then show in section \ref{S4} that the special case of $q$-th reduction of KP fits in our framework.

\subsection{Behavior at the poles and isomonodromic times}

In this paragraph, we review classical results from the theory of linear differential systems. A $d \times d$ invertible matrix $\mathbf\Psi(x)$ solution to $\hbar\,\partial_x \mathbf{\Psi}(x) = \mathbf{L}(x)\mathbf{\Psi}(x)$ can have singularities only at poles of $\mathbf{L}(x)$. For any $p \in \mathcal{P}$, it can be put locally around $x = p$ in the form\footnote{When $p = \infty$, the factors $(x - p)$ should be replaced by $1/x$.}:
\beq
\label{asy}\mathbf{\Psi}(x) = \widetilde{\mathbf{\Psi}}_{p}(x)\exp\Big(\mathbf{B}_p\ln(x - p) + \mathbf{A}_{p}(x)\Big)\mathbf{C}_p,\qquad \mathbf{A}_{p}(x) = \sum_{k = 1}^{m_p} \frac{\mathbf{A}_{p;k}}{(x - p)^k},\qquad \widetilde{\Psi}_{p}(x) \mathop{\sim}_{x \rightarrow p} \mathbf{1}_{d},
\eeq
where $\mathbf{A}_p(x)$ and $\mathbf{B}_p$ are Jordanized matrices. Such asymptotics can only be valid in an angular sector near $x = p$, and the constant matrix $\mathbf{C}_p$ depends on the sector. $\mathbf{B}_p$ describes the monodromy around $p$ of the right hand side of \eqref{asy}.

Imagine that $\mathbf{L}(x)$ depends smoothly on parameters $\vec{t} = (t_{\alpha})_{\alpha}$, generically called "times". One can always define a matrix $\mathbf{M}_{\alpha}(x) = \hbar\,\partial_{t_{\alpha}} \mathbf{\Psi}(x)\,\mathbf{\Psi}(x)^{-1}$, so that $\mathbf{\mathbf{\Psi}}(x)$ satisfy on top of \eqref{1} the compatible systems:
\beq
\forall \alpha,\qquad \hbar\,\partial_{t_{\alpha}}\mathbf{\Psi}(x) = \mathbf{M}_{\alpha}(x)\mathbf{\Psi}(x).
\eeq
Requiring that $\mathbf{M}_{\alpha}(x)$ be rational is equivalent to requiring that the local monodromies do not depend on $\vec{t}$. If $\partial_{t_{\alpha}} \mathbf{B}_p \equiv 0$ for any $p \in \mathcal{P}$, we say that $t_{\alpha}$ is an \emph{isomonodromic} time. Integrable systems in Lax form provide examples of such rational compatible differential systems. A second realization of this setting in the realm of formal series in $\vec{t}$ can be achieved by deformation of any given $\mathbf{L}(x)$ and solution $\mathbf{\Psi}(x)$ (independent of parameters) \cite[Chapter 5]{BBT}. The latter might not be the restriction of an integrable system in Lax form (for $\mathbf{\Psi}(x,\vec{t})$ might not be defined as a function of $\vec{t}$). Our formalism applies equally well to the two cases.

\subsection{Isomonodromic Tau function}

In this section, we assume that $\mathbf{L}(x)$ depends on a family of isomonodromic times $\vec{t} = (t_{\alpha})_{\alpha}$. 
If there is more than one time, we first need a remark. Let us define:
\bea
\Upsilon_{\alpha}(\vec{t}) & = & -\sum_{p \in \mathcal{P}} \Res_{x \rightarrow p} \dd x\,\Tr\big[\mathbf{\Psi}^{-1}(x)(\partial_x \mathbf{\Psi}(x))\,e^{-\mathbf{A}_{p}(x)}(\partial_{t_{\alpha}}e^{\mathbf{A}_{p}(x)})\big] \nonumber \\
& = & - \sum_{p \in \mathcal{P}} \Res_{x \rightarrow p} \dd x\,\Tr\big[\widetilde{\mathbf{\Psi}}^{-1}_p(x)(\partial_x \widetilde{\mathbf{\Psi}}_p(x))\,e^{-\mathbf{A}_{p}(x)}(\partial_{t_{\alpha}}e^{\mathbf{A}_{p}(x)})\big] \nonumber \\
\label{tauded} & = & -\sum_{p \in \mathcal{P}} \Res_{x \rightarrow p} \sum_{a = 1}^{d} \big[\mathcal{W}_1(\sheet{x}{a}) \,\big(e^{-\mathbf{A}_p(x)} \partial_{t_{\alpha}} e^{\mathbf{A}_p(x)}\big)_{a,a}\big].
\eea

\begin{lemma}
\beq
\forall \alpha,\beta,\qquad \partial_{t_{\beta}}\Upsilon_{\alpha}(\vec{t}) = \partial_{t_{\alpha}} \Upsilon_{\beta}(\vec{t}).
\eeq
\end{lemma}
\textbf{Proof.} The definition of $\Upsilon_{\alpha}$ and this result is due to Jimbo, Miwa and Ueno for integrable systems in Lax form and diagonal $\mathbf{A}_{p,k}$ (see also \cite{BBT}). It was generalized to non-diagonal $\mathbf{A}_{p,k}$ in \cite{BertolaMarchal}. The proof is essentially the same. \hfill $\Box$

\begin{definition}
\label{defta}We define the \emph{isomonodromic Tau function} as a function $\mathcal{T}(\vec{t})$ (or as a power 
series in $\vec{t}$), such that:
\beq
\partial_{t_{\alpha}}\ln \mathcal{T}(\vec{t}) = \Upsilon_{\alpha}(\vec{t}).
\eeq
It is defined up to a constant independent of $\vec{t}$.
\end{definition}

Tau functions play an important role in the theory of integrable systems and its applications, and they have been extensively studied, we refer to \cite{BBT} and references therein.

\subsection{Case of an integrable system: expansion of the Tau function}

If $\mathbf{L}$ depends on isomonodromic times $\vec{t}$, an isomonodromic Tau function $\mathcal{T}(\vec{t})$ has been defined in Definition~\eqref{defta}. It is a consequence of Corollary~\ref{W1cora} and the formula \eqref{tauded} for the isomonodromic Tau function that:
\begin{corollary}
If $\mathbf{A}_{p} = \hbar^{-1}\,\mathbf{A}_{p}^{[0]}$ is diagonal for any pole $p$, we have an expansion of the form:
\beq
\ln \mathcal{T}(\vec{t}) = \sum_{k \geq -2} \hbar^{k}\,F^{[k]}(\vec{t}),
\eeq
where:
\beq
\partial_{t_{\alpha}} F^{[k]}(\vec{t}) = -\sum_{p\in{\cal P}}\Res_{x \rightarrow p} \sum_{a = 1}^d \big[\dd x\,\mathcal{W}_1^{[k + 1]}(\sheet{x}{a})\,\partial_{t_{\alpha}} (\mathbf{A}_p^{[0]}(x))_{a,a}\big].
\eeq
\hfill $\Box$
\end{corollary}

\begin{corollary}
\label{co42} In particular, if the TT property holds, then only even powers of $\hbar$ appear:
\beq
\ln \mathcal{T}(\vec{t}) = \sum_{g \geq 0} \hbar^{2g - 2}\,F^{(g)}(\vec{t}),
\eeq
where
\beq
\partial_{t_{\alpha}} F^{(g)}(\vec{t}) = -\sum_{p\in{\cal P}}\Res_{z \rightarrow p}  \big[\omega^{(g)}_1(z)\,f_\alpha(z)\big],
\eeq
with
\beq
\frac{\dd f_\alpha(z)}{\dd x(z)} = \partial_{t_{\alpha}}y(z)\big|_{x(z)}.
\eeq
\end{corollary}
\noindent\textbf{Proof.} 
Indeed, when there is an expansion of topological type, we have ${\cal W}^{[2g-1]}_1(\sheet{x}{a})\,\dd x = \omega^{(g)}_1(z^a(x))$.
 \hfill $\Box$

\subsection{Compatibility of the insertion operator with isomonodromic deformations}

The definition of Picard-Vessiot rings is easily generalized to a family of compatible differential systems $\hbar\partial_x \mathbf \Psi(x) = \mathbf L(x)\mathbf \Psi(x)$ and $\hbar\partial_{t_\alpha} \mathbf \Psi(x) = \mathbf M_\alpha(x)\mathbf \Psi(x)$. We amend Definition~\ref{defdeltay} of insertion operators:
\begin{definition}
We say that an insertion operator $\delta$ is \emph{compatible} if it commutes with all $\partial_{t_{\alpha}}$, i.e. if it satisfies:
\beq
\hbar \partial_{t_\alpha}   \mathbf U(\sheet{y}{a}) 
= \delta^a_y \mathbf M_\alpha(x) + [\mathbf M_\alpha(x),\mathbf U(\sheet{y}{a}) ] + \left[ \frac{\mathbf M_\alpha(x)-\mathbf M_\alpha(y)}{x-y},\mathbf P(\sheet{y}{a}) \right].
\eeq
\end{definition}
The existence of an insertion operator compatible with all times is not something obvious, but if it exists it is quite useful. For the $q$-th reduction of KP, we construct in \S~\ref{sec:compaq} a compatible differential operator, which enables to prove the $O(\hbar^{n - 2})$ axiom of the TT property.

\section{Application to finite reductions of KP}\label{secminmodels}
\label{S4}

In this section, we show an important application of the former formalism, namely to the $q$-th reductions of the KP hierarchy. They are related to the Drinfeld-Sokolov hierarchies \cite{DrinfeldSokolov}, and they contain as a more special case the $(p,q)$ models exemplified in Section~\ref{S6}. They appear in one of the formulation of $2d$ quantum gravity \cite{doug}, and conjecturally describe the algebraic critical points which can arise in hermitian multi-matrix models. In physics, the $(p,q)$ models are expected to describe thermodynamic observables in the coupling of Liouville theory to the $(p,q)$ minimal models of conformal field theory \cite{dFZJ}, the latter corresponding to the classification of finite representations of the conformal Virasoro symmetry of central charge $c = 1 - 6(p - q)^2/pq$ \cite{PagesJaunes}. The $q$-th reduction of KP is related to perturbations of this coupled theory by primary operators. 

\subsection{Pseudo-differential approach to the $q$-th reduction of KP}

Let $t$ be a $1$-dimensional variable, and $\mathcal{C}^{\infty}$ denote an algebra of smooth functions of $t$. Let $\mathbb{D} = \mathcal{C}^{\infty}[\hbar\partial_{t},\hbar^{-1}\partial^{-1}_{t}]$ be the graded algebra of pseudodifferential operators. Let $\mathbb{D}_+ = \mathcal{C}^{\infty}(\mathbb{R})[\hbar\,\partial_{t}]$ its subalgebra of differential operators, graded by the degree. We say that $D \in \mathbb{D}$ is monic of degree $r \geq 0$ if
$$D = \hbar^r\partial_{t}^{r} + \sum_{k = -\infty}^{r - 1} a_k(t)(\hbar\,\partial_{t})^k.$$ 
We then recall that there exists a unique pseudodifferential operator denoted $D^{1/r}$, which is monic of degree $1$ and satisfies $(D^{1/r})^r = D$. We denote $D_+$, the projection of any $D \in \mathbb{D}$ to $\mathbb{D}_+$.

The \emph{string equation} is a relationship
\beq
\label{string}[P,Q] = \hbar
\eeq
between differential operators $P$ and $Q$. It can be written as the compatibility condition of two differential equation for a function $\psi(x,t)$:
\beq
\label{asso}x\psi(x,t) = Q\psi(x,t),\qquad -\hbar\,\partial_x \psi(x,t) = P\psi(x,t).
\eeq
We call \eqref{asso} the \emph{associated linear system}.

Let $(p,q)$ be a couple of positive integers distinct from $(1,1)$. The \emph{$(p,q)$ model} is a hierarchy of $1$-dimensional nonlinear differential equations for a sequence of functions $u(t)$, $u_k(t)$ for $k \in \ldbrack 1,p - 3 \rdbrack$, and $v_l(t)$ for $l \in \ldbrack 1,q - 3 \rdbrack$, ensuing by looking\footnote{The choice $u_{q - 1} = v_{p - 1} = 0$ can always be achieved by a redefinition of the variable $t$. And then $u_{q - 2}/q = v_{p - 2}/p$ follows from the string equation, and we denote $u = - u_{q-2}/q = -v_{p-2}/p$.} for a solution of a string equation of the form:
\bea
\label{sqi}
P = \sum_{k = 0}^{p} v_k(t)\,(\hbar\,\partial_t)^{k},  \qquad  v_p\equiv 1, \, v_{p-1}=0,\, v_{p-2}=- p u, \\
Q = \sum_{l = 0}^{q} u_l(t)\,(\hbar\,\partial_t)^l,\qquad    u_q\equiv 1, \, u_{q-1}=0,\, u_{q-2}=- q u.
\eea
We thus have:
\beq
\label{63}\left\{\begin{array}{l} P = (\hbar\,\partial_t)^p - pu(t)\,(\hbar\,\partial_t)^{p - 2} + \sum_{l = 0}^{p - 3} v_l(t)\,(\hbar\,\partial_t)^l \\
Q = (\hbar\,\partial_t)^{q} - qu(t)\,(\hbar\,\partial_t)^{q - 2} + \sum_{k = 0}^{q - 3} u_k(t)\,(\hbar\,\partial_t)^l \end{array}\right. .
\eeq
When $P$ and $Q$ assume the form \eqref{sqi}, it is well-known that:
\begin{theorem} \cite{DrinfeldSokolov,dFZJ}
The most general solution of \eqref{string} is of the form:
\beq
P = \sum_{l = 0}^{p} t_l\,(Q^{l/q})_+,\qquad Q = \sum_{k = 0}^{q} \widetilde{t}_k\,(P^{k/p})_+.
\eeq
for some constants $t_l$ and $\widetilde{t}_k$ (with $t_p = 1$ and $\widetilde{t}_q = 1$).
\end{theorem}
For coprime $(p,q)$, the $(p,q)$ model is defined by the choice $P = (Q^{p/q})_+$. The string equation $[P,Q]=\hbar$ usually implies a non-linear equation for $u(t)$.

\vspace{0.2cm}

\noindent \textbf{Example of PDEs} for the $(p,q) = (3,2)$ model. Let us denote $\dot{u}(t) = \partial_t u(t)$. We have:
\beq
P=(\hbar\partial_t)^3-3u\hbar\partial_t + v
\qquad
Q=(\hbar\partial_t)^2-2u
\eeq
and the string equation implies
\beq
v = -\frac{3}{2}\,\hbar\,\dot{u} + t_1
\eeq
for some constant $t_1$, and the Painlev\'e I equation for $u(t)$:
\beq
-\frac{1}{2}\,\hbar^2\,\ddot{u} + 3 u^2 = t.
\eeq

\subsection{Constructing the Lax pair by "Folding"}

In this paragraph we show that the associated linear system is an integrable system in Lax form, i.e. it can be written:
\beq
\label{asso2}\hbar\,\partial_x \mathbf{\Psi}(x,t) = \mathbf{L}(x,t)\mathbf{\Psi}(x,t),\qquad \hbar\,\partial_t \mathbf{\Psi}(x,t) = \mathbf{M}(x,t)\mathbf{\Psi}(x,t),
\eeq
for a matrix
\beq
\label{IBWL}\mathbf{\Psi}(x,t) = \left(\begin{array}{ccc} \psi_1(x,t) & \cdots & \psi_q(x,t) \\  (\hbar\,\partial_t)\psi_1(x,t) & \cdots & (\hbar\,\partial_t)\psi_{q}(x,t) \\ \vdots & & \vdots \\ (\hbar\,\partial_t)^{q - 1}\psi_1(x,t) & \cdots & (\hbar\,\partial_t)^{q - 1}\psi_q(x,t) \end{array}\right).
\eeq
where the $\psi_j(x)$ are independent solutions of the associated linear system \eqref{asso}.

It is easy to achieve the second equation with the companion matrix:
\beq
\label{Mform}\mathbf{M}(x,t) = \left(\begin{array}{ccccc} & 1 & & & \\ & & 1 & & \\ & & & \ddots & \\ & & & & 1 \\ x - u_0(t) & -u_1(t) & \cdots & -u_{q - 2}(t) & -u_{q-1}(t) \end{array}\right)
\eeq
where we recall that $u_{q-2}=-q u$, and up to a redefinition of time $t$ we can chose $u_{q-1}\equiv 0$.
We now construct the matrix $\mathbf{L}(x,t)$ to realize the first equation. Naively, $\partial_x\partial_t^{k}\psi$ can be expressed by the action of a differential operator of order $(p + k)$ on $\psi$. But, if we want to write an equation like \eqref{asso2} with $\mathbf{L}(x,t)$ having coefficients which are functions of $x$ -- and not differential operators --, only derivatives of order smaller than $(q - 1)$ are allowed. To bypass this restriction, we can use the first member of \eqref{asso} to express any $q$-th order derivative of $\psi$ in terms of derivatives of lower order. This can be systematized with the notion of folding operators.
\begin{definition}
We define for any integer $l$ the \emph{folding operators}:
\beq
F_l(x,t) = \sum_{j \geq 0} F_{l,j}(x,t)\,(\hbar\,\partial_t)^j \in \mathbb{D}_+[x],
\eeq
by the following recursion:
\beq
F_0(x,t) = 1,\qquad F_{l + 1}(x,t) = (\hbar\,\partial_t) F_l(x,t) + F_{l,q - 1}(x,t)(x - Q).
\eeq
They have the property that for every solution $\psi_l$ of \eqref{asso}
\beq
\forall\,i\in\mathbb Z_+\, ,\,\,\, \forall\,l=1,\dots,q\, ,  \qquad (\hbar\partial_t)^i\,\psi_l(x,t)  = \sum_{j=0}^{q-1} F_{i,j}(x,t)\,(\hbar \partial_t)^j\,\psi_l(x,t)
\eeq
in other words they express any time derivative in terms of only up to order $q-1$ derivatives.
\end{definition}
Notice that $F_l(x,t) = (\hbar\,\partial_t)^{l}$ for $l \in \ldbrack 0,q - 1 \rdbrack$, but:
\beq
F_{q}(x,t) = (\hbar\,\partial_t)^{q} + x - Q = x - qu(t)\,(\hbar\,\partial_t)^{q - 2} - \sum_{k = 0}^{q - 3} u_k(t)\,(\hbar\,\partial_t)^{k}.
\eeq
\begin{lemma}
For any integer $l$, $F_{l,j}(x,t) \equiv 0$ whenever $j \geq q$. Besides, for every solution $\psi$ of \eqref{asso}
\beq
\label{cod}-\hbar\,\partial_x\psi(x,t) = P\psi(x,t) = \Big(\sum_{l = 0}^{p} v_l(t)F_l(x,t)\Big)\psi(x,t)
\eeq
\end{lemma}
\textbf{Proof.} Since $Q$ is monic of degree $q$, the last term in \eqref{cod} prevents $F_l(x,t)$ to have terms of degree higher than $(q - 1)$, as one can show by recursion. Then, recall that $(x - Q)\psi(x,t) = 0$, so these operators satisfy $(\hbar\,\partial_t)^l\psi(x,t) = F_{l}(x,t)\psi(x,t)$, hence \eqref{cod}. \hfill $\Box$.
\begin{definition}
For any integer $k$, we define the operators:
\beq
L_k(x,t) = \sum_{j \geq 0} L_{k,j}(x,t)\,(\hbar\,\partial_t)^{j} \in \mathbb{D}_+[x]
\eeq
by the following recursion:
\beq
L_0(x,t) = - \sum_{l = 0}^{p} v_l(t) F_l(x,t),\qquad L_{k + 1}(x,t) = (\hbar\,\partial_t) L_k(x,t) + L_{k,q - 1}(x,t)(x - Q).
\eeq
\end{definition}
We have similarly:
\begin{lemma}
For any integer $k$, $L_{k,j}(x,t) \equiv 0$ whenever $j \geq q$. \hfill $\Box$
\end{lemma}
We are now in position to conclude:
\begin{proposition}
The first equation of \eqref{asso2} is achieved with $\mathbf{L}(x,t) = (L_{k,j}(x,t))_{0 \leq k,j \leq q - 1}$. \hfill $\Box$
\end{proposition}
In particular, the string equation is equivalent to the compatibility condition of this system:
\beq\label{Laxeqpq}
 [\mathbf{M}(x,t),\mathbf{L}(x,t)] = \hbar\,\partial_t\mathbf{L}(x,t) - \hbar\,\partial_x\mathbf{M}(x,t).
\eeq
By a gauge transformation, one can chose $u_{q - 1}(t) \equiv 0$, i.e. $\mathbf{M}(x,t)$ traceless and therefore $\det \mathbf{\Psi}(x,t)$  independent of $t$. If an initial condition $\mathbf{\Psi}(x,t_0)$ is invertible, $\mathbf{\Psi}(x,t)$ remains invertible for all $t$.

\vspace{0.2cm}

\noindent{\bf Example of folding} for the $(3,2)$ model. We have:
\beq
P=(\hbar \partial_{t})^3 - 3 u \,\hbar \partial_{t} - \frac{3}{2}\,\hbar\,\dot u + t_1
\qquad , \qquad
Q=(\hbar \partial_{t})^2 - 2u.
\eeq
for which the string equation $[P,Q]=\hbar$ implies the Painlev\'e I equation for $u(t)$:
$-\frac{1}{2}\,\hbar^2 \ddot  u + 3 u^2 = t$.
The first folding operators are 
\beq
F_1=\hbar \partial_{t}
\quad ,\,\,
F_2=x+2u
\quad ,\,\,
F_3 = x\,\hbar \partial_{t} +2u\hbar \partial_{t} +2\hbar\dot u
\quad ,\,\,
F_4= x^2+4ux+4u^2+4\hbar^2\dot u\, \partial +2\hbar^2\ddot u.
\eeq
This gives
\bea
L_0 & = & -F_3+3uF_1+(\frac{3}{2}\,\hbar\dot u-t_1)F_0, \nonumber \\
L_1 & = & -F_4+3uF_2+3\hbar\dot u F_1+(\frac{3}{2}\,\hbar\dot u-t_1)F_1+\frac{3}{2}\hbar^2\ddot u F_0,
\eea
and consequently
\beq
\mathbf L(x,t)=\begin{pmatrix}
-\frac{1}{2}\hbar\dot u - t_1 & -x+u \cr
-(x-u)(x+2u)-\frac{1}{2}\hbar^2\ddot u & \frac{1}{2}\hbar\dot u-t_1
\end{pmatrix}
\eeq
and
\beq
\mathbf M(x,t)=\begin{pmatrix}
0 & 1 \cr
x+2u & 0
\end{pmatrix}.
\eeq

\subsection{Semiclassical spectral curve and formal $\hbar$ expansion}

We consider formal solutions of the string equation, i.e. $u_k$ and $v_l$ which have a formal series expansion in $\hbar$. Let us denote:
\beq
u_k(t) = \sum_{m \geq 0} \hbar^m\,u_k^{[m]}(t),\qquad v_l(t) = \sum_{m \geq 0} \hbar^{m}\,v_l^{[m]}(t).
\eeq
\begin{lemma} $u_k^{[0]}(t)$ and $v_l^{[0]}(t)$ can be obtained by replacing $\hbar\,\partial_{t}$ by a variable $z \in \widehat{\mathbb{C}}$. Namely, one defines
\beq\label{defxypqlim}
\left\{\begin{array}{rcl} X(z) & := & \sum_{k = 0}^{q} u_k^{[0]}(t)\,z^{l} \\
Y(z) & := & \sum_{l = 0}^{p} v_l^{[0]}(t)\,z^{k}\end{array}\right. .
\eeq
(which are the $\hbar \to 0$ semiclassical limit of $Q$ and $P$). The leading order in $\hbar$ of the string equation becomes a Poisson bracket:
\beq
\label{dsanu}\partial_z Y(z)\partial_t X(z) - \partial_z X(z) \partial_t Y(z) = 1,
\eeq
which gives an algebraic constraint on $u_k^{[0]}$ and $v_l^{[0]}$.
\end{lemma}
\textbf{Proof.} 
The leading order of $[P,Q]$ is:
\beq
\hbar = [P,Q] =\sum_{k,l}  \hbar\,l\,v^{[0]}_l\,\dot u^{[0]}_k\,\partial_{t}^{k+l-1} - \sum_{k,l} \hbar\,k\,u^{[0]}_k\, \dot v^{[0]}_l\,\partial_{t}^{k+l-1} + O(\hbar^2),
\eeq
i.e. this means that
\beq
Y'(z)\dot X(z)- X'(z)\dot Y(z) = 1.
\eeq
 \hfill $\Box$

\bl
A solution of \eqref{dsanu} is obtained as follows
\beq\label{xyfguz2}
X(z)  =(u^{[0]})^{q/2}\,f\big(z\,(u^{[0]})^{-1/2}\big)
\qquad , \qquad
Y(z)  =(u^{[0]})^{p/2}\,g\big(z\,(u^{[0]})^{-1/2}\big),
\eeq
where $u^{[0]} = (t/\rho)^{\frac{2}{p+q-1}}$, and the functions $f$ and $g$ satisfy:
\beq\label{fgpoisson}
q f(\zeta) g'(\zeta) - p g(\zeta) f'(\zeta) = (p+q-1)\rho,
\eeq
and $\rho$ is chosen such that at large $\zeta$ the solution of \eqref{fgpoisson} behaves as $f(\zeta)=\zeta^q(1-q u^{[0]} \zeta^{-2}+ O(\zeta^{-3}))$ and $g(\zeta)=\zeta^p\big(1-p u^{[0]} \zeta^{-2}+ O(\zeta^{-3})\big)$. We call it the \emph{homogeneous solution}.
\el
\textbf{Proof.} 
The result is claimed in \cite{dFZJ}. Let us prove it directly. If we assume the form \eqref{xyfguz2}, and write $\zeta=z\,(u^{[0]})^{-1/2}$, then we have
\beq
\partial_{z}X = (u^{[0]})^{(q-1)/2}\,f'(\zeta),\qquad 
\partial_{t}X = \frac{1}{2}\,\partial_{t} u^{[0]}\,\left( q\,(u^{[0]})^{(q-2)/2}\,f(\zeta) - (u^{[0]})^{(q-3)/2}\,f'(\zeta)\right),
\eeq
\beq
\partial_{z} Y = (u^{[0]})^{(p-1)/2}\,g'(\zeta),\qquad 
\partial_{t} Y = \frac{1}{2}\,\partial_{t}u^{[0]}\,\left( p\,(u^{[0]})^{(p-2)/2}\,g(\zeta) - (u^{[0]})^{(p-3)/2}\,g'(\zeta)\right).
\eeq
It follows:
\beq
1 = \partial_{t} X\,\partial_{z} Y - \partial_{t} Y\,\partial_{z} X = \frac{1}{2}\,\partial_{t}u^{[0]}\,\,(u^{[0]})^{(p+q-3)/2}\, (q f g' - p g f'),
\eeq
which is satisfied if $u^{[0]} = (t/\rho)^{\frac{2}{p+q-1}}$ and $q fg'-p g f' = (p+q-1)\rho $. \hfill $\Box$

\bl
If $p + q \geq 4$, this implies when $\zeta \rightarrow \infty$ that:
\bea
f(\zeta) &=& g(\zeta)^{q/p} - \frac{\rho}{p}\,\zeta^{1-p} \left( 1+ u^{[0]} \left(p-2+\frac{2}{p+q+1}\right)\zeta^{-2} + O( \zeta^{-3}) \right), \nonumber \\
g(\zeta) &=& f(\zeta)^{p/q} -  \frac{\rho}{q}\,\zeta^{1-q} \left( 1+ u^{[0]} \left(q-2+\frac{2}{p+q+1}\right)\zeta^{-2} + O( \zeta^{-3}) \right). \nonumber
\eea
In particular:
\beq
f = (g^{q/p})_+
\quad , \quad
g = (f^{p/q})_+.
\eeq

\el

\noindent{\bf Proof.} Write $f=g^{q/p}\,h$, the equation then gives:
\beq
-p\,\frac{h'}{h} = \frac{(p+q-1)\,\rho}{fg} = \frac{(p+q-1)\rho}{\zeta^{p+q}}\,\big(1+(p+q)u^{[0]}\zeta^{-2}+O(\zeta^{-3})\big),
\eeq
and upon integration:
\beq
\ln h = \frac{\rho}{p}\,\zeta^{1-p-q}\,\Big(1 + \frac{(p+q)(p+q-1)\,u^{[0]}}{p+q+1}\,\zeta^{-2} + O(\zeta^{-3})\Big).
\eeq
Then, using $p + q \geq 4$ to ensure that $2(p + q - 1) > p + q + 1$, we can exponentiate:
\beq
h = 1 + \frac{\rho}{p}\,\zeta^{1-p-q}\,\Big(1 + \frac{(p+q)(p+q-1)\,u^{[0]}}{p+q+1}\,\zeta^{-2} + O(\zeta^{-3})\Big).
\eeq
We then multiply by $g^{q/p} = \zeta^q \big(1-qu^{[0]}\zeta^{-2}+O(\zeta^{-3})\big)$ and get
\beq
f = g^{q/p} + \frac{\rho}{p}\,\zeta^{1-p}\,\Big(1 + (p-2+\frac{2}{p+q+1})\,u^{[0]}\,\zeta^{-2} + O(\zeta^{-3})\Big).
\eeq
We have the same proof for $g$.
\hfill $\Box$

\subsubsection{Special solutions}

In the $(p,q)$  model, we have $P = (Q^{p/q})_+$ and similarly $Q = (P^{q/p})_+$. Therefore, at the semiclassical limit, we find $Y(z) = (X^{p/q}(z))_+$ and $X(z) = (Y^{q/p}(z))_+$. The relation \eqref{dsanu} can be solved explicitly in the case $p = (2m + 1)q \pm 1$ for some integer $m$ \cite{dFZJ}:
\beq
\left\{\begin{array}{rcl} f(\zeta) & = & \sum_{n = 0}^{m} \frac{\Gamma(n + 1)}{\Gamma(p/q + 1)\Gamma(n - p/q + 1)}\,T_{p - 2nq}(\zeta)  \\ g(\zeta) & = & T_{q}(\zeta) \end{array}\right. ,\qquad \rho = 2p.
\eeq
where $T_{l}(2\cos \theta) = 2\cos(l\theta)$ are the Chebyshev polynomials of the first kind. In particular, for the so-called "unitary" models $p = q + 1$, we find: 
\beq
\left\{\begin{array}{rcl} f(\zeta) & = & T_{q + 1}(\zeta) \\ g(\zeta) & = & T_{q}(\zeta) \end{array}\right. ,\qquad \rho = 2(q + 1).
\eeq

\subsection{Semi-classical spectral curve}

\bp
In the semiclassical limit $\hbar\to 0$, the eigenvalues of $\mathbf M(x,t)$ and $\mathbf L(x,t)$ are given by the functions $x(z)$ and $y(z)$ defined in \eqref{defxypqlim}, by:
\beq
z={\rm eigenvalue\,of}\,\,\mathbf M^{[0]}(x,t) \quad \Longleftrightarrow \quad x=X(z)=\sum_{k=0}^q u_k^{[0]}(t)\,z^k.
\eeq
\beq
y={\rm eigenvalue\,of}\,\,\mathbf L^{[0]}(X(z),t) \quad \Longleftrightarrow \quad y=Y(z)=\sum_{l=0}^p v_l^{[0]}(t)\,z^k.
\eeq
The leading order spectral curve, i.e. the locus of eigenvalues of $\mathbf L^{[0]}(x(z),t)$, is a genus 0 algebraic plane curve.
\ep
\noindent {\bf Proof.} Since $\mathbf M(x,t)$ is a companion matrix, its characteristic polynomial is
\beq
0= \det\left( z\,\mathbf 1_q - \mathbf M(x,t)\right)
= x-\sum_{k=0}^q u_k(t)\,z^k,
\eeq
therefore in the limit $\hbar\to 0$, the eigenvalues of $\mathbf M^{[0]}(x,t)$ are the  $z$'s such that $X(z)=x$:
\beq
\sum_{k=0}^q u_k^{[0]}(t)\,z^k = x = X(z),
\eeq
where $X(z)$ is the function introduced in \eqref{defxypqlim}. It follows that in the limit $\hbar\to 0$, $\hbar \partial_t \psi(x,t) \sim z\,\psi(x,t)\,(1+O(\hbar))$. The eigenvalues $y$ of $\mathbf L(x,t)$, by definition are such that
\beq
y\,\psi(x,t) = -\hbar\,\partial_x \psi(x,t) = \sum_{l=0}^p v_l(t)\,(\hbar\partial_t)^l\,\psi(x,t),
\eeq
and thus in the $\hbar\to 0$ limit, the eigenvalues of $\mathbf L^{[0]}(x,t)$ are such that
\beq
y=Y(z) = \sum_{l=0}^p v_l^{[0]}(t)\,z^l.
\eeq
The spectral curve $P(x,y) = \det(y\,\mathbf 1_q - \mathbf L^{[0]}(x,t))$ is a polynomial of $x$ and $y$, monic of degree $q$ in $y$, which vanishes if and only if $y$ is an eigenvalue of $\mathbf L^{[0]}(x)$, i.e. if and only if there exists some $z$ such that $x=X(z)$ and $y=Y(z)$. Therefore $P(x,y)$ is proportional to the resultant of the polynomials $X(z)-x$ and $Y(z)-y$:
\bea
(-1)^q\,\mathcal{P}(x,y) &=& {\rm Resultant}(X(z)-x,Y(z)-y) \cr 
&=& 
\det{\begin{pmatrix}
1 & u^{[0]}_{q-1} & u^{[0]}_{q-2} & \dots & u^{[0]}_{1} & u^{[0]}_{0}-x & & & & \cr
& 1 & u^{[0]}_{q-1} & u^{[0]}_{q-2} & \dots & u^{[0]}_{1} & u^{[0]}_{0}-x & & &  \cr
& & \ddots &  &  &  &  & \ddots & &   \cr
& & & 1 & u^{[0]}_{q-1} & u^{[0]}_{q-2} & \dots & u^{[0]}_{1} & u^{[0]}_{0}-x &   \cr
1 & v^{[0]}_{p-1}  & \dots & v^{[0]}_{1} & v^{[0]}_{0}-y & & & & & \cr
& 1 & v^{[0]}_{p-1}  & \dots & v^{[0]}_{1} & v^{[0]}_{0}-y & & & &  \cr
& & \ddots &   &  &  & \ddots & & &   \cr
& & & \ddots &   &  &  & \ddots & &    \cr
 & & & & 1 & v^{[0]}_{p-1}  & \dots & v^{[0]}_{1} & v^{[0]}_{0}-y   \cr
\end{pmatrix}}. \cr \nonumber 
\eea
As mentioned above, it admits a parametric solution:
\beq
\mathcal{P}(X(z),Y(z))=0
\eeq
with $x$ and $y$ polynomials of $z$. This means that there is a holomorphic map $z\mapsto (X(z),Y(z))$ from the Riemann sphere $\widehat{\mathbb C}$ to the spectral curve (the locus of $\mathcal{P}(x,y)=0$ in $\mathbb C\times \mathbb C$). In particular this implies that the spectral curve is an algebraic plane curve of genus $\mathfrak{g} = 0$. \hfill $\Box$

\subsection{Asymptotic expansion and TT property}
\label{Sasyn}

As in Section~\ref{S32}, we look for asymptotics of the form:
\beq
\mathbf \Psi(x,t) \sim \mathbf V(x,t)\,\widehat{\mathbf \Psi}(x,t)\,\,\e^{\frac{1}{\hbar}\,\mathbf S(x,t)},
\eeq
where:
\begin{itemize}
\item[$\bullet$] $\mathbf S(x,t)={\rm diag}(S(z_a))_{1 \leq a \leq q}$ is such that $\partial_t S_a(z)|_{X(z) = x} = z^i$ are the eigenvalues of $\mathbf M^{[0]}(x,t)$, where $z = z_a$ is related to $x$ by
\beq
x=X(z) = z^q -q u^{[0]}(t)\,z^{q-2} +  \sum_{k=0}^{q-2} u_k^{[0]}(t)\,z^k.
\eeq
Thanks to \eqref{dsanu}, it also satisfies:
\beq
\partial_{x} S_a(z) = Y(z_a)
\eeq
where $Y(z_a)$ are the eigenvalues of $\mathbf L^{[0]}(x,t)$.
\item[$\bullet$] $\mathbf V(x,t)$ is a matrix whose columns are eigenvectors of both $\mathbf M^{[0]}(x,t)$ and $\mathbf L^{[0]}(x,t)$, normalized such that $\mathbf V^{-1}\,\partial_{x} \mathbf V(x,t)$ has a vanishing diagonal. Since $\mathbf{M}^{[0]}(x,t)$ is a companion matrix, ${\mathbf{V}}(x,t)$ can be found rather explicitly, as a Vandermonde matrix, with columns normalized by a factor $1/\sqrt{X'(z_a)}$:
\beq
{\mathbf{V}}_{a,b}(x,t) = \frac{(z_b(x))^{a-1}}{\sqrt{X'(z_a)}} \quad {\rm where} \,\, x = {X}(z_b) =  \sum_{k=0}^{q} u_k^{[0]}(t)\,z_b^k.
\eeq
Its inverse is
\beq
({\mathbf{V}}^{-1})_{a,b} = \frac{({X}(z_a(x))z_a(x)^{-b})_+}{\sqrt{X'(z_a(x))}} = \frac{\sum_{k=b}^{q} u_k^{[0]}(t)\,\,z_a(x)^{k-b}}{\sqrt{X'(z_a(x))}}.
\eeq
It satisfies:
\bea
\label{eqVxij}
{\rm if}\,a\neq b & \quad ({\mathbf{V}}^{-1}\,\partial_x{{\mathbf{V}}})_{a,b} & = \frac{\sqrt{X'(z_b)}}{\sqrt{X'(z_a)}}\,\frac{1}{z_a-z_b} = O(x^{-1/q}), \\
\label{eqVxii}
{\rm if}\,a=b & \quad ({\mathbf{V}}^{-1}\,\partial_x{{\mathbf{V}}})_{a,a} & = 0, \\
\label{eqVtij}
{\rm if}\,a\neq b & \quad ({\mathbf{V}}^{-1}\,\partial_t{\mathbf{V}})_{a,b} & = \frac{\partial_{t} X(z_b)}{\sqrt{X'(z_a)\,X'(z_b)}}\,\frac{1}{z_a-z_b} = O(x^{-2/q}), \\
\label{eqVtii}
{\rm if}\,a= b & \quad ({\mathbf{V}}^{-1}\,\partial_t {\mathbf V})_{a,a} & = -\,\frac{1}{2}\,\frac{\dot X'(z_a)}{X'(z_a)}  = O(x^{-2/q}).
\eea
\item[$\bullet$] The matrix $\widehat{\mathbf \Psi}(x,t)=\mathbf 1_q + O(\hbar)$ has a formal asymptotic series as $\hbar\to 0$. From $\hbar \partial_t \mathbf \Psi\cdot\mathbf \Psi^{-1}=\mathbf M = \mathbf M^{[0]} - \mathbf e_q (\mathbf u-\mathbf u^{[0]})^T$, where $\mathbf e_q=(0,0,\dots,0,1)$ and $\mathbf u=(u_0,\dots,u_{q-1})$, we get the equation for $\widehat{\mathbf \Psi}$ involving the diagonal matrix $\mathbf{Z} = {\rm diag}(z_1,\ldots,z_q)$ of eigenvalues  of $\mathbf{M}^{[0]}$:
\beq\label{eqpsihatpqM}
[\mathbf Z,\widehat{\mathbf \Psi}] = {\mathbf{V}}^{-1}\,\mathbf e_q\,(\mathbf u-\mathbf u^{[0]})^t\,{\mathbf{V}}\,\widehat{\mathbf \Psi} + {\mathbf{V}}^{-1}\,\hbar\partial_t {\mathbf{V}}\,\,\widehat{\mathbf \Psi} + \hbar \partial_t \widehat{\mathbf \Psi},
\eeq
i.e.
\beq\label{eqpsihatpqMij}
 (z_a-z_b)\,\widehat{\mathbf \Psi}_{a,b} = \sum_{l=1}^q\,\,\frac{\sum_{k=0}^{q-2} (u_k-u_k^{[0]})\,z_l^{k}}{\sqrt{X'(z_a)\,X'(z_b)}}\,\,\widehat{\mathbf \Psi}_{l,b} +\hbar \sum_{l=1}^q ({\mathbf{V}}^{-1}\partial_t {\mathbf{V}})_{a,l}\,\widehat{\mathbf \Psi}_{l,b} + \hbar\partial_t \widehat{\mathbf \Psi}_{a,b}.
\eeq
This equation uniquely determines $\widehat{\mathbf \Psi}=\mathbf{1}_{q}+O(\hbar)$ as its asymptotic expansion in powers of $\hbar$. In fact it also uniquely determines $\widehat{\mathbf \Psi}= \mathbf{1}_{q} +O(x^{-1/q})$ as an asymptotic series at large $x$, in powers of $x^{1/q}$. From $\hbar \partial_x \mathbf \Psi\cdot\mathbf \Psi^{-1}=\mathbf L $ we also get an ODE for $\widehat{\mathbf \Psi}$:
\beq\label{eqpsihatpqL}
\mathbf V^{-1}\mathbf L\mathbf V\, \widehat{\mathbf \Psi} - \widehat{\mathbf \Psi} \,\mathbf \Lambda^{[0]}\,  =  \hbar {\mathbf{V}}^{-1}\partial_x {\mathbf{V}}\,\,\widehat{\mathbf \Psi} + \hbar \partial_x \widehat{\mathbf \Psi}.
\eeq
\end{itemize}

We observe that the semiclassical spectral curve has genus $0$. Therefore, we will be able to apply Theorem~\ref{topro} if we can show:
\begin{itemize}
\item[$\bullet$]  the existence of a $\hbar \leftrightarrow -\hbar$ symmetry. This is a technical but simple check done in \S~\ref{sysm}.
\item[$\bullet$] that the $n$-point correlators $\mathcal{W}_n(x_1,\ldots,x_n)$ are $O(\hbar^{n - 2})$ after a suitable gauge transformation. This is a non-trivial property of $(p,q)$ models, that we establish in \S~\ref{sysm2} by constructing an insertion operator $\delta_{x}^a$ which is compatible with $\partial_{t}$. 
\item[$\bullet$] the pole property, i.e. that $\omega_n^{(g)}$ have poles only at ramification points, established in \S~\ref{Spropole}.
\end{itemize}
The consequences of Theorem~\ref{topro} for the $(p,q)$ models are gathered in Section~\ref{Sto}.

\subsection{$\hbar \leftrightarrow -\hbar$ symmetry}
\label{sysm}
The goal of this subsection is that the $(p,q)$ models admit conjugated solutions in the terminology of \S~\ref{SSym}:
\begin{theorem}
\label{theoa} For any invertible solution $\mathbf{\Psi}(x,t)$ of \eqref{asso2} with coupling constant $\hbar$, there exists a solution $\mathbf{\Phi}(x,t)$ of \eqref{asso2} with coupling constant $-\hbar$, such that $\mathbf{\gamma}(x,t) = \mathbf{\Phi}(x,t)\mathbf{\Psi}^{T}(x,t)$ is independent of $x$.
\end{theorem}
This theorem is proved below, but in order to do so, we need some intermediate results and definitions:
 
We first introduce a conjugation operator:
\begin{definition}
There is a unique antilinear operator $\dagger\,:\,\mathbb{D} \rightarrow \mathbb{D}$, such that:
\begin{itemize}
\item[$\bullet$] for any $f \in \mathcal{C}^{\infty} \subseteq \mathbb{D}$, $f^{\dagger} = f$.
\item[$\bullet$] $(\hbar\partial_t)^{\dagger} = -(\hbar\partial_t)^{\dagger}$.
\item[$\bullet$] for any $D_1,D_2 \in \mathbb{D}$, $(D_1D_2)^{\dagger} = D_2^{\dagger}D_1^{\dagger}$.
\end{itemize}
\end{definition}
In particular, if $P,Q \in \mathbb{D}_+$ satisfy $[P,Q] = \hbar$, then $[P^{\dagger},Q^{\dagger}] = -\hbar$. Moreover, if $P$ and $Q$ are differential operators of the form \eqref{63}, so are $P^{\dagger}$ and $Q^{\dagger}$. To summarize, $\dagger$ puts in correspondence the models with coupling constant $\hbar$ and $-\hbar$.

The linear system associated to $(P^{\dagger},Q^{\dagger})$ is:
\beq
\label{qiqo}x\phi(x,t) = Q^{\dagger}\phi(x,t),\qquad \hbar\,\partial_x \phi(x,t) = P^{\dagger}\phi(x,t).
\eeq
If $\phi_1(x,t),\ldots,\phi_q(x,t)$ denotes a family of solutions of \eqref{qiqo}, we can define a matrix:
\beq
\mathbf{\Phi}(x,t) = \left(\begin{array}{ccc} \phi_1(x,t) & \cdots & \phi_q(x,t) \\  (\hbar\,\partial_t)\phi_1(x,t) & \cdots & (\hbar\,\partial_t)\phi_{q}(x,t) \\ \vdots & & \vdots \\ (\hbar\,\partial_t)^{q - 1}\phi_1(x,t) & \cdots & (\hbar\,\partial_t)^{q - 1}\phi_q(x,t) \end{array}\right),
\eeq
As before, we can represent \eqref{qiqo} in Lax form, and we denote $\mathbf{L}_{-\hbar}(x,t)$ and $\mathbf{M}_{-\hbar}(x,t)$ the corresponding Lax matrices:
\beq
-\hbar\,\partial_x \mathbf{\Phi}(x,t) = \mathbf{L}_{-\hbar}(x,t)\mathbf{\Phi}(x,t),\qquad -\hbar\,\partial_t \mathbf{\Phi}(x,t) = \mathbf{M}_{-\hbar}(x,t)\mathbf{\Phi}(x,t).
\eeq
The following result gives a correspondence between solutions of the associated linear systems of $(P,Q)$ and $(P^{\dagger},Q^{\dagger})$.
\begin{proposition}
Let $\psi_1,\ldots,\psi_{q}$ be a basis of solutions of \eqref{asso}, $\Psi(x,t)$ as defined in \eqref{IBWL},  and define:
\bea
\Delta(x) & = & \det \mathbf{\Psi}(x,t), \\
\Delta_{i_0 - 1,j_0}(x,t) & = & \det \big[(\hbar\,\partial_t)^{i - 1}\psi_{j}(x,t)\big]_{1 \leq i,j \leq q}^{i \neq i_0,\,\,j \neq j_0}, \\
\widetilde{\phi}_{j}(x,t) & = & \Delta_{q - 1,j}(x,t).
\eea
then $(\widetilde{\phi}_{j}(x,t))_{1 \leq j \leq q}$ is a basis of solutions of \eqref{qiqo}.
\end{proposition}
The proof of this proposition relies on a technical result:
\begin{lemma}
Let $j \in \ldbrack 1,q \rdbrack$. With the convention $\Delta_{-1,j} \equiv 0$, we have for any $i \in  \ldbrack 0,q - 1 \rdbrack$,
\beq
\label{3221}\hbar\,\partial_t \Delta_{i,j}(x,t) = \Delta_{i - 1,j}(x,t) +  (-1)^{q  - j}(u_i(t) - \delta_{i,0}x)\Delta_{q - 1,j}(x,t),
\eeq
and for any $k \in \ldbrack 1,q + 1 \rdbrack$,
\beq
\label{3222}\hbar\,\partial_t \Delta_{q - k,j}(x,t) = \Big(\sum_{l = 1}^{k} (-1)^{l + 1} (\hbar\,\partial_t)^{k - l}[u_{q - l + 1}(t)\Delta_{q - 1,j}(x,t)]\Big) + \delta_{k,q + 1}(-1)^{q} x\Delta_{q - 1,j}(x,t).
\eeq
\end{lemma}
\textbf{Proof.} By multilinearity, we can differentiate the minors $\Delta_{i,j}$ line by line:
\beq
\hbar\,\partial_t \Delta_{i,j} = \det\left(\begin{array}{c} \psi_m \\
\vdots \\ (\hbar\,\partial_t)^{i - 2} \psi_m \\
(\hbar\,\partial_t)^{i} \psi_m \\
\hline
(\hbar\,\partial_t)^{i + 1} \psi_m \\
\vdots \\
(\hbar\,\partial_t)^{q - 1} \psi_m
\end{array}\right)_{m \neq j} \!\!\!\!\! + \det\left(\begin{array}{c} \psi_m \\
\vdots \\
(\hbar\,\partial_t)^{i - 1} \psi_m \\
\hline
(\hbar\,\partial_t)^{i + 1} \psi_m \\
\vdots \\
(\hbar\,\partial_t)^{q - 2} \psi_m \\
(\hbar\,\partial_t)^{q} \psi_m
\end{array}\right)_{m \neq j} \!\!\!\!\! + \sum_{\substack{k = 0 \\ k \neq i}}^{q - 2} \det\left(\begin{array}{c} \psi_m \\
\vdots \\
(\hbar\,\partial_t)^{k - 2} \psi_m \\ (\hbar\,\partial_t)^{k} \psi_m \\
(\hbar\,\partial_t)^{k} \psi_m \\
\vdots \\
\hline
\vdots \\
(\hbar\,\partial_t)^{q - 1} \psi_m
\end{array}\right)_{m \neq j}.
\eeq
The non-zero contributions arise only from the terms where:
\begin{itemize}
\item[$\bullet$] the $i$-th line is differentiated. We then recognize the definition of $D_{i - 1,j}(x,t)$.
\item[$\bullet$] the $(q - 1)$-th line is differentiated. Since $\psi_1,\ldots,\psi_q$ are solutions of \eqref{asso}, we can replace $(\hbar\partial_t)^{q}\psi_m$ by a $x\psi_m - \sum_{l = 0}^{q - 2} u_l(t)\,(\hbar\,\partial_t)^{k}$. By subtraction of the other lines, we may keep in the latter only the term involving a derivative of order $i$-th, which was absent from the minor. We thus recreate a minor $D_{q - 1,j}(x,t)$, with a prefactor $(x\delta_{i,0} - u_{i}(t))$, and up to a sign $(-1)^{q - i}$ taking into account the ordering of the lines.
\end{itemize} 
We therefore arrive to \eqref{3221}, and \eqref{3222} follows by recursion. In particular, we obtain at the last step of the recursion ($k = q + 1$):
\bea
0 = \Delta_{-1,j}(x,t) & = &\Big(\sum_{l = 1}^{q + 1} (-1)^{l + 1} (\hbar\,\partial_t)^{k - l} u_{q - l + 1}(t) + (-1)^qx\Big)\Delta_{q - 1,j}(x,t) \nonumber \\
& = & (-1)^{q}(x - Q^{\dagger})\Delta_{q - 1,j}(x,t).
\eea
Accordingly, $\widetilde{\phi}_j(x,t) \equiv \Delta_{q - 1,j}(x,t)$ provides a solution of \eqref{qiqo} for any $j \in \ldbrack 1,q \rdbrack$. To show that $(\widetilde{\phi}_j)_j$ is a basis, we define the matrix $\mathbf{\Phi}(x,t) = [(\hbar\,\partial)^{i - 1}\widetilde{\phi}_{j}]_{1 \leq i,j \leq q}$ and compute its determinant. Thanks to \eqref{3221}, we may write:
\beq
\det \widetilde{\mathbf{\Phi}} = \det\left(\begin{array}{c} \Delta_{q - 1,m} \\
\hbar\,\partial_t \Delta_{q - 1,m} \\ \vdots \\ (\hbar\,\partial_t)^{q - 1}\Delta_{q - 1,m} \end{array}\right)_{1 \leq m \leq j} = \det\left(\begin{array}{c}\Delta_{q - 1,m} \\ \Delta_{q - 2,m} + (u_{q - 1}(t) - x\delta_{q,1})\Delta_{q - 1,m} \\ \vdots \\ (\hbar\,\partial_t)^{q - 1} \Delta_{q - 1,m} \end{array}\right),
\eeq
and upon subtracting the first line in the second line, we can replace the second line by $[\Delta_{q - 2,m}]_{1 \leq m \leq q}$. We find recursively that the $i$-th line can be replaced by $\Delta_{q - i,m}$, and thus:
\bea
\det \widetilde{\mathbf{\Phi}} & = & \det [\Delta_{q - k,j}]_{1 \leq j,k \leq q} = (\det \Psi)^q\,\det\Big[(-1)^{j - 1}\frac{\Delta_{k - 1,j}}{\det \Psi}\Big]_{1 \leq j,k \leq q} \nonumber \\
& = & (\det \Psi)^{q - 1}.
\eea
So, $(\widetilde{\phi}_j)_{j}$ is a basis of solutions of \eqref{qiqo} if and only if $(\psi_j)_j$ is a basis of solutions of \eqref{asso} \hfill $\Box$

\vspace{0.2cm}

In order to obtain Theorem~\ref{theoa}, we exploit the freedom to choose a normalization of $\phi_j(x,t)$ depending on $x$. As we shall see, an appropriate choice is:
\beq
\phi_j(x,t) = (-1)^j \frac{\widetilde{\phi}_{j}(x,t)}{\det \mathbf{\Psi}(x)} = (-1)^{j} \frac{\Delta_{q - 1,j}(x,t)}{\Delta(x)} = (-1)^{q - 1}\mathbf{\Psi}^{-1}_{j,q - 1}(x,t),
\eeq
and we define the matrix:
\beq
\mathbf{\Phi}(x,t) = \left(\begin{array}{ccc} \phi_1(x,t) & \cdots & \phi_q(x,t) \\ -\hbar\,\partial_t \phi_1(x,t) & \cdots & -\hbar\,\partial_t \phi_q(x,t) \\ \vdots & & \vdots \\ (-\hbar\,\partial_t)^{q}\phi_1(x,t) & \cdots & (-\hbar\partial_t)^{q} \phi_q(x,t)\end{array}\right).
\eeq
It remains to show that:
\beq
\label{eqref}\mathbf{C}_{i,j}(x,t) = \sum_{k = 1}^q [(\hbar\,\partial_{t})^{i - 1} \phi_{k}(x,t)]\,[(\hbar\,\partial_t)^{j - 1}\psi_k(x,t)],\qquad i,j \in \ldbrack 1,q \rdbrack
\eeq
does not depend on $x$. For this purpose, we first observe:
\beq
\label{ini}\forall j \in \ldbrack 1,q \rdbrack,\qquad \mathbf{C}_{1,j} = \sum_{k = 1}^q  (-1)^q \mathbf{\Psi}^{-1}_{k,q - 1} \mathbf{\Psi}_{i - 1,k} = (-1)^{q - 1}\delta_{i,q}.
\eeq
Besides, from the very structure of \eqref{eqref}, we observe:
\beq
\forall i,j \in \ldbrack 1,q - 1 \rdbrack,\qquad  \hbar\,\partial_{t}\mathbf{C}_{i,j}
 = \mathbf{C}_{i,j + 1} - \mathbf{C}_{i + 1,j},
 \eeq
and when $j = q$, we use the fact that $\psi_j$ is solution to the system \eqref{asso} to write:
\beq
\label{sau}\forall i \in \ldbrack 1,q - 1 \rdbrack,\qquad \hbar\,\partial_t \mathbf{C}_{i,q - 1} = - \mathbf{C}_{i + 1,q - 1} - \sum_{l = 0}^{q - 2} (u_l(t) - \delta_{l,0}x)\mathbf{C}_{i,l + 1}.
\eeq
Considering \eqref{ini} as an initial condition for \eqref{sau}, we obtain by recursion that $\mathbf{C}_{i,j} = 0$ whenever $i + j \leq q$. Hence, $\sum_{l = 0}^{q - 2} \delta_{l,0}\mathbf{C}_{i,l + 1}$ always vanish. This implies that the recursion relation \eqref{sau} does not depend on $x$. Since $\mathbf{C}_{i,j}$ is determined uniquely from \eqref{sau} with the constant initial condition \eqref{ini}, we conclude that $\mathbf{C}$ does not depend on $x$, which completes the proof of Theorem~\ref{theoa}.

\subsection{The $\hbar^{n - 2}$ property}
\label{sysm2}

We are going to construct a suitable insertion operator allowing to prove the $\hbar^{n - 2}$ property. 

\subsubsection{A useful decomposition}

The very special form \eqref{Mform} of the matrix $\mathbf{M}(x,t)$ in $(p,q)$ models allows a decomposition:
\begin{lemma}
\label{thep}$\mathbf{P}(\sheet{x}{a}) = \mathbf{A}(\sheet{x}{a}) + x\,\mathbf{B}(\sheet{x}{a}) + \hbar \,\mathbf{C}(\sheet{x}{a})$ where $\mathbf{A}$ and $\mathbf{B}$ do not depend on $\hbar$ and have the properties:
\bea
\label{comut1} [\mathbf{A}(\sheet{x}{a},t),\mathbf{A}(\sheet{y}{b},t)] & = & 0 \\
\label{comut2} [\mathbf{B}(\sheet{x}{a},t),\mathbf{B}(\sheet{y}{b},t)] & = & 0 , \\
\label{comut3}[\mathbf{A}(\sheet{x}{a},t),\mathbf{B}(\sheet{y}{b},t)] & = & [\mathbf{A}(\sheet{y}{b},t),\mathbf{B}(\sheet{x}{a},t)],
\eea
and $\mathbf{C}$ depends on $\hbar$, is $O(1)$, and is expressible in terms of matrix elements of $\mathbf{P}(x,t)$ and their time derivatives.
\end{lemma}
\textbf{Proof.} The projectors $\mathbf{P}$ satisfy the evolution equation:
\beq
\label{ucom}\hbar\,\partial_t \mathbf{P}(\sheet{x}{a},t) = [\mathbf{M}(x,t),\mathbf{P}(\sheet{x}{a},t)].
\eeq
We have:
\beq
M_{l,m}(x,t) = \delta_{m,l + 1} + \delta_{l,q}\big(x\,\delta_{m,1} - u_{m - 1}(t)\big),
\eeq
hence:
\beq
[\mathbf{M}(x,t)\,\mathbf{P}(\sheet{x}{a},t)]_{l,n} = P_{l + 1,n}(\sheet{x}{a},t) + \delta_{l,d}\Big(x\,P_{1,n}(\sheet{x}{a},t) - \sum_{m = 1}^q u_{m - 1}\,P_{m,n}(\sheet{x}{a},t)\Big),\nonumber
\eeq
\beq
[\mathbf{P}(\sheet{x}{a},t)\,\mathbf{M}(x,t)]_{l,n} = P_{l,n - 1}(\sheet{x}{a},t) + \big(x\,\delta_{n,1} - u_{n - 1}(t)\big)P_{l,q}(\sheet{x}{a},t). \nonumber
\eeq
Omitting to precise the variables, \eqref{ucom} implies the relations:
\bea
1 \leq l < d & \quad & \hbar\,\partial_t P_{l,1} = P_{l + 1,1} - (x - u_0)P_{l,q}, \nonumber \\
1 \leq l < d,\,\, 1 < n \leq d & \quad & \hbar\,\partial_t P_{q,n} = P_{l + 1,n} - P_{l,n - 1} + u_{n - 1}\,P_{l,q}, \nonumber \\
1 \leq n \leq d & \quad & \hbar\,\partial_t P_{q,n} = x\,P_{1,n} - \sum_{l = 1}^d u_{l - 1}\,P_{l,n} - P_{q,n - 1} - (x\,\delta_{n,1} - u_{n - 1})P_{q,q}. \nonumber
\eea
These relations give an expression of the elements $P_{l,n}$ in terms of the elements $P_{k,q}$ of the last column and their time derivatives. If we introduce:
\beq
\Gamma_{1} = \Gamma_{q} = 0,\qquad \Gamma_{k} = P_{k,q}\,\,\,\mathrm{if}\,\,\,k \in \ldbrack 2,q - 1\rdbrack,
\eeq
we find for elements above and on the diagonal:
\beq
1 \leq l \leq n \leq d,\qquad P_{l,n} = \Gamma_{q + l - n} + \sum_{m = n}^{q - 1} u_{m}\,\Gamma_{m + l  - n} -  \sum_{m = 0}^{q - n - 1} \hbar\,\partial_t P_{l + m,n + m + 1},
\eeq
and for elements below the diagonal:
\beq
1 \leq n < l \leq q,\qquad P_{l,n} = x\,\Gamma_{l - n} - \sum_{m = 0}^{n - 1} u_m\,\Gamma_{m + l - n} + \sum_{m = 0}^{n - 1} \hbar\,\partial_{t} P_{l - m - 1,n - m}.
\eeq
Consequently, we may write:
\beq
\mathbf{P} = \mathbf{A} + x\,\mathbf{B} - \hbar\,\mathbf{C},
\eeq
with:
\bea
1 \leq l \leq n \leq q & \qquad & A_{l,n} = \Gamma_{q + l - n} + \sum_{m = n}^{q - 1} u_m\,\Gamma_{m + l - n}, \\
1 \leq n < l \leq q & \qquad & A_{l,n} = -\sum_{m = 0}^{n - 1} u_m\,\Gamma_{m + l - n}, \\
\label{Gamd} 1 \leq l,n \leq d & \qquad & B_{l,n} = \Gamma_{l - n}, \\
1 \leq l \leq n \leq d & \qquad & C_{l,n} = -\sum_{m = 0}^{q - n - 1} \partial_t M_{l + m,n + m + 1}, \\
1 \leq n < \leq l \leq d & \qquad & C_{l,n} =  \sum_{m = 0}^{n - 1} \partial_t M_{l -m - 1,n - m}.
\eea
We now prove the commutation relations. We claim that, for any $\theta \in \mathbb{C}$ generic, the matrix
\beq
\mathbf{G}_{\theta}(\sheet{x}{a},t) = \mathbf{A}(\sheet{x}{a},t) + \theta\,\mathbf{B}(\sheet{x}{a},t)
\eeq
has a basis of eigenvectors which independent of $x$ and $a$. This will imply:
\beq
[\mathbf{G}_{\theta}(\sheet{x}{a},t),\mathbf{G}_{\theta}(\sheet{y}{b},t)] = 0,
\eeq
from which the relations \eqref{comut1}-\eqref{comut3} can be deduced by identification of the coefficients of $\theta$. Let $(\zeta_{i})_{1 \leq i \leq q}$ be the roots of:
\beq
\label{polu} X^{q} + \sum_{m = 0}^{q - 1} u_m\,X^{m} = \theta.
\eeq
For generic $\theta$, the roots are simple, so that the column vector $\mathbf{v}_i(z) = (\zeta_i^j)_{0 \leq j \leq q - 1}$ form a basis of $\mathbb{C}^q$. Let us set:
\beq
\lambda_i = (\mathbf{G}_{\theta}\mathbf{v}_i)_{1} = \sum_{m = 1}^q A_{1,m}\,\zeta_i^{m}.
\eeq
Considering the second line:
\beq
(\mathbf{G}_{\theta}\mathbf{v}_i - \lambda_i\,\mathbf{v}_i)_{2} = \theta\,B_{2,1} + \sum_{m = 1}^{q} A_{2,m}\,\zeta_i^{m - 1} - \sum_{m = 1}^q A_{1,m}\,\zeta_i^{m},
\eeq
but since $B_{2,1} = \Gamma_1$, $A_{2,1} = -u_0\Gamma_1$ and $A_{1,d} = \Gamma_1$, using the polynomial equation \eqref{polu} for $\zeta_i$, it must vanish. If we proceed to the $k$-th line, we have:
\bea
(\mathbf{G}_{\theta}\mathbf{v}_i - \lambda_i\,\mathbf{v}_i)_k & = & \theta\sum_{m = 1}^{k - 1} B_{k,m}\,\zeta_i^{m - 1} + \sum_{m = 1}^{q} A_{k,m}\,\zeta_i^{m - 1} - \sum_{m = q - k + 2}^{q} A_{1,m}\,\zeta_i^{m + k - 2} \\
& = & \sum_{m = 1}^{k - 1} (\theta\,B_{k,m} + A_{k,m})\zeta_i^{m - 1} + \sum_{m = k}^{q} (A_{k,m} - A_{1,m - k + 1})z^{m - 1} \nonumber \\
&& - \sum_{m = q - k + 2}^{q} A_{1,m}\,\zeta_i^{m + k - 2}. \nonumber
\eea
Using:
\bea
1 \leq m < k \leq q & \qquad & \theta\,B_{k,m} + A_{k,m} = \theta\,\Gamma_{k - m} - \sum_{n = 0}^{m - 1} u_n\,\Gamma_{k - m + n}, \\
1 \leq k \leq m \leq q & \qquad & A_{k,m} - A_{1,m - k + 1} = -\sum_{n = 1}^{k - 1} u_{m + n - k}\,\Gamma_n, \\
1 \leq m \leq q && A_{1,m} = \Gamma_{d - m + 1} + \sum_{n = 1}^{q - m} u_{m + n - 1}\,\Gamma_{n},
\eea
we may collect the terms relative to a given $\Gamma_m$ and we obtain:
\beq
(\mathbf{G}_{\theta}\mathbf{v}_i - \lambda_i\,\mathbf{v}_i)_k = \Big(\sum_{n = 1}^{k - 1} \Gamma_n\,\zeta_i^{k - n - 1}\Big)\Big(\theta - \sum_{m = 0}^{q - 1} u_m\,\zeta_i^{m} - \zeta_i^{q}\Big) = 0.
\eeq
This concludes the proof. \hfill $\Box$

\vspace{0.2cm}

\subsubsection{Main argument of the proof}

\noindent Thanks to the decomposition of Lemma~\ref{thep}, we can prove:
\begin{corollary}
If we choose $\mathbf{U}(\sheet{y}{a}) = \mathbf{B}(\sheet{y}{a},t) + \hbar\,\mathbf{V}(\sheet{y}{a},t)$ to define an insertion operator, then
\beq
\delta_y^a\mathbf{P}(\sheet{x}{b},t) \in O(\hbar),
\eeq
and is expressible in terms of $\mathbf{V}(\sheet{y}{a},t)$, and matrix elements of $\mathbf{P}$ and their time derivatives.
\end{corollary}
\textbf{Proof.} From \eqref{Pder}, we have:
\bea
\delta_{y}^a\mathbf{P}(\sheet{x}{a}) & = & \frac{1}{x - y}\Big[\mathbf{A}(\sheet{y}{a},t) + x\,\mathbf{B}(\sheet{y}{a},t) + \hbar\,\mathbf{C}(\sheet{y}{a},t)\,,\, \mathbf{A}(\sheet{x}{b},t) + x\,\mathbf{B}(\sheet{x}{b},t) + \hbar\,\mathbf{C}(\sheet{x}{b},t)\Big] \nonumber \\
& & + \hbar\,[\mathbf{V}(\sheet{y}{a},t),\mathbf{P}(\sheet{x}{b},t)],
\eea
and using the commutation relations \eqref{comut1}-\eqref{comut3}, we obtain:
\bea
\delta_{y}^a\mathbf{P}(\sheet{x}{a},t) & = & \frac{\hbar}{x - y}\big\{-[\mathbf{P}(\sheet{x}{b},t),\mathbf{C}(\sheet{y}{a},t)] + [\mathbf{P}(\sheet{y}{a},t),\mathbf{C}(\sheet{x}{b},t)]\big\} + \hbar\,[\mathbf{B}(\sheet{y}{a},t),\mathbf{C}(\sheet{x}{b},t)] \nonumber \\
& & + \frac{\hbar^2}{(x - y)^2}\,[\mathbf{C}(\sheet{x}{b},t),\mathbf{C}(\sheet{y}{a},t)] + \hbar\,[\mathbf{V}(\sheet{y}{a},t),\mathbf{P}(\sheet{x}{a},t)].
\eea
\hfill $\Box$

\begin{corollary}
\label{cococo}If $\delta_{y}^a$ is a compatible insertion operator such that $\mathbf{U}(\sheet{y}{a},t) = \mathbf{B}(\sheet{y}{a},t) + \hbar\,\mathbf{V}(\sheet{y}{a})$ and $\mathbf{V}$ depends on $\hbar$, is of order $1$ and is expressible in terms of matrix elements of $\mathbf{P}(\sheet{x}{a})$ and their time derivatives, then:
\beq
\delta_{y_1}^{a_1}\cdots\delta_{y_k}^{a_k}\mathbf{P}(\sheet{x}{a}) \in O(\hbar^{k}),
\eeq
and:
\beq
\mathcal{W}_n(\sheet{x_1}{a_1},\ldots,\sheet{x_n}{a_n}) \in O(\hbar^{n - 2}).
\eeq
\end{corollary}
\textbf{Proof.} If $\delta_{y}^a$ commutes with $\partial_t$, we also have for any $k \geq 0$:
\beq
\label{under}\delta_{y}^a \partial_t^k \mathbf{P}(\sheet{x}{b}) \in O(\hbar).
\eeq
Since $\delta_{y}^a$ itself is expressible in terms of elements of the matrices $\mathbf{P}$ and their time derivatives, we can apply repeatedly \eqref{under} to show that each application of the insertion operator to $\mathbf{P}(\sheet{x}{a})$ increases at least by one the order in $\hbar$. Now, starting from the expression \eqref{W2r} of $\mathcal{W}_2$ and by successive applications of the insertion operator to compute $\mathcal{W}_n$ according to \eqref{djq}, we obtain that \mbox{$\mathcal{W}_n \in O(\hbar^{n - 2})$}. \hfill $\Box$

\subsubsection{Existence of a compatible insertion operator}
\label{sec:compaq}
It is possible to construct explicitly an insertion operator which commutes with $\partial_t$:
\begin{theorem}
\label{compat}The choices:
\bea
\label{sol} U(\sheet{x}{a},t)_{k,m} & = & \sum_{l = 0}^{k - m - 1} {m + l -1 \choose l} (\hbar\,\partial_t)^{l} P_{k - m - l,q}(\sheet{x}{a},t) = B_{k,m}(\sheet{x}{a},t) + O(\hbar), \\
\label{usol}\delta_y^a u_{k}(t) & = & P_{1,k}(\sheet{y}{a},t) - \delta_{k,1} P_{q,q}(\sheet{y}{a},t) + \sum_{m = k}^q u_{m}(t)\,U_{m + 1,k}(\sheet{y}{a},t),
\eea
where we used the convention $u_q(t) = -1$, define the unique insertion operator which commutes with $\partial_t$.
\end{theorem}
\textbf{Proof.} The commutativity of $\delta_{y}^a$ and $\partial_t$ is equivalent to:
\beq
\delta_y^a \partial_t\mathbf{\Psi}(x,t) = \partial_t \delta_y^a \mathbf{\Psi}(x,t),
\eeq
that is:
\beq
\label{cons2}\delta_{y}^a\mathbf{M}(x,t) = [\mathbf{U}(\sheet{y}{a},t),\mathbf{M}(x,t)] + \hbar\,\partial_t \mathbf{U}(\sheet{y}{a},t) + \Big[\mathbf{P}(\sheet{y}{a},t),\frac{\mathbf{M}(x,t) - \mathbf{M}(y,t)}{x - y}\Big].
\eeq
With the expression \eqref{Mform} of $\mathbf{M}(x,t)$ for $(p,q)$ models, we compute:
\bea
\frac{\mathbf{M}(x,t) - \mathbf{M}(y,t)}{x - y} & = & \mathbf{E}_{q,1} \\
\delta_{y}^a M_{k,m}(x,t) & = & -\delta_{k,q}\,\delta_{y}^a u_{m - 1}(t).
\eea
The equation \eqref{cons2} gives a strong constraints upon the matrix $\mathbf{U}(\sheet{y}{a},t)$. For instance, it cannot be zero since:
\beq
[\mathbf{E}_{q,1},\mathbf{P}(\sheet{y}{a},t)]_{k,m} = \delta_{k,q}\mathbf{P}_{1,m}(\sheet{y}{a},t) - \delta_{m,1}\mathbf{P}_{k,q}(\sheet{y}{a},t).
\eeq
We compute:
\bea
[\mathbf{U}(\sheet{y}{a},t),\mathbf{M}(x,t)]_{k,m} & = & U_{k,m - 1}(\sheet{y}{a},t) + \big(x\,\delta_{m,1} - u_{m - 1}(t)\big)U_{k,q}(\sheet{y}{a},t) - U_{k + 1,m}(\sheet{y}{a},t) \nonumber \\
& & + \delta_{m,q}\Big(x\,U_{1,m}(\sheet{y}{a},t) - \sum_{l = 1}^q u_{l - 1}(t)\,U_{l,m}(\sheet{y}{a},t)\Big).
\eea
The condition \eqref{cons2} is an affine function of $x$. With the choice $U_{k,q} = U_{1,m} = 0$ for any $k,m \in \ldbrack 1,q$, the coefficient of $x$ vanish. The remaining constraint reads:
\bea
-\delta_{k,q}\delta_y^a u_{m - 1}(t) & = & U_{k,m - 1}(\sheet{y}{a},t) - U_{k + 1,m}(\sheet{y}{a},t) - \delta_{k,q}\sum_{l = 1}^q u_{l - 1}(t)\,U_{l,m}(\sheet{y}{a},t) \nonumber \\
& & + \hbar\,\partial_t U_{k,m}(\sheet{y}{a},t) - \delta_{k,q} P_{1,m}(\sheet{y}{a},t) + \delta_{m,1}P_{k,q}(\sheet{y}{a},t).
\eea
Omitting the dependence in $y$, $a$ and $t$, we have for $k \neq q$:
\beq
\label{teha}U_{k + 1,m} = U_{k,m - 1} +  \delta_{m,1}P_{k,q} - \hbar\,\partial_t U_{k,m}.
\eeq
The solution at leading order in $\hbar$ is:
\beq
U_{k,m} = \left\{\begin{array}{ll} P_{k - m,q} + O(\hbar) & m > k \\ O(\hbar) & m \leq k \end{array}\right. ,
\eeq
which coincides with the definition of the matrix $\mathbf{B}$ in \eqref{Gamd}. Eqn.~\eqref{teha} can be solved recursively, and we find that its unique solution is given by \eqref{sol}. To define completely an insertion operator, it remains to specify how it acts on the functions $u_{k}(t)$. The commutativity condition prescribes \eqref{usol}. \hfill $\Box$

Although we did not make use of this property, we show for completeness that insertion operators pairwise commute:
\begin{lemma}
For any $a,b \in \ldbrack 1,q \rdbrack$, we have $[\delta_{x}^a,\delta_y^b] = 0$.
\end{lemma}
\textbf{Proof.} This condition is equivalent to:
\beq
\label{comuq}\delta_{x}^a\mathbf{U}(\sheet{y}{b},t) - \delta_{y}^b\mathbf{U}(\sheet{x}{a},t) + [\mathbf{U}(\sheet{x}{a},t),\mathbf{U}(\sheet{y}{b},t)] = 0.
\eeq
Since $\mathbf{U}(\sheet{x}{a},t) = \mathbf{B}(\sheet{x}{a},t) + O(\hbar)$ and $\delta_{x}^a\mathbf{U}(\sheet{y}{b}) \in O(\hbar)$ owing to Lemma~\ref{cococo}, the commutation relation \eqref{comut2} implies \eqref{comuq} at leading order. 
\hfill $\Box$

\subsection{The pole property}
\label{Spropole}
We need to prove that $\omega_n^g$ has poles only at ramification points, in particular, no pole at $\infty$ or at double zeroes. For this purpose, we will use the observations of \S~\ref{Sasyn}.

\subsubsection{Double points}
\label{Secqsq}
\begin{lemma}
In the $q$-th reduction of KP, for any $n,g$, $\omega_n^g$ are regular at preimages in $\mathcal{S}^{[0]}$ of double points.
\end{lemma}

\noindent \textbf{Proof.} We remind that this property is not obvious because equations \eqref{squ1} and \eqref{squ2}, which allow the computation of the WKB expansion of $\mathbf \Psi(x,t) = \mathbf V \widehat{\mathbf \Psi} \,\e^{\mathbf{S}/\hbar}\,\mathbf C$, may have a pole $1/(\lambda_a^{[0]}(x,t)-\lambda_b^{[0]}(x,t))$, i.e. at the double points. However, this analysis was performed for the differential equation with respect to $x$. But now, we have a second differential equation
\beq
\label{eq:difft}\hbar\,\partial_{t}\mathbf{\Psi}(x,t) = \mathbf{M}(x,t)\mathbf{\Psi}(x,t),
\eeq
from which we can perform a similar WKB analysis. One notices that solving \eqref{eqpsihatpqMij} for $\widehat{\mathbf{\Psi}}(x,t) = \mathbf{1}_{q} + \sum_{k \geq 1} \hbar^{k}\,\widehat{\mathbf{\Psi}}^{[k]}(x,t)$ recursively, the only denominators are of the form $1/(z_a-z_b)$, and thus  the only poles that are produced are when $x \rightarrow \alpha$ such that $z^{a}(\alpha) = z^{b}(\alpha)$ for $a \neq b$, i.e. when $z$ goes to a ramification point. The conclusion is that poles at double points in $x$ (and thus at preimages of double points in $z \in \mathcal{S}^{[0]}$) do not occur.  \hfill $\Box$

\subsubsection{Behavior at $z\to \infty$}

\begin{lemma}
The $q$-th reduction of KP satisfies Assumption~\ref{asumptionpolesL}.
\end{lemma}

\noindent\textbf{Proof.} 
We now expand ${\mathbf \Psi}$  at large $x$ as
\beq
\mathbf \Psi = \mathbf V \widehat{\mathbf \Psi} \,\e^{\mathbf{S}/\hbar}\,\mathbf C,
\eeq
where:
\beq
\partial_x S_i = \Lambda^{[0]}_i(x) = Y(z_i),\qquad \mathbf V^{-1}\partial_x\mathbf V = O(x^{-1/q}),\qquad \widehat{\mathbf{\Psi}} = \mathbf{1}_{q} + O(x^{-1/q}).
\eeq
Moreover, as in Section~\ref{S32}, the equation $\hbar\,\partial_x\mathbf\Psi = \mathbf L\,\mathbf\Psi$ implies that there is also a large $x$ expansion of the form:
\beq
\mathbf \Psi = \widetilde{\mathbf V} \widetilde{\mathbf \Psi} \,\e^{\widetilde{\mathbf S}/\hbar}\,\mathbf C,
\eeq
where $\partial_x \widetilde{\mathbf S}={\rm Diag}(\Lambda_i(x))$, and $\widetilde{\mathbf V}^{-1}\partial_x \widetilde{\mathbf V} = O(x^{-1/q})$ and 
 $\widetilde{\mathbf{\Psi}} = \mathbf{1}_{q} + O(x^{-1/q})$.
This implies that:
\beq
\mathbf \Lambda = \mathbf \Lambda^{[0]}+O(x^{-1/q}),
\eeq
and thus the pole property of Assumption~\ref{asumptionpolesL} is satisfied. This implies that, for any $g,n \neq (1,0)$, the $\omega_n^{(g)}(z_1,\ldots,z_n)$ are regular when $z_i=\infty$.

\subsection{Tau function}

It is well known that for $(p,q)$ model we have \cite{dFZJ}:
\bt
\label{btue}\beq
\hbar^2\,\partial_t^2 \ln\Tau(t) = u(t).
\eeq
\et

\subsection{Application of the topological recursion}
\label{Sto}

Theorem~\ref{topro}, and in particular Corollary~\ref{toproc} (since our spectral curve has genus 0), implies that the correlators have the expansion:
\beq
W_n(\sheet{x_1}{a_1},\ldots,\sheet{x_n}{a_n})\dd x_1,\ldots \dd x_n = \sum_{g \geq 0} \hbar^{2g-2+n}\,\omega_n^{(g)}(z^{a_1}(x_1),\dots,z^{a_n}(x_n)),
\eeq
where the $\omega_n^{(g)}(z_1,\ldots,z_n)$ are computed by the topological recursion. The initial data is:
\beq
\omega_{1}^{(0)} = - Y(z)\dd X(z),\qquad \omega_{2}^{(0)}(z_1,z_2) = \frac{\dd z_1\dd z_2}{(z_1 - z_2)^2}.
\eeq
To justify the second equation, we know from Corollary~\ref{cor43} that $\omega_{2}^{(0)} \in \mathcal{B}(\mathcal{S}^{[0]})$, and there is a unique such object on a genus $0$ curve, which can be written as in the second equation in any uniformization variable $z$.

In particular, we can retrieve the expansion of the Tau function with Corollary~\ref{co42}.
\beq
\ln\Tau = \sum_{g \geq 0} \hbar^{2g-2}\,F^{(g)},
\eeq
Since $Y'\dot{X} - X'\dot{Y} = 1$, we find that $\partial_{t} Y|_{X(z)} = - \dd z/\dd X$, hence:
\beq
\label{parFg} \partial_t F^{(g)} = \Res_{z\to\infty} z\,\omega_{1}^{(g)}(z).
\eeq
Remember that $\mathcal{T}$ is defined up to a multiplicative constant, so the constant of integration to get $F^{(g)}$ from \eqref{parFg} is irrelevant here. A direct integration can be done explicitly for $F^{(0)}$ \cite{Dubrovin} and $F^{(1)}$ \cite{EKK2}, but the formulas are complicated to state. In simple examples, it is more efficient to rely on \eqref{parFg}.

\subsubsection*{Case of the homogeneous solution}

For the homogeneous solution,  we have
\beq
X(z) = (u^{[0]})^{q/2}\,f(\zeta), \qquad
Y(z) = (u^{[0]})^{p/2}\,g(\zeta),
\qquad \zeta = z\,\,(u^{[0]})^{-1/2},
\eeq
and where $u^{[0]}(t)=(t/\rho)^{\frac{2}{p+q-1}}$. By homogeneity of the topological recursion (see \cite{EOFg,EORev}) this implies:
\beq
\omega_{n}^{(g)}(z_1,\dots,z_n) = (u^{[0]})^{(2-2g-n)(p+q)/2}\,\,\check{\omega}_{n}^{(g)}(\zeta_1,\dots,\zeta_n) = (t/\rho)^{(2-2g-n)(p+q)/(p+q-1)}\,\,\check{\omega}_{n}^{(g)}(\zeta_1,\dots,\zeta_n).
\eeq 
where $\check{\omega}_{n}^{(g)}$ is computed as if $u^{[0]}$ was equal to $1$. In particular for $n=0$
\beq
\forall g \neq 1,\qquad F^{(g)}(t) = t^{(2-2g)(p+q)/(p+q-1)}\,\,F^{(g)}(1).
\eeq
For $F^{(1)}$, we have:
\beq
\partial_{t} F^{(1)} = \Res_{z \rightarrow \infty} z\,\omega_{1}^{(g)}(\zeta) = (u^{[0]})^{-(p + q - 1)/2}\big\{\Res_{\zeta \rightarrow \infty} \zeta\,\omega_{1}^{(1)}(\zeta)\big\} = \frac{\rho}{t}\big\{\Res_{\zeta \rightarrow \infty} \zeta\,\omega_{1}^{(1)}(\zeta)\big\},
\eeq
therefore:
\beq
F^{(1)}(t) = c\,\ln t,\qquad c = \rho\,\Res_{\zeta \rightarrow \infty} \zeta\,\omega_{1}^{(1)}(\zeta).
\eeq
where the arbitrary integration constant was set to $0$ for $t = 1$.

\vspace{0.2cm}

For the homogeneous solution, we observe that the $\hbar\to 0$ expansion coincides with a $t\to\infty$ expansion:
\beq
\Tau = \exp\Big(\sum_{g \geq 0} \hbar^{2g-2} F^{(g)}(t)\Big) = t^{c}\exp\Big(\sum_{g \geq 0} (\hbar\,t^{-(p+q)/(p+q-1)})^{2g-2}\,F_g(1)\Big).
\eeq
We see that $\hbar$ can be absorbed in a redefinition of the variable $t$. We also have:
\beq
u(t) = \hbar^2\,\partial_t^2 \ln\Tau = t^{\frac{2}{p+q-1}}\,\sum_{g \geq 0} (\hbar\,t^{-(p+q)/(p+q-1)})^{2g}\,u^{\{g\}}(1),
\eeq
where
\beq
u^{\{g\}}(1) = \frac{(p+q)(2 - 2g)\big((p + q)(2 - 2g) - 1\big)}{(p+q-1)^2}\,F_g(1).
\eeq
In particular we see that
\beq
u^{\{0\}}(1)= \rho^{-2/(p + q - 1)},\qquad
F^{(0)}(1) = \frac{1}{2}\,\frac{(p+q-1)^2}{(p+q)(p+q+1)}\,\rho^{-2/(p + q - 1)}.
\eeq
 
\section{Examples}
\label{S6}
The $q$-th reductions of KP, and in particular the $(p,q)$ models describe universal behavior -- provably or conjecturally -- in statistical physics, random matrix theory, and integrable systems. For those reasons, many of them have received names referring to the problems where they appear. The $(1,2)$ model is known to appear when studying the double scaling limit of random matrices at a generic edge of the spectral density, and is related to the Airy process \cite{AiryPS}. The $(3,2)$ model was shown, first in physics \cite{GM90,doug}, then rigorously \cite{FIK92}, to describe generating series of random maps with generic critical weights, and thus was called "pure gravity". The $(4,3)$ (resp. the $(6,5)$ model) is expected to describe the generating series of random maps carrying an Ising model (resp. $3$-Potts model) with non-generic critical weights, and in fact, the theory we developed allows a proof of those conjectures \cite{BEcartespq}.

All the $(p,q)$ models are conjectured to describe the double-scaling limit in random matrices around an edge $a$ where the spectral density behaves like $|x - a|^{p/q}$. This is also relevant for systems of vicious walkers via Dyson Brownian motion \cite{DysonBM}, and this is related to $2d$ quantum gravity for reasons dating back to \cite{BIPZ}. This has been proven so far in a handful of case (see e.g. \cite{Kuijl} and references therein), but mainly for $q = 2$ cases -- which correspond to the Gelfand-Dikii hierarchies \cite{GelDikii}. This conjecture is based on an ansatz \cite{GM90} for the convergence of operators $\hat{P}$ and $\hat{Q}$ -- interpreted as differentiation and multiplication in the vector space generating by orthogonal polynomials -- which has not been justified rigorously so far. Our methods do not provide a proof that double scaling limits exist. However, once this existence is granted and it is characterized in terms of a Lax pair, it can actually prove that the semiclassical expansion of the limit laws are computed by the topological recursion. Moreover, if the semiclassical spectral curve of the Lax pair can be identified with a blow-up of the large $N$ spectral curve of the matrix model when parameters become critical, it shows -- combining the results of \cite{EOFg} and \cite{BG11} that the semiclassical expansion of the double-scaling limit does coincide with a limit of coefficients in an off-critical $1/N$ expansion when approaching criticality. This crossover is expected and we are able to justify it only relying on loop equations, i.e. by algebraic methods. We refer to \cite{BETW,BEMarchal} for applications relying on those ideas.

In the remaining of the text, we illustrate some $(p,q)$ models, by describing the non-linear PDEs they generate, the spectral curves and the first few coefficients in the $\hbar \rightarrow 0$ expansion of the correlators and of the Tau function. 

\subsection{$(p,q) = (3,2)$: pure gravity}

Here we chose $q=2$ and $p=3$
\beq
Q= (\hbar\partial_{t})^2 -2u,\qquad P = (\hbar\partial_{t})^3 -3 u\,\hbar\partial_{t} -\frac{3}{2}\hbar \dot u + v.
\eeq
The string equation $[P,Q]=\hbar$ implies that $\dot v=0$ and the Painlev\'e I equation for $u(t)$:
\beq
- \frac{1}{2}\,\hbar^2\,\ddot u + 3u^2   = t,\qquad v=t_1.
\eeq
It has the $\hbar$ expansion:
\beq\label{expu32fromPainleve}
u = \sqrt{\frac{t}{3}} - \frac{\hbar^2}{48}\,t^{-2} - \frac{49\,\hbar^4}{2^9 3^{3/2}}\,t^{-9/2}  - \frac{5^2\,7^2\,\hbar^6}{2^{11}3^2}\,t^{-7}+ O(\hbar^8).
\eeq
The Lax pair is given by
\beq
\mathbf M(x,t) = \begin{pmatrix}
0 & 1 \cr
x+2u & 0
\end{pmatrix},
\eeq
and
\beq
\mathbf L(x,t)=\begin{pmatrix}
\frac{1}{2}\hbar\,\dot u(t) - t_1  & x-u \cr
(x-u)(x+2u)+\frac{1}{2}\,\hbar^2\ddot u & -\frac{1}{2}\hbar\,\dot u - t_1
\end{pmatrix}.\eeq
The spectral curve is:
\beq
\det(y\,\mathbf 1_{2} - \mathbf L(x,t)) 
= (y+t_1)^2 - (x+2u)(x-u)^2 - \frac{1}{2}\,\hbar^2\,\ddot u\,(x-u)- \frac{1}{4}\,\hbar^2\dot u^2.
\eeq
To leading order in $\hbar$, the eigenvalues of $\mathbf L^{[0]}(x,t)$ are thus:
\beq
y= -t_1 \pm  (x-u^{[0]})\sqrt{x+2u^{[0]}},
\eeq
and they are parametrized by:
\beq
\left\{\begin{array}{l}
X(z) = z^2-2u^{[0]} \cr
Y(z) = z^3 - 3 u^{[0]}\,z - t_1
\end{array}
\right.
\qquad {\rm with}\quad u^{[0]} = \sqrt{\frac{t}{3}}.
\eeq
Notice that with $\zeta = (u^{[0]})^{-1/2}z$, we recover the Chebyshev polynomials:
\beq
\left\{\begin{array}{l}
X(z) = u^{[0]}\,(\zeta^2-2)  = u^{[0]}\, T_2(\zeta)\cr
Y(z) = (u^{[0]})^{3/2}\,(\zeta^3 - 3 \zeta) - t_1 = (u^{[0]})^{3/2}\,T_3(\zeta) - t_1
\end{array}
\right. .
\eeq
Applying the topological recursion gives the coefficients of expansion of the correlators:
\bea
\omega_{1}^{(0)}(z)& = & -Y(z)\dd X(z) = - 2\,(z^4-3 u^{[0]} \,z^2-t_1 z)\,\dd z,  \nonumber \\
\omega_{2}^{(0)}(z_1,z_2) & = & \frac{\dd z_1\,\dd z_2}{(z_1-z_2)^2}, \nonumber \\
\omega_{3}^{(0)}(z_1,z_2,z_3) & = & \frac{-1}{6u^{[0]}}\,\,\,\frac{\dd z_1\,\dd z_2\,\dd z_3}{z_1^2\,z_2^2\,z_3^2}, \nonumber \\
\omega_{4}^{(0)}(z_1,\ldots,z_4) & = & \frac{1}{36\,(u^{[0]})^3}\,\,\,\frac{\dd z_1\,\dd z_2\,\dd z_3\,\dd z_4} {z_1^2\,z_2^2\,z_3^2\,z_4^2}\,\left(1+\sum_{i = 1}^4 \frac{3u^{[0]}}{z_i^2}\right), \nonumber
\eea
\bea
\omega_{5}^{(0)}(z_1,\ldots,z_5) & = & \frac{-1}{72\,(u^{[0]})^5}\,\,\,\frac{\dd z_1\,\dd z_2\,\dd z_3\,\dd z_4\,\dd z_5}{z_1^2\,z_2^2\,z_3^2\,z_4^2\,z_5^2}\,\left(1+\sum_{i = 1}^5 \frac{3u^{[0]}}{z_i^2}+\sum_{i = 1}^5 \frac{5\,(u^{[0]})^2}{z_i^4} + \sum_{i<j} \frac{6(u^{[0]})^2}{z_i^2\,z_j^2}\right), \nonumber \\
\omega_{1}^{(1)}(z) & = & -\,\frac{1}{144\,(u^{[0]})^2}\,\frac{\dd z}{z^4}\,(z^2+3u^{[0]}), \nonumber \\
\omega_{2}^{(1)}(z_1,z_2) &=& \frac{1}{864\,(u^{[0]})^4}\,\frac{\dd z_1\,\dd z_2}{z_1^2\,z_2^2}\,\left(2+ 6u^{[0]}(z_1^{-2}+z_2^{-2}) + 9(u^{[0]})^2\,z_1^{-2}\,z_2^{-2}\right. \nonumber \\
& &  \left. + 15(u^{[0]})^2(z_1^{-4}+z_2^{-4}) \right), \nonumber \\
\omega_{1}^{(2)}(z) & = & -\,\frac{7}{2^{10}3^5\,(u^{[0]})^7}\,\frac{\dd z}{z^{10}}\,\big(4z^8+12 u^{[0]}\,z^6+36\,(u^{[0]})^2\,z^4+87\,(u^{[0]})^3\,z^2+135\,(u^{[0]})^4\big), \nonumber \\
\omega_{1}^{(3)}(z) & = & -\,\frac{7}{2^{15}3^9\,(u^{[0]})^{12}}\,\frac{\dd z}{z^{16}}\!\left(1400 z^{14} + 4200 u^{[0]} z^{12} + 12600 (u^{[0]})^2 z^{10} + 34740 (u^{[0]})^3 z^8\right. \nonumber \\
&& \left. + 85860 (u^{[0]})^4 z^6 + 181764 (u^{[0]})^5 z^4 + 297297 (u^{[0]})^6 z^2 + 289575 (u^{[0]})^7\right). \nonumber
\eea
The expansion of the Tau function $\ln\Tau = \sum_{g \geq 0} \hbar^{2g-2} F^{(g)}$ is obtained from:
\beq
\partial_t F^{(g)} = \Res_{z\to\infty} z\,\omega_1^{(g)}(z) = 6u^{[0]}\dot{u}^{[0]} \Res_{z \to \infty} \omega_{1}^{(g)}(z)
\eeq
and the solution $u = u^{[0]} + \sum_{g \geq 1} \hbar^{2g}\,u^{\{g\}}$ from $u^{\{g\}} = \partial_{t}^2 F^{(g)}$. We emphasized that $1 = 6u^{[0]}\dot{u}^{[0]}$ to facilitate the integration. That gives:
\beq
\begin{array}{lcl}
\partial_t F^{(1)} = \frac{6 u^{[0]} \dot u^{[0]}}{144\,(u^{[0]})^2} = \frac{\dot u^{[0]}}{24\,u^{[0]}} & \,\, \Rightarrow \,\, & F^{(1)} = \frac{\ln u^{[0]}}{24} = \frac{1}{48}\ln(t/3) \\
&\,\, \Rightarrow\,\, & u^{\{1\}} = \frac{-1}{48\,t^2} \\
\partial_t F^{(2)} = \frac{7\cdot 6u^{[0]}\dot u^{[0]}}{2^83^5(u^{[0]})^7} = \frac{7\,\dot u^{[0]}}{2^73^4(u^{[0]})^6}
& \,\, \Rightarrow \,\,  & F^{(2)} = \frac{-7}{2^73^45\,(u^{[0]})^5} = \frac{-7}{2^73^{3/2}5t^{5/2}} \\
 & \,\, \Rightarrow \,\,  & u^{\{2\}} = \frac{-49}{2^93^{3/2}t^{9/2}}. \\
\partial_t F^{(3)} = \frac{7\cdot 1400\cdot 6u^{[0]}\dot u^{[0]} }{2^{15}3^9(u^{[0]})^{12}} = \frac{5^27^2\dot u^{[0]} }{2^{11}3^8(u^{[0]})^{11}}
 & \,\, \Rightarrow \,\, & F^{(3)} = \frac{ - 5\cdot 7^2}{2^{12}3^8(u^{[0]})^{10}} = \frac{ - 5\cdot 7^2}{2^{12}3^3t^5} \\
 & \,\, \Rightarrow \,\, & u^{\{3\}} = \frac{ - 5^27^2}{2^{11}3^2t^7}.
 \end{array}
\eeq
These results agree with the direct $\hbar$ expansion of the solution of the Painlev\'e I equation \eqref{expu32fromPainleve}.

\subsection{$(p,q) = (2,3)$}

Here, we consider pure gravity again, but exchange the role of $P$ and $Q$, namely we chose $p=2$ and $q=3$. This gives the $3 \times 3$ Lax pair:
\beq
\mathbf M(x,t) = \left(\begin{array}{ccc}
0 & 1 &0   \cr
0 &0  & 1   \cr
x+\frac{3}{2}\hbar \dot{u}-t_1 & 3u & 0 \cr
\end{array}\right),
\eeq
\beq
\mathbf L(x,t) = \left(\begin{array}{ccc}
2u & 0 & -1   \cr
t_1-x+ \frac{1}{2}\,\hbar\dot{u} & -u  & 0   \cr
\frac{1}{2}\,\hbar^2\ddot{u} & t_1-x-\frac{1}{2}\,\hbar\dot{u} & -u \cr
\end{array}\right).
\eeq
The spectral curve is:
\beq
\det\big(y\,\mathbf{1}_{3}-\mathbf{L}(x,t)\big)
= y^3 - 2 u^2 y -2 u^3 +(x-t_1)^2 +\frac{1}{2}\,\hbar^2(y \ddot{u}-\frac{1}{2}\,\dot{u}^2+u\ddot{u}).
\eeq
To leading order the spectral curve is thus:
\beq
y^3 - 2(u^{[0]})^2 y  + (x-t_1)^2 - 2(u^{[0]})^3 = 0,
\eeq
which admits the parametrization:
\beq
\left\{\begin{array}{l}
X(z) = (u^{[0]})^{3/2}\,T_3(\zeta)  = z^3-3u^{[0]}z\cr
Y(z) = -u^{[0]}\,T_2(\zeta) = 2u^{[0]} - z^2 \cr
\end{array}\right.
\qquad u^{[0]} = (t/3)^{1/2}.
\eeq
The ramification points are at $\zeta=a_\pm=\pm 1$, they correspond to $X(a_\pm)=\mp 2\,(u^{[0]})^{3/2}$.
The local Galois conjugate near $a=\pm 1$ is:
\beq
\sigma_a(\zeta) = \frac{-1}{2}\,\left( \zeta - a \sqrt{12-3 \zeta^2}\right).
\eeq
The topological recursion gives (we denote $\zeta=(u^{[0]})^{-1/2}z$) for the expansion of the correlators:
\bea
\omega_1^{(0)}(z) &=& -Y(z)\dd X(z) =  3\,(u^{[0]})^{5/2}\,(\zeta^2-2)(\zeta^2-1)\,\dd \zeta, \nonumber \\
\omega_2^{(0)}(z_1,z_2) & = & \frac{\dd\zeta_1\,\dd\zeta_2}{(\zeta_1-\zeta_2)^2}, \nonumber \\ 
\omega_3^{(0)}(z_1,z_2,z_3) &=& \frac{- \dd \zeta_1\,\dd \zeta_2\,\dd\zeta_3}{12\,(u^{[0]})^{5/2}}\,\Big( \frac{1}{(\zeta_1-1)^2(\zeta_2-1)^2(\zeta_3-1)^2}  + \frac{1}{(\zeta_1+1)^2(\zeta_2+1)^2(\zeta_3+1)^2} \Big), \nonumber \\
\omega_1^{(1)}(z) & = & \frac{-\dd\zeta}{288\,(u^{[0]})^{5/2}}\,\Big(\frac{5-3\zeta+\zeta^2}{(\zeta-1)^4}+ \frac{5+3\zeta+\zeta^2}{(\zeta+1)^4}\Big), \nonumber \\
\omega_1^{(2)}(z) & = & \frac{-\dd \zeta}{2^{19}3^5\,(u^{[0]})^{15/2}}\,\left\{\frac{1}{(\zeta-1)^{10}}\Big(7168 \zeta^8 -61957 \zeta^7 +246834 \zeta^6-602251 \zeta^5 \right. \nonumber \\
&& +1016572 \zeta^4-1271499\zeta^3+1218226 \zeta^2-862277 \zeta+369664 \Big) \nonumber \\
&& 
+ \frac{1}{(\zeta+1)^{10}}\Big(7168 \zeta^8 +61957 \zeta^7 +246834 \zeta^6+602251 \zeta^5 \nonumber \\
&& \left. +1016572 \zeta^4+1271499\zeta^3+1218226 \zeta^2+862277 \zeta+369664 \Big) \right\}. \nonumber
\eea
It is necessary to compute $\omega_{2}^{(1)}$ in order to obtain $\omega_{1}^{(2)}$, but we omitted its expression for conciseness. The expansion of the Tau function $\ln\Tau = \sum_{g \geq 0} \hbar^{2g-2} F^{(g)}$ and the solution $u = u^{[0]} + \sum_{g \geq 1} \hbar^{2g}\,u^{\{g\}}$ from $u^{\{g\}} = \partial_{t}^2 F^{(g)}$. We may use $6u^{[0]}\dot{u}^{[0]} = 1$ to perform the integration. That gives:
\beq
\begin{array}{lcl}
\partial_t F^{(1)} = \frac{6\,(u^{[0]})^{3/2}\dot{u}^{[0]}}{144\,(u^{[0]})^{5/2}} = \frac{\dot{u}^{[0]}}{24\,u^{[0]}}
& \,\, \Rightarrow \,\, & F^{(1)} = \frac{\ln u^{[0]}}{24} = \frac{1}{48}\,\ln(t/3), \\
& \,\, \Rightarrow \,\, & u^{\{1\}} = \frac{-1}{48\,t^2}. \\
\partial_t F^{(2)} = \,\frac{6\,(u^{[0]})^{3/2}\dot{u}^{[0]}\,7168}{2^{18}3^5\,(u^{[0]})^{15/2}} = \frac{7\,\dot{u}^{[0]}}{2^73^4\,(u^{[0]})^6}
& \,\, \Rightarrow \,\, & F^{(2)} = \frac{-7}{2^73^45\,(u^{[0]})^5} = \frac{-7}{2^73^{3/2}5\,t^{5/2}}, \\
& \,\, \Rightarrow \,\, & u^{\{2\}} = \frac{-49}{2^93^{3/2}\,t^{9/2}}.
\end{array}
\eeq
This again perfectly agrees with the direct $\hbar$ expansion of the solution of the Painlev\'e I equation \eqref{expu32fromPainleve}, and this agrees with the $(3,2)$ model, as an illustration of the $(p,q) \to (q,p)$ duality.

\subsection{$(p,q) = (4,3)$: Ising model}
\label{secexIsing}

The model is defined by:
\beq
Q = (\hbar\partial_{t})^3 - 3 u\,\hbar\partial_{t} +u_0,\qquad P = (\hbar\partial_{t})^4 - 4 u\,(\hbar\partial_{t})^2  + v_1\,\hbar\partial_{t} + v_0.
\eeq
where $u,u_0,v_1,v_0$ are functions of $t$.
The string equation implies
\beq
u_0=- \frac{3}{2}\,\hbar\,\dot u - 3w + t_1,
\eeq
where $w$ is a function of $t$, and:
\beq
v_1 = -4 w - 4\,\hbar\,\dot u,\qquad v_0 = 2u^2  -\frac{5}{3}\,\hbar^2\,\ddot u - 2 \hbar\,\dot w  + t_2,
\eeq
where $w$ satisfies
\beq
12 uw - 2 \hbar^2\,\ddot w  =  t_3,
\eeq
and then $u(t)$ satisfies
\beq
\frac{1}{6}\,\hbar^4\, \ddddot{u} - 3\,\hbar^2 u \ddot{u} - \frac{3}{2}\,\hbar^2 {\dot u}^2 + 4 u^3 + 6 w^2  = t,
\eeq
where $t_1,t_2,t_3$ are integration constants. A particular choice is $t_1=t_2=t_3=0$ and $w=0$, in which case we have
\beq
\frac{1}{6}\,\hbar^4\,\ddddot{u} - 3 \hbar^2 u\ddot u - \frac{3}{2}\,\hbar^2 {\dot u}^2 + 4 u^3  = t.
\eeq
The first few orders of expansion are:
\beq\label{asympu43fromPainleve}
u = \frac{1}{2}\,(2t)^{1/3} - \frac{1}{24}\,\frac{\hbar^2}{t^2} - \frac{1925}{1458}\,\frac{\hbar^4}{(2t)^{13/3}}- \frac{509575}{13122}\,\frac{\hbar^4}{(2t)^{20/3}}+O(\hbar^8),
\eeq
and from the relation $\hbar^2\partial_{t}^2\ln Z = u$:
\beq
\ln Z = \frac{9}{224}\,\frac{(2t)^{7/3}}{\hbar^2} + \frac{1}{24}\,\ln{t} - \frac{55}{1296}\,\frac{\hbar^2}{(2t)^{7/3}}- \frac{29975}{81648}\,\frac{\hbar^4}{(2t)^{14/3}}+O(\hbar^6).
\eeq
The Lax pair is:
\bea
\mathbf M(x,t) & = & \left(\begin{array}{ccc}
 0 & 1 & 0   \cr
 0 & 0 & 1   \cr
x+\frac 3 2\,\hbar\,\dot u + 3w -t_1 & 3u & 0 \cr
\end{array}\right), \nonumber \\
\mathbf L(x,t) & = & \begin{pmatrix}
2u^2 + t_2 & x-t_1-w &  -u \cr
(t_1-x-3w)u & -u^2 + t_2  & x-t_1-w\cr
(x-t_1)^2 + 2(x-t_1) w   -3 w^2 & -2(t_1-x+3w)u& -u^2 + t_2 \cr
\end{pmatrix}  \\
& & + \hbar\,\begin{pmatrix}
 \dot{w} - \frac{1}{6}\,\hbar \ddot{u} & \frac{1}{2}\,\dot{u} & 0 \cr
\frac{5}{2}\,u\dot{u} + \hbar^2\,\ddot{w} - \frac{1}{6}\,\hbar^3 \dddot{u} & \frac{1}{3}\,\hbar\,\ddot{u} & -\frac{1}{2}\,\dot{u} \cr
9 u \dot{w} + \hbar\big(\frac{7}{4}\,\dot{u}^2 + \frac{5}{2}\,u\ddot{u}\big) + \hbar\,\ddot{w} - \frac{1}{6}\,\hbar^2\,\ddddot{u} & -u \dot{u} + \hbar\ddot{w} - \frac{1}{6}\,\hbar^2 \dddot{u} & - \dot{w} - \frac{1}{6}\,\hbar\,\ddot{u} \cr
\end{pmatrix}.
\nonumber
\eea
In the particular case where $t_1=t_2=t_3=w=0$, we have:
\beq
\mathbf L(x,t) =  \begin{pmatrix}
2u^2 - \frac{1}{6}\,\hbar^2 \ddot{u} & x + \frac{1}{2}\,\hbar\,\dot{u} &  -u \cr
-ux + \frac{5}{2}\,\hbar\,u\dot{u} - \frac{1}{6}\,\hbar^3\,\dddot{u} & -u^2 + \frac{1}{3}\,\hbar^2\,\ddot{u} & x - \frac{1}{2}\,\hbar\,\dot{u} \cr
x^2 + \hbar^2\big(\frac{7}{4}\dot{u}^2 + \frac{5}{2}\,u\ddot{u}\big) - \frac{1}{6}\,\hbar^4\,\ddddot{u} & 2ux - \hbar\,u\dot{u} + \frac{1}{6}\,\hbar^3 \,\dddot{u} & -u^2 - \frac{1}{6}\,\hbar^2\,\ddot{u}  \cr
\end{pmatrix}.
\eeq
The spectral curve is:
\bea
\det(y\,\mathbf{1}_{3}-\mathbf L(x,t)) & = & y^3 - \Big(3u^4-\frac{1}{6}\,\hbar\,\dot{u}u^3-3\hbar^2\,u^2 \ddot{u} +\frac{1}{12}\,\hbar^2(\ddot{u}^2+2 u \ddddot{u})\Big)y - x^4+ tx^2 + 2u^6  \nonumber \\
& &  + \hbar^2(u^3\dot{u}^2-3 u^4\ddot{u})+\hbar^4\Big(-\frac{7}{16}\,\dot{u}^4 + \frac{1}{4}\,u\dot{u}\ddot{u}+\frac{3}{4}\, u^2 \ddot{u}^2 -\frac{1}{2}\,u^2 \dot{u}\dddot{u}+\frac{1}{6}\,u^3 \ddddot{u}\Big) \nonumber \\
&& + \hbar^6 \Big(\frac{1}{108}\,\ddot{u}^3 + \frac{1}{36}(- \dot{u}\ddot{u}\dddot{u}+ u\dddot{u}^2) + \frac{1}{24}\,\dot{u}^2 \ddddot{u} -\frac{1}{18}\,u\ddot{u}\ddddot{u}\Big).
\eea
To leading order the spectral curve is thus:
\beq
y^3 - 3(u^{[0]})^4 y    = x^4-4(u^{[0]})^3 x^2 +2(u^{[0]})^6,
\eeq
i.e. in terms of Chebyshev polynomials:
\beq
T_3\big(y/(u^{[0]})^2\big) = T_4\big(x/(u^{[0]})^{3/2}\big),
\eeq
which admits the parametrization:
\beq
\left\{\begin{array}{l}
X(z) = (u^{[0]})^{3/2}\,T_3(\zeta)  = z^3-3u^{[0]}z\cr
Y(z) = (u^{[0]})^{2}\,T_4(\zeta) =z^4-4u^{[0]}z^2+2(u^{[0]})^2 \cr
\end{array}\right.
\qquad u^{[0]} = (t/4)^{1/3}.
\eeq
The ramification points are at $\zeta=a_\pm=\pm 1$, they correspond to $X(a_\pm)=\mp 2$.
The local Galois conjugate near $a=\pm 1$ is:
\beq
\sigma_a(\zeta) = \frac{-1}{2}\,\left( \zeta - a \sqrt{12-3 \zeta^2}\right).
\eeq
The topological recursion gives (we denote $\zeta=z/\sqrt{u^{[0]}}$) for the expansion of the correlators:
\bea
\omega_1^{(0)}(z) &=& -Y(z)\dd X(z) =  -3\,(u^{[0]})^{7/2}\,(\zeta^4-4\zeta^2+2)(\zeta^2-1)\,\dd \zeta, \nonumber \\
\omega_2^{(0)}(z_1,z_2) & = & \frac{\dd\zeta_1\,\dd\zeta_2}{(\zeta_1-\zeta_2)^2}, \nonumber \\ 
\omega_3^{(0)}(z_1,z_2,z_3) &=& \frac{-\dd \zeta_1\,\dd \zeta_2\,\dd\zeta_3}{24\,(u^{[0]})^{7/2}}\,\Big( \frac{1}{(\zeta_1-1)^2(\zeta_2-1)^2(\zeta_3-1)^2}  + \frac{1}{(\zeta_1+1)^2(\zeta_2+1)^2(\zeta_3+1)^2} \Big), \nonumber \\
\omega_1^{(1)}(z) & = & \frac{-\dd\zeta}{576\,(u^{[0]})^{7/2}}\,\Big(\frac{7+7\zeta+3\zeta^2}{(\zeta+1)^4}+ \frac{7-7\zeta+3\zeta^2}{(\zeta-1)^4}\Big), \nonumber \\
\omega_{1}^{(2)}(z) & = & \frac{-5\dd\zeta}{2^{13}3^5\,(u^{[0]})^{21/2}}\,\frac{1}{(\zeta^2 - 1)^{10}}\Big(791 + 10831\zeta^2 + 5642\zeta^4 + 8010\zeta^6 - 5060\zeta^8 \nonumber \\
& & + 6556\zeta^{10}  - 4098\zeta^{12} + 1982\zeta^{14} - 539\zeta^{16} + 77\zeta^{18}\Big), \nonumber \\
\omega_{1}^{(3)}(z) & = & \frac{-5\dd\zeta}{2^{19}3^9\,(u^{[0]})^{35/2}}\,\frac{1}{(\zeta^2 - 1)^{16}}\Big(1534020 + 51852480\zeta^2 +  139051115\zeta^4 \nonumber \\
& & + 126732801\zeta^6 + 14026336\zeta^8 + 136206860\zeta^{10}  - 165273597\zeta^{12} + 227618305\zeta^{14} \nonumber \\
& &  - 221591820\zeta^{16} + 175823400\zeta^{18} - 107773575\zeta^{20} + 51069755\zeta^{22} - 17959320\zeta^{24} \nonumber \\
& & + 4465420\zeta^{26} - 701415\zeta^{28} + 53955\zeta^{30}\Big). \nonumber
\eea
The computation of $\omega_1^{(2)}$ (resp. $\omega_1^{(3)}$) required the knowledge of $\omega_2^{(1)}$ (resp. the knowledge of $\omega_{4}^{(0)}$, $\omega_{3}^{(1)}$ and $\omega_{2}^{(2)}$), but since their expression is lengthy we do not copy them here. The expansion of the Tau function $\ln\Tau = \sum_{g \geq 0} \hbar^{2g-2} F^{(g)}$ and the solution $u = u^{[0]} + \sum_{g \geq 1} \hbar^{2g}\,u^{\{g\}}$ from $u^{\{g\}} = \partial_{t}^2 F^{(g)}$. We can use $12(u^{[0]})^2\dot{u}^{[0]} = 1$ to perform the integration. That gives:
\beq
\begin{array}{lcl}
\label{jij} \partial_t F^{(1)} = 12\,(u^{[0]})^{5/2}\dot{u}^{[0]}\,\frac{1}{2^5 3\,(u^{[0]})^{7/2}} = \frac{\dot{u}^{[0]}}{8\,u^{[0]}},
& \,\,\Rightarrow \,\, & F^{(1)}= \frac{\ln{u_0}}{8}=\frac{\ln{(t/4)}}{24}. \\
& \,\, \Rightarrow \,\, & u^{\{1\}} = -\frac{1}{24t^2}, \\
\partial_{t} F^{(2)} = 12\,(u^{[0]})^{5/2}\dot{u}^{[0]}\,\frac{5\cdot 7 \cdot 11\,}{2^{13}3^5\,(u^{[0]})^{21/2}} = \frac{5\cdot 7 \cdot 11}{2^{11}3^4\,(u^{[0]})^{8}} & \,\, \Rightarrow \,\, & F^{(2)} = -\frac{5\cdot 11}{2^{11}3^4\,(u^{[0]})^7} = -\frac{55}{1296\,(2t)^{7/3}}. \\
& \,\, \Rightarrow \,\, &  u^{\{2\}} = -\frac{1925}{1458\,(2t)^{13/3}}, \\
\partial_{t} F^{(3)} = 12\,(u^{[0]})^{5/2}\dot{u}^{[0]}\,\frac{5^2 11\cdot 109}{2^{19}3^7\,(u^{[0]})^{35/2}} = \frac{5^211 \cdot 109\,\dot{u}^{[0]}}{2^{17}3^6\,(u^{[0]})^{15}} & \,\,\Rightarrow \,\, & F^{(3)} = - \frac{5^211\cdot 109}{2^{18}3^67\,(u^{[0]})^{14}} = -\frac{29975}{81648\,(2t)^{14/3}} \\
& \,\,\Rightarrow \,\, & u^{\{3\}} = -\frac{509575}{13122\,(2t)^{20/3}}. \nonumber
\end{array}
\eeq
This matches \eqref{asympu43fromPainleve}.

\appendix
\numberwithin{equation}{section}

\section{Proof of Lemma \ref{L0a}} \label{appproofdeltas}

If $\delta_{y}^{a}$ is an insertion operator, we now prove the following formulae. For any $n \geq 1$, any $a,b,a_1,\ldots,a_n \in \ldbrack 1,d \rdbrack$,
\bea
\delta_{y}^{a}\mathbf{K}(x_1,x_2) & = & -\mathbf{K}(x_1,y)\mathbf{E}_a\mathbf{K}(y,x_2),  \nonumber \\
\delta_{y}^{a} \mathbf{P}(\sheet{x}{b}) & = & \Big[\frac{\mathbf{P}(\sheet{y}{a})}{x - y} + \mathbf{U}(\sheet{y}{a}),\mathbf{P}(\sheet{x}{b})\Big], \nonumber \\
\delta_{y}^{a} \mathbf{L}(x) & = & \Big[\frac{\mathbf{P}(\sheet{y}{a})}{x - y} + \mathbf{U}(\sheet{y}{a}),\mathbf{L}(x)\Big] - \frac{\mathbf{P}(\sheet{y}{a})}{(x - y)^2}, \nonumber \\
\delta_{y}^{a} \Tr \mathbf{L}(x) & = &  - \frac{1}{(x - y)^2}, \nonumber \\
\delta_{y}^{a} \ln\det \mathbf{\Psi}(x) & = &   \frac{1}{x - y} + \Tr U(\sheet{y}{a}), \nonumber \\
\delta_{y}^{a} \ln\left(\frac{\det\mathbf{\Psi}(x)}{\det\mathbf{\Psi}(z)}\right) & = &   \frac{1}{x - y} -\frac{1}{z-y},  \nonumber \\
\delta_{y}^{a} \mathcal{W}_{n}(\sheet{x_1}{a_1},\ldots,\sheet{x_n}{a_n}) & = & \mathcal{W}_{n + 1}(\sheet{y}{a},\sheet{x_1}{a_1},\ldots,\sheet{x_n}{a_n}).
\eea

\smallskip

\noindent\textbf{Proof of Lemma~\ref{L0a}.} First we have by the Leibniz rule $\delta_{y}^{b} (\mathbf{\Psi}^{-1}(x) \mathbf{\Psi}(x))=0$, which leads to:
\beq
\delta_{y}^{a} \mathbf{\Psi}^{-1}(x) = - \mathbf{\Psi}^{-1}(x)(\delta_{y}^a\mathbf{\Psi}(x))\mathbf{\Psi}^{-1}(x) = \frac{\mathbf{\Psi}^{-1}(x)\mathbf{P}(\sheet{y}{a})}{y - x}-\mathbf{\Psi}^{-1}(x)\mathbf{U}(\sheet{y}{a}).
\eeq
Then, we compute
\bea
\delta_{y}^{a} \mathbf{K}(x_1,x_2) & = & \frac{1}{x_1 - x_2}\,\delta_{y}^{b}[\mathbf{\Psi}^{-1}(x_1)\mathbf{\Psi}(x_2)] \nonumber \\
& = & \frac{1}{x_1 - x_2}\Big(\frac{\mathbf{\Psi}^{-1}(x_1)\mathbf{P}(\sheet{y}{a})\mathbf{\Psi}(x_2)}{y - x_1} + \frac{\mathbf{\Psi}^{-1}(x_1)\mathbf{P}(\sheet{y}{a})\mathbf{\Psi}(x_2)}{x_2 - y} \cr
&& + \mathbf{\Psi}^{-1}(x_1)\mathbf{U}(\sheet{y}{a})\mathbf{\Psi}(x_2) - \mathbf{\Psi}^{-1}(x_1)\mathbf{U}(\sheet{y}{a})\mathbf{\Psi}(x_2)\Big) \nonumber \\
& = & - \frac{\mathbf{\Psi}^{-1}(x_1)\mathbf{\Psi}(y)}{x_1 - y}\mathbf{E}_a\,\frac{\mathbf{\Psi}^{-1}(y)\mathbf{\Psi}(x_2)}{y - x_2} = -\mathbf{K}(x_1,y)\mathbf{E}_a\mathbf{K}(y,x_2).
\eea
notice that $\mathbf U$ disappears in this computation.
Similarly,
\bea
\delta_{y}^{a} \mathbf{P}(\sheet{x}{b}) 
&=& (\delta_{y}^{a} \mathbf{\Psi}(x))\mathbf{E}_b\mathbf{\Psi}^{-1}(x) +\mathbf{\Psi}(x)\mathbf{E}_b(\delta_{y}^a\mathbf{\Psi}^{-1}(x))  \nonumber \\
&=& \frac{\mathbf{P}(\sheet{y}{a})\mathbf{\Psi}(x)\mathbf{E}_b\mathbf{\Psi}^{-1}(x)}{x - y} -\frac{\mathbf{\Psi}(x)\mathbf{E}_b\mathbf{\Psi}^{-1}(x)\mathbf{P}(\sheet{y}{a})}{x - y}  \nonumber \\
&& + \mathbf{U}(\sheet{y}{a})\mathbf{\Psi}(x)\mathbf{E}_b\mathbf{\Psi}^{-1}(x) -\mathbf{\Psi}(x)\mathbf{E}_b\mathbf{\Psi}^{-1}(x)\mathbf{U}(\sheet{y}{a})  \nonumber \\
\label{311} &= & \frac{[\mathbf{P}(\sheet{y}{a}),\mathbf{P}(\sheet{x}{b})]}{x - y} + [\mathbf{U}(\sheet{y}{a}),\mathbf{P}(\sheet{x}{b})].
\eea
Then we have
\bea
\delta_{y}^{a} \mathbf{L}(x) 
&=& (\delta_{y}^{a} \left( \hbar\,\partial_x \mathbf \Psi(x)\,\,\mathbf\Psi^{-1}(x) \right) \cr
&=& \hbar\,\partial_x\,\left( \delta_{y}^{a} \mathbf \Psi(x) \right)\,\,\mathbf\Psi^{-1}(x)  -  \hbar\,\partial_x\,\mathbf\Psi(x)\,\,\delta_{y}^a\left(\mathbf\Psi^{-1}(x) \right) \cr
&=& \hbar\,\partial_x\,\left( \left( \frac{\mathbf P(\sheet{y}{a})}{x-y} + \mathbf U(\sheet{y}{a})\right)   \mathbf \Psi(x) \right)\,\,\mathbf\Psi^{-1}(x)  -  \mathbf L(x)\,\,\delta_{y}^a\left(\mathbf\Psi^{-1}(x) \right) \cr
&=& -\,\frac{\hbar}{(x-y)^2}\,\mathbf P(\sheet{y}{a}) + \left[  \frac{\mathbf P(\sheet{y}{a})}{x-y} + \mathbf U(\sheet{y}{a}), \mathbf L(x)\right].
\eea
To compute the action of $\delta_{y}^{a}$ on the correlators, we consider $n = 1$ separately:
\bea
\delta_{y}^{b} \mathcal{W}_1(\sheet{x}{b}) 
& = & \delta_{y}^{b} \left( \lim_{z\to x} \mathbf K_{a,a}(x,z) - \frac{1}{x-z} \right) \cr
& = &  \lim_{z\to x} \delta_{y}^{b} \mathbf K_{a,a}(x,z)  \cr
& = & -\, \lim_{z\to x} \mathbf  K_{a,b}(x,y)\, \mathbf  K_{b,a}(y,z)  \cr
& = & -\,  \mathbf  K_{a,b}(x,y)\, \mathbf  K_{b,a}(y,x) = {\cal W}_2(x,y).
\eea
Then, for $n \geq 2$, we can use formula \eqref{eq:dee}:
\bea
&& \delta_{y}^{a} \mathcal{W}_n(x_1,\ldots,x_n)  \cr
& = & (-1)^{n + 1}\sum_{\sigma = n\textrm{-cycle}} \delta_{y}^{a}\Big[\prod_{i = 1}^n \mathbf{K}_{a_i,a_{\sigma(i)}}(x_i,x_{\sigma(i)})\Big] \nonumber \\
& = & (-1)^{n + 2} \sum_{\sigma = n\textrm{-cycle}} \sum_{j = 1}^n \mathbf{K}_{a_j,a}(x_j,y)\mathbf{K}_{a,a_{\sigma(j)}}(y,x_{\sigma(j)})\prod_{i \neq j} \mathbf{K}_{a_{i},a_{\sigma(i)}}(x_i,x_{\sigma(i)}) \nonumber \\
& = & (-1)^{n + 2}\sum_{\substack{\sigma = (n + 1)\textrm{-cycle} \\ y = x_{n + 1},\,\, a_{n + 1} = a}} \prod_{i = 1}^n \mathbf{K}_{a_i,a_{\sigma(i)}}(x_i,x_{\sigma(i)}) \nonumber \\
& = & \mathcal{W}_{n + 1}(\sheet{y}{a},\sheet{x_1}{a_1},\ldots,\sheet{x_n}{a_n}).
\eea
\hfill $\Box$

\section{Proof of Theorem~\ref{topro}}
\label{apptoprec}

We assume that all ramification points are simple (see \cite{BEthink} for the case or higher ramifications), the embedding of the curve $\mathcal{S}^{[0]} \rightarrow \mathbb{C}^2$ by the functions $(x,y)$ is regular, and that TT is satisfied. We shall prove the topological recursion using the linear (Theorem~\ref{linearL}) and quadratic (Theorem~\ref{quda}) loop equations only. This is already done in \cite{EOFg,BEO}, but we present here a self-contained proof. Contrarily to \cite{BEO} which is more general, we take advantage here that the semiclassical spectral curve $\mathcal{S}^{[0]}$ is a compact Riemann surface of genus $\mathfrak{g}$, to identify more precisely the possible holomorphic term in \eqref{eq340}.

From the TT hypothesis, we have that every $\omega^{(g)}_n$ with $(g,n)\neq(0,1),(0,2)$ has poles only at the ramification points.
We have called $\mathbf r = \{r_1,\dots,r_m\}$ the set of ramification points.
Let $r\in\mathbf r$ be a ramification point, by definition and assumption there are exactly two indices $a\neq b$ such that $z^a(r)=z^b(r)$, and we define the local Galois involution $\sigma_r$ in a vicinity of $r$, as the map $z^a(x)\mapsto z^b(x)$. Let $J=\{2,\dots,n\}$ and $\mathbf z_J= (z_j)_{j\in J}$, and define:
\beq
\tilde{\cal Q}^{(g)}_{n}(z,z';\mathbf{z}_J) := \omega^{(g-1)}_{n+2}(z,z',\mathbf z_J) + \sum'_{h+h'=g,\, I\dot{\cup}I'=J} \omega^{(h)}_{1+|I|}(z,\mathbf z_I)\,\omega^{(h')}_{1+|I'|}(z',\mathbf z_{I'})
\eeq
and
\beq
{\cal Q}^{(g)}_{n}(z,z';\mathbf{z}_J) := \omega^{(g-1)}_{n+2}(z,z',\mathbf z_J) + \sum_{h+h'=g,\, I\dot{\cup} I'=J} \omega^{(h)}_{1+|I|}(z,\mathbf z_I)\,\omega^{(h')}_{1+|I'|}(z',\mathbf z_{I'}),
\eeq
where $\sum'$ means that we exclude the cases $(h,I)=(0,\emptyset)$ and $(h,I)=(g,J)$, i.e.
\beq
{\cal Q}^{(g)}_{n}(z,z';\mathbf{z}_J)  = \tilde{\cal Q}^{(g)}_{n}(z,z';\mathbf{z}_J) + \omega^{(0)}_1(z)\,\omega^{(g)}_{n+1}(z',\mathbf z_J)+ \omega^{(g)}_{n+1}(z,\mathbf z_J)\,\omega^{(0)}_1(z').
\eeq

\bl
\label{lB1}Near a ramification point $r$, we have:
\beq
\label{eq24}\sum_{a<b} {\cal Q}^{(g)}_{n}(z^a,z^b;\mathbf{z}_J) = {\cal Q}^{(g)}_{n}(z,\sigma_r(z);\mathbf{z}_J) + {\rm analytical\,\,at}\,\,z\to r.
\eeq

\el
\noindent {\bf Proof of the lemma.}
To simplify notations, we can always label $1$ and $2$ the sheets meeting at the ramification point $r$. I.e. if $z=z^1$, we have $\sigma_r(z)=z^2$.
Let us decompose the sum over indices as:
\bea
\sum_{1 \leq a < b \leq d} {\cal Q}^{(g)}_{n}(z^a,z^b;\mathbf{z}_J) & = & {\cal Q}^{(g)}_{n}(z^1,z^2;z_J)  +\sum_{2< b \leq d} {\cal Q}^{(g)}_{n}(z^1,z^b;\mathbf{z}_J)  \\
& & +\sum_{2<b \leq d} {\cal Q}^{(g)}_{n}(z^2,z^b;\mathbf{z}_J) + \sum_{2<a<b \leq d} {\cal Q}^{(g)}_{n}(z^a,z^b;\mathbf{z}_J).  \nonumber
\eea
The linear loop equation implies that:
\beq
{\cal Q}^{(g)}_{n}(z^1,z^b;\mathbf{z}_J)+{\cal Q}^{(g)}_{n}(z^2,z^b;\mathbf{z}_J) = - \sum_{2 < a \leq d} {\cal Q}^{(g)}_{n}(z^a,z^b;\mathbf{z}_J),
\eeq
and thus:
\beq
\sum_{1 \leq a<b \leq d} {\cal Q}^{(g)}_{n}(z^a,z^b;\mathbf{z}_J) = {\cal Q}^{(g)}_{n}(z^1,z^2;\mathbf{z}_J) -\sum_{2<a,b \leq d} {\cal Q}^{(g)}_{n}(z^a,z^b;\mathbf{z}_J)  + \sum_{2<a<b \leq d} {\cal Q}^{(g)}_{n}(z^a,z^b;\mathbf{z}_J).
\eeq
The last two lines have no poles at the ramification point, hence the announced result. \hfill $\Box$

\vspace{0.2cm}
\noindent\textbf{Remark}. Since the analytic term in $r$ in \eqref{eq24} is a quadratic differential in $z$ invariant under Galois involution, it must actually have a double zero at $r$.

\vspace{0.2cm}

\bt
The $\omega^{(g)}_n$'s satisfy the topological recursion:
\bea
\omega^{(g)}_{n+1}(z_1,\mathbf z_J) = \frac{1}{2} \sum_{r\in \mathbf r} \Res_{z\to r} \frac{\int_{\sigma_r(z)}^z \omega_2^{(0)}(z_1,\cdot)}{\omega_1^{(0)}(z)-\omega_1^{(0)}(\sigma_r(z))}
\tilde{\cal Q}^{(g)}_{n}(z,\sigma_r(z);\mathbf{z}_J) + {\rm holomorphic}(z_1).
\eea

\et

\noindent {\bf Proof.} First, Lemma~\ref{lB1} together with the quadratic loop equation imply that
${\cal Q}^{(g)}_{n}(z,\sigma_r(z);z_J)$ has no pole at the ramification point $r$. This means that:
\beq
\tilde{\cal Q}^{(g)}_{n}(z,\sigma_r(z);\mathbf{z}_J) = -\omega^{(0)}_1(z)\,\omega^{(g)}_{n+1}(\sigma_r(z),\mathbf z_J)- \omega^{(g)}_{n+1}(z,\mathbf z_J)\,\omega^{(0)}_1(\sigma_r(z)) + {\rm analytical\,\,at}\,\,r.
\eeq
Moreover, using again the linear loop equation we have that
\beq
\omega^{(g)}_{n+1}(\sigma_r(z),\mathbf z_J)
= -\omega^{(g)}_{n+1}(z,\mathbf z_J) + {\rm analytical\,\,at}\,\,r,
\eeq
and thus
\beq
\tilde{\cal Q}^{(g)}_{n}(z,\sigma_r(z);\mathbf{z}_J) = \big[\omega^{(0)}_1(z)-\omega^{(0)}_1(\sigma_r(z))\big]\,\omega^{(g)}_{n+1}(z,\mathbf z_J) + {\rm analytical\,\,at}\,\,r.
\eeq
According to the previous remark, the remainder has actually a double zero at $z = r$. We remind that $\omega_1^{(0)} = y\dd x$, and since we assume that the embedding of $\mathcal{S}^{[0]}$ in $\mathbb{C}^2$ by $(x,y)$ is regular, $\dd y(r) \neq 0$. Combined with the assumption that $x$ has simple ramification points, this implies that $\big[\omega_1^{(0)}(z) - \omega_{1}^{(0)}(\sigma_{r}(z))\big]$ has exactly a double zero at $z = r$. Therefore, we find:
\bea
&& \frac{1}{2} \sum_{r\in \mathbf r} \Res_{z\to r} \frac{\int_{\sigma_r(z)}^z \omega_2^{(0)}(z_1,\cdot)}{\omega_1^{(0)}(z)-\omega_1^{(0)}(\sigma_r(z))}
\tilde{\cal Q}^{(g)}_{n}(z,\sigma_r(z);\mathbf{z}_J) \\
&=&  \frac{1}{2} \sum_{r\in \mathbf r} \Res_{z\to r} \left(\int_{\sigma_r(z)}^z \omega_2^{(0)}(z_1,\cdot)\right)\omega^{(g)}_{n+1}(z,\mathbf z_J) \cr
&=&  \frac{1}{2} \left\{ \sum_{r\in \mathbf r} \Res_{z\to r} \left(\int_{o}^z \omega_2^{(0)}(z_1,\cdot) \right)\omega^{(g)}_{n+1}(z,\mathbf z_J)
- \Res_{z\to r} \left(\int^{\sigma_r(z)}_o \omega_2^{(0)}(z_1,\cdot)\right)\omega^{(g)}_{n+1}(z,\mathbf z_J), \nonumber
\right\}
\eea
where $o$ is an arbitrary base point on the spectral curve. We rename the integration variable in the second term $z\to \sigma_r(z)$, and get:
\bea
&& \frac{1}{2} \sum_{r\in \mathbf r} \Res_{z\to r} \frac{\int_{\sigma_r(z)}^z \omega_2^{(0)}(z_1,\cdot)}{\omega_1^{(0)}(z)-\omega_1^{(0)}(\sigma_r(z))}
\tilde{\cal Q}^{(g)}_{n}(z,\sigma_r(z);\mathbf{z}_J) \\
&=&  \frac{1}{2} \sum_{r\in \mathbf r} \left\{\Res_{z\to r} \left(\int_{o}^z \omega_2^{(0)}(z_1,\cdot) \right)\omega^{(g)}_{n+1}(z,\mathbf z_J)
- \Res_{z\to r} \left(\int^{z}_o \omega_2^{(0)}(z_1,\cdot)\right)\omega^{(g)}_{n+1}(\sigma_r(z),\mathbf z_J)\nonumber
\right\}
\eea
using again the linear loop equation, i.e. that $\omega^{(g)}_{n+1}(\sigma_r(z),\mathbf z_J)= - \omega^{(g)}_{n+1}(z,\mathbf z_J) + {\rm analytical\,\,at}\,\,r$, we arrive to
\beq
 \frac{1}{2} \sum_{r\in \mathbf r} \Res_{z\to r} \frac{\int_{\sigma_r(z)}^z \omega_2^{(0)}(z_1,\cdot)}{\omega_1^{(0)}(z)-\omega_1^{(0)}(\sigma_r(z))}
\tilde{\cal Q}^{(g)}_{n}(z,\sigma_r(z);\mathbf{z}_J) 
=  \sum_{r\in \mathbf r} \Res_{z\to r} \left(\int_{o}^z \omega_2^{(0)}(z_1,\cdot) \right)\omega^{(g)}_{n+1}(z,\mathbf z_J).
\eeq
Now, observe that $\omega^{(g)}_{n+1}(z,\mathbf z_J)$ has poles only at the ramification points, whereas $\omega_2^{(0)}(z,z_1)$ has a pole only at $z=z_1$ (a double pole). We may move the integration contours from surrounding the poles of $\omega^{(g)}_{n+1}(z,\mathbf z_J)$ to surrounding the poles of $\omega_2^{(0)}(z,z_1)$, i.e. using the Riemann bilinear identity:
\bea
& & \frac{1}{2} \sum_{r\in \mathbf r} \Res_{z\to r} \frac{\int_{\sigma_r(z)}^z \omega_2^{(0)}(z_1,\cdot)}{\omega_1^{(0)}(z)-\omega_1^{(0)}(\sigma_r(z))}
\tilde{\cal Q}^{(g)}_{n}(z,\sigma_r(z);\mathbf{z}_J) \\
&=&  - \Res_{z\to z_1} \left(\int_{o}^z \omega_2^{(0)}(z_1,\cdot) \right)\omega^{(g)}_{n+1}(z,\mathbf z_J) \nonumber \\
& & + \frac{1}{2{\rm i}\pi}\sum_{i=1}^{\mathfrak{g}} \left\{\Big(\oint_{{\cal A}_i} \omega^{(0)}_2(z_1,\cdot)\Big)\Big(\oint_{{\cal B}_i} \omega^{(0)}_{n+1}(\cdot,\mathbf z_J)\Big)  - \Big( \oint_{{\cal B}_i} \omega^{(0)}_2(z_1,\cdot) \Big)\Big(\oint_{{\cal A}_i} \omega^{(0)}_{n+1}(\cdot,\mathbf z_J)\Big)\right\}, \nonumber
\eea
where the cycles ${\cal A}_i, {\cal B}_j$ are chosen to form a basis of $2\mathfrak{g}$ non-contractible cycles on $\mathcal{S}^{[0]}$, with canonical intersections $ {\cal A}_i\cap {\cal B}_j = \delta_{i,j}$. Observe that $\left(\int_{o}^z \omega_2^{(0)}(z_1,\cdot) \right)$ has a simple pole at $z_1 = z$ with residue $1$, so the first term is:
\beq
 - \Res_{z\to z_1} \left(\int_{o}^z \omega_2^{(0)}(z_1,\cdot) \right)\omega^{(g)}_{n+1}(z,\mathbf z_J)  = \omega_{n + 1}^{(g)}(z_1,\mathbf{z}_J).
 \eeq
Since $\omega_2^{(0)}\in {\cal B}({\cal S}^{[0]})$ (from Corollary \ref{cor43}), we also know that $\oint_{{\cal A}_i} \omega^{(0)}_2(z_1,\cdot)$ and $\oint_{{\cal B}_i} \omega^{(0)}_2(z_1,\cdot)$ are holomorphic forms of $z_1$, and thus we have obtained the decomposition:
\beq
 \frac{1}{2} \sum_{r\in \mathbf r} \Res_{z\to r} \frac{\int_{\sigma_r(z)}^z \omega_2^{(0)}(z_1,\cdot)}{\omega_1^{(0)}(z)-\omega_1^{(0)}(\sigma_r(z))}
\tilde{\cal Q}^{(g)}_{n}(z,\sigma_r(z);\mathbf{z}_J) 
=  \omega^{(g)}_{n+1}(z_1,\mathbf z_J)  + {\rm holomorphic\,}(z_1).
\eeq
This finishes the proof of Theorem~\ref{topro}. \hfill $\Box$



\newpage


\newcommand{\etalchar}[1]{$^{#1}$}
\providecommand{\bysame}{\leavevmode\hbox to3em{\hrulefill}\thinspace}
\providecommand{\MR}{\relax\ifhmode\unskip\space\fi MR }
\providecommand{\MRhref}[2]{%
  \href{http://www.ams.org/mathscinet-getitem?mr=#1}{#2}
}
\providecommand{\href}[2]{#2}

\end{document}